\documentclass[11pt]{article}
\usepackage[T1]{fontenc}
\usepackage{lmodern,microtype,amsmath,amssymb,amsthm,mathtools}
\usepackage[margin=.88in]{geometry}
\usepackage[colorlinks=true,allcolors=blue]{hyperref}
\usepackage{enumitem}
\setlist{itemsep=4pt,topsep=5pt}
\newcommand{\E}{\mathbb E}
\newcommand{\Prb}{\mathbb P}
\newcommand{\one}{\mathbf 1}

\newcommand{\unif}{\mathbf u}
\newcommand{\rv}[1]{\boldsymbol{#1}}
\newcommand{\rG}{\rv{G}}
\newcommand{\emm}{\mathrm{e}}
\newcommand{\rGhat}{\widehat{\rG}}
\newcommand{\rsigma}{\rv{\sigma}}
\newcommand{\rtau}{\rv{\tau}}
\newcommand{\re}{\rv{e}}
\newcommand{\reprime}{\rv{e}'}
\newcommand{\rB}{\rv{B}}
\newcommand{\rRoots}{\rv{\mathcal R}}
\newcommand{\rs}{\rv{s}}
\newcommand{\rS}{\mathbf{S}}

\newcommand{\rTime}{\rv{\mathcal T}}

\newcommand{\rTree}{\rv{T}}
\newcommand{\rz}{\rv{z}}
\newcommand{\rMarginal}{\rv{\mu}}
\newcommand{\rLike}{\rv{\Lambda}}
\newcommand{\rDensity}{\rv{Z}}

\newcommand{\rVolume}{\rv{V}}
\newcommand{\rLikelihood}{\rv{\Lambda}}

\newcommand{\exc}{{\mathtt{exc}}}
\newcommand{\cG}{\mathcal{G}}

\usepackage[color]{changebar}
\makeatletter
\newif\if@inbar
\@inbarfalse
\newcommand{\@currentbarcolor}{none}

\newcommand{\startmarkcolor}[1]{%
  \if@inbar
    \def\@tempa{#1}%
    \ifx\@tempa\@currentbarcolor\relax
    \else
      \cbend
      \cbcolor{#1}%
      \def\@currentbarcolor{#1}%
      \cbstart
    \fi
  \else
    \cbcolor{#1}%
    \def\@currentbarcolor{#1}%
    \cbstart
    \@inbartrue
  \fi
}

\newcommand{\stopmark}{%
  \if@inbar
    \cbend
    \@inbarfalse
    \def\@currentbarcolor{none}%
  \fi
}

\newtheorem{theorem}{Theorem}[section]
\newtheorem{lemma}[theorem]{Lemma}

\newtheorem*{thmer}{Theorem~\ref{th:er}}
\newtheorem*{lempsimonotonicity}{Lemma~\ref{lem:psimonotonicity}}
\newtheorem*{leminputs}{Lemma~\ref{lem:inputs}}
\newtheorem*{lemtamemarginals}{Lemma~\ref{lem:tamemarginals}}
\newtheorem*{lemforestpair}{Lemma~\ref{lem:forestpair}}
\newtheorem*{lemrecursion}{Lemma~\ref{lem:recursion}}
\theoremstyle{definition}
\newtheorem{definition}[theorem]{Definition}
\newtheorem{algorithm}[theorem]{Algorithm}
\newtheorem{remark}[theorem]{Remark}
\title{Sampling colourings of $G(n,d/n)$ down to non-reconstruction}
\author{Andreas Galanis\thanks{Department of Computer Science, University of Oxford.
  Email: \texttt{andreas.galanis@cs.ox.ac.uk}.}
  \and Xusheng Zhang\thanks{Department of Computer Science, Durham University.
  Email: \texttt{xusheng.zhang@durham.ac.uk}.}}
\date{30 November 2026}

\begin{document}
\maketitle
\vspace{-1.5em}
\setcounter{tocdepth}{1}
\begin{abstract}
\begin{NoHyper}
\let\thefootnote\relax\footnotetext{For the purpose of Open Access, the authors have applied a CC BY copyright licence to any Author Accepted Manuscript version arising from this submission. All data is provided in full in the results section of this paper.}
\end{NoHyper}
A tantalising conjecture for many canonical random CSPs  is that sampling solutions uniformly is as easy as finding a solution. For the search problem in particular, there has been a long line of works connecting tractability with  the non-reconstruction threshold; yet,  sampling algorithms that match the non-reconstruction threshold have remained elusive.

We resolve this for proper colourings of $G(n,d/n)$. For all sufficiently
large  $d$ and any integer $q\ge d/\log d$, we give a randomised
algorithm for nearly uniform sampling of $q$-colourings of $G(n,d/n)$, matching the best known condition on $q$
for finding a single proper colouring. With high probability over $G$, the algorithm returns a colouring
in expected time at most $n^{8/5}$ whose distribution is within
$n^{-\omega(1)}$ in total variation of the uniform distribution over colourings of $G$.

The algorithm inserts edges in a random order and repairs the colouring
in a local neighbourhood utilising the tree-like structure.  The main difficulty is that
local repairs introduce bias and the weights correcting this bias
can accumulate large fluctuations. To bound the latter, for each edge insertion we simultaneously update a ``helper'' pair of
vertices to obtain an exact weighted update whose mean weight over the edges is independent of the resampling neighbourhood. To control
the accumulated weight, we study the marginals at the endpoints
of the inserted edge and the helper pair. Starting from Sly's
non-reconstruction results, we derive higher-order
moment bounds that strengthen concentration of the root marginal on Poisson trees. We then transfer these bounds to graph
neighbourhoods containing polynomially many excess edges (beyond the tree) using new stability estimates.
Together with partition function estimates of Bapst, Coja-Oghlan
and Efthymiou, these bounds allow us to control the accumulated weight and make
the final rejection step efficient.
\end{abstract}

\section{Introduction}

A central question for random constraint satisfaction problems is whether
sampling a solution uniformly can be as easy as finding one. A classical
example is the problem of colouring the sparse random graph $G(n,d/n)$,
where each pair of vertices is joined independently with probability
$d/n$. For any fixed $\eta>0$ and sufficiently large fixed $d$, a simple
greedy algorithm finds a proper $q$-colouring once
$q\ge(1+\eta)d/\log d$~\cite{AM}. Such an algorithm, however, gives little
control over the distribution of the colouring it produces. Despite
substantial progress in sampling algorithms, the best known guarantees
for nearly uniform sampling on $G(n,d/n)$ require roughly $d$ colours,
leaving a logarithmic gap between search and sampling.

While the sampling threshold $q\approx d$ is the right answer for worst-case settings, its significance for sampling on average-case instances is far less clear and going beyond that has been a major open problem. This throttle on the range of $q$ is in fact present in various random constraint satisfaction problems, where the best known sampling algorithms have their origins in worst-case analysis settings. Part of the difficulty of lowering $q$ is that even adapting these algorithms to the average-case setting still requires highly non-trivial work to account for the presence of atypical structures (which are not present for example on graph classes with more explicit structure, e.g., graphs of max degree $d$). But there is another source of difficulty: after the worst-case threshold, subtle phenomena emerge that typically impose a conceptual bottleneck to the applicability of such algorithms.

To explain the difficulty in lowering the value of $q$ in the case of colourings, it is useful to consider the case of sampling colourings of (worst-case) graphs of max-degree $d$, where it is conjectured that  the sampling threshold is $q\geq d+1$ for any $d\geq 3$. There has been startling progress in recent years towards this conjecture by trying to analyse the mixing time of Glauber dynamics, see \cite{Jerrum,SS,Vigoda,DF03,GMP05,DFHV13,CDMPP} for some early approaches. Recent progress includes spectral independence and entropy
factorisation~\cite{CGSV,FGYZ,CLV,BCCPSV}, coupling and
local-uniformity arguments~\cite{CarlsonVigoda,CF25,JMV26}, and
correlation-decay and local-to-global
methods~\cite{CLMM,CL26,LO26,Yang26}. On the other hand, once the number of colours satisfies $2<q<d$, Glauber dynamics can be exponentially slow and, in fact, \cite{GSV} showed that, for even $q<d$, sampling colourings is NP-hard on triangle-free $d$-regular graphs.\footnote{Such graphs can be coloured efficiently with $O(d/\log d)$ colours \cite{Johansson96,Molloy19}.} The threshold $q\geq d+1$ captures a key correlation-decay property on the $d$-regular tree, known as uniqueness: fix an arbitrary colouring at the vertices of height $h$ that is compatible with the tree; does the marginal at the root converge to the vector $\unif=(1/q,\hdots,1/q)$, uniformly as $h\rightarrow \infty$? For colourings, uniqueness holds precisely when
$q\ge d+1$~\cite{Jonasson}. More generally, uniqueness is a correlation decay property that underpins the existence of sampling algorithms in various worst-case settings.

From this picture on graphs of max degree $d$, it should be clear now why going below $q\approx d$ presents a conceptual
obstacle: aside from handling atypical structures in $G(n,d/n)$ such as large degree vertices, the bigger problem is that we can no longer expect to control the influence of \emph{every} boundary condition, making the applicability of Glauber analysis tools very hard. Nevertheless, the  conjecture for $G(n,d/n)$ is that the appropriate correlation-decay condition for sampling is captured by typical boundaries rather than worst-case, a property  known as non-reconstruction (see~\cite{MosselSurvey}). The relevant tree for $G(n,d/n)$ is a Galton--Watson tree with Poisson$(d)$ offspring distribution. Roughly, for non-reconstruction, we ask that the convergence of the root's marginal to uniform holds with high probability over the tree and the choice of the random colouring, see Section~\ref{sec:tree} and Lemma~\ref{lem:sly} for precise definitions. The main distinction from uniqueness is that the boundary is
sampled from a typical colouring of the tree, rather than chosen arbitrarily.
Sly's seminal results~\cite{Sly} locate the transition at
$q=(1+o_d(1))\frac{d}{\log d}$; we will in fact revisit his analysis in Section~\ref{sec:tree} (Lemma~\ref{lem:moments}) and obtain stronger quantitative bounds for the decay that
will be important in our setting.

The non-reconstruction threshold already plays a central role in 
the search problem, i.e., finding a colouring. The chromatic number of $G(n,d/n)$ has been studied
extensively and is close to $d/(2\log d)$ for large $d$~\cite{AN,CV}.
Thus there is a factor-two gap between the existence
of colourings and the range where polynomial-time algorithms are known
to find them, and the latter range agrees with the non-reconstruction
threshold (to leading order). Despite the simplicity of the algorithms
achieving the bound $q\geq d/\log d$, improving the condition on $q$ has proved
remarkably difficult. The shattering results of Achlioptas and
Coja-Oghlan~\cite{AC} provide a structural explanation: for $q$ between $d/\log d$ and $d/(2\log d)$ almost all colourings lie in exponentially many groups,
each containing an exponentially small fraction of the colourings,
with distinct groups separated by linear Hamming distance. This
separation motivates the conjecture that the factor-two gap is
unavoidable for poly-time algorithms, although a general
computational lower bound remains open.

On a more positive note, non-reconstruction also holds on the random graph for $q\geq (1+o_d(1))\frac{d}{\log d}$~\cite{MRT}. Here one draws a
uniform proper colouring and considers the conditional colour
distribution at a typical vertex after revealing the colours outside
its neighbourhood of radius $\ell$.  Thus the correlation decay suggested by the tree remains
available on the graph itself, and it is therefore conceivable that
efficient sampling is possible throughout this range, despite the fact
that the best known sampling algorithms have so far remained at $q\approx d$.

Our main contribution is to resolve this question for proper colourings of $G(n,d/n)$. For
sufficiently large $d$, our algorithm samples nearly uniform
$q$-colourings for every integer $q\ge d/\log d$, matching the
regime for finding a colouring. To state our result more precisely, for a $q$-colourable graph $G$, let $\mu_G$ be the
uniform distribution on its proper $q$-colourings. An
\emph{$\varepsilon$-sample from $\mu_G$} is a random proper colouring
whose distribution is within total variation distance $\varepsilon$
of $\mu_G$. Throughout, $d$ and $q$ are fixed, and \emph{with high
probability} (whp) refers to a probability tending to one as
$n\to\infty$ over the input graph. For a fixed input $G$, the
expectation and tail probability for the running time $\rTime_G$ are
over the algorithm's random choices.

\newcommand{\statethmer}[1][]{
Let $d$ be sufficiently large and $q\ge d/\log d$ be an integer.
There is a randomised algorithm $\mathcal{A}$ such that the following
holds whp over $G\sim G(n,d/n)$ with $\varepsilon=n^{-\omega(1)}$.#1

The algorithm $\mathcal{A}$ on input $G$ returns an
$\varepsilon$-sample from $\mu_G$, and its running time $\rTime_G$
satisfies $\E[\rTime_G]\le n^{8/5}$ and
$\Prb(\rTime_G>n^{8/5})=n^{-\Omega(\log n)}$.
}
\begin{theorem}\label{th:er}
\statethmer[\footnote{\label{fn:omega}Here and elsewhere, $\omega(1)$ denotes
a deterministic function $g(n)\to\infty$, which may depend only on
$d,q$.}]
\end{theorem}

The proof of Theorem~\ref{th:er} is given in Section~\ref{sec:mainproof}.

\begin{remark}\label{rem:runtime}
The running time in Theorem~\ref{th:er} can be sharpened to $n^{3/2+\delta}$ in place of $n^{8/5}$ for every fixed $\delta>0$,  provided $d$ is sufficiently large depending on $\delta$.
We use the exponent $8/5$ to keep the parameters fixed throughout the
proof; see also the running-time discussion in Section~\ref{sec:overview}.
\end{remark}

\begin{remark}\label{rem:nae}
Our approach can be adapted to sampling for random
NAE-$k$-SAT, where in a satisfying  assignment every clause must contain both a satisfied and an
unsatisfied literal. This applies for sufficiently large fixed $k$ and
any fixed density $\alpha<\alpha_{\mathrm{rec}}(k)$, where
$\alpha_{\mathrm{rec}}(k)=(1+o_k(1))2^{k-1}\log k/k$
is the reconstruction threshold. We outline the required modifications
in Section~\ref{sec:extension-other-models}, where we also discuss the applicability of our approach to other symmetric models.
\end{remark}

\subsection{Further related work and our approach}
Before discussing our approach in more detail, we discuss further related work on non-reconstruction and sampling graph colourings.

\paragraph{Non-reconstruction and random CSPs.}
Non-reconstruction and related correlation-decay properties play a central
role in studying random CSPs. Gerschenfeld and
Montanari~\cite{GM} gave conditions under which reconstruction on a random
graph is equivalent to reconstruction on its limiting tree. This
connection was developed further in~\cite{MRT}, with subsequent results for random graph
colourings~\cite{CEJ} and broader families of random CSPs~\cite{CKM}.
Related correlation-decay
estimates justify local descriptions of marginals and estimates for
partition functions~\cite{CP,CKPZ}, and have also been used to analyse
metastability regimes~\cite{CGGRSV}.

More directly relevant to our analysis are the non-reconstruction
results for colourings on trees due to Sly~\cite{Sly} (see also~\cite{EfthGW} for related results). As we explain later, for the algorithm we require stronger concentration of the conditional root
marginal. On regular trees, there is an established approach to
obtaining doubly exponentially small tail probabilities in the depth, developed
by Martinelli, Sinclair and Weitz~\cite{MSW04,MSW07} and subsequently
by Sly and Zhang~\cite{SZ}. Applying this approach to Poisson trees
has some complications because the sizes and branching patterns of the
neighbourhoods fluctuate. We instead use higher moments, which yield
the required concentration and allow us to approach the decay rate
of the associated tree recursion.

\paragraph{Glauber dynamics on random graphs.}
For sampling colourings, the standard approach is to use Glauber dynamics, which iteratively updates the colour of a uniformly chosen vertex. A central difficulty that arises for $G(n,d/n)$ is that the maximum degree
grows with $n$, so available mixing-time bounds for bounded-degree graphs do not give a sampling algorithm for
constant $q$. Nevertheless, Mossel and Sly~\cite{MSG10} proved polynomial mixing time on $G(n,d/n)$
with high probability for a sufficiently large number of colours
depending only on $d$. Efthymiou~\cite{EfthMCMC} reduced the colour
requirement to a constant multiple of $d$, and Efthymiou, Hayes,
\v{S}tefankovi\v{c} and Vigoda~\cite{EHSV} subsequently proved
polynomial mixing for $q\ge(\alpha+\eta)d$, where
$\alpha\approx1.763$ is the solution to $\alpha=\exp(1/\alpha)$ and
$\eta>0$ is fixed. Recent results of Li and Yang~\cite{LY26} and
Schiller~\cite{Schiller26} bring the colour requirement down to
$(1+o(1))d$ as $d\to\infty$. 

These results approach the uniqueness benchmark but leave a logarithmic
gap to non-reconstruction. At $q\approx d/\log d$, even an average-degree vertex can have
neighbours using every colour other than its current one, leaving no
alternative colour available to a single-vertex update. Its neighbours
must then be recoloured before it can change its colour. Non-reconstruction does
not rule out this local obstruction, and working around this on random graphs has proved to be quite challenging. On regular trees,
Sly and Zhang~\cite{SZ}, building on earlier work in
\cite{MSW07,BVVW}, proved rapid mixing throughout the known non-reconstruction
regime. However, transferring such conclusions from trees to sparse random
graphs remains a substantial difficulty; successful transfers of tree
estimates are known in certain uniqueness and low-temperature
regimes of the ferromagnetic Ising, Potts and random-cluster
models~\cite{MS,BG23,BGPZ27,CGGMSZ,BGPZ26}. These arguments require
additional structural or spatial-mixing hypotheses, and often exploit
monotonicity present in the ferromagnetic setting, either directly or through the random-cluster representation.

\paragraph{Samplers via edge insertions.}
Another line of work, initiated by Efthymiou~\cite{Efth12}, constructs a
sample by adding edges one at a time via local update rules. This is also
appealing because it resembles the ``cavity'' approach, which studies
how adding a constraint changes the distribution of solutions~\cite{CP,CKPZ}.
The starting graph is simple
enough to sample from directly, and each new edge is handled by
recolouring part of the current graph using the local update rule. Efthymiou~\cite{Efth} developed
this approach for $G(n,d/n)$ with $q\ge(1+\eta)d$, and it was later explored for the Potts and random-cluster models by 
Blanca, Galanis, Goldberg, \v{S}tefankovi\v{c}, Vigoda and
Yang~\cite{BGGSVY}. Later work of
Efthymiou~\cite{EfthSym} extended the approach to a broader family of
random constraint problems and further refined the treatment of
cycles. 

These algorithms are frequently referred to as ``weak'' samplers because the TV-error guarantee is bounded by $n^{-c}$ for some fixed $c>0$ depending
on the model parameters, rather than by an arbitrary prescribed
inverse power of $n$. A difficulty that arises with the local update rules is
that cycles create dependencies which prevent the updates from
preserving the target distribution exactly. The analysis controls
the resulting approximation error over all insertions, but the exponent
in this error bound is fixed by the model parameters.

\paragraph{Overview of our approach.} Our algorithm refines the ``weak''-sampler approach in two directions, first reducing the
number of colours down to $d/\log d$ and then suppressing the error down to $n^{-\omega(1)}$.
For an input colouring, we introduce an exact weighted edge insertion using an auxiliary pair of vertices that we refer to as a helper; we correct the bias of
each local repair using a so-called score for the insertion. The helper ensures that, conditioned on the input colouring, the mean
score, averaged over the inserted pair and the helper pair, does
not depend on the neighbourhood used for resampling. The product of
these scores corrects the bias over the entire sequence of insertions,
and a final rejection step turns the weighted construction into a
sampler.

The main challenge is to control fluctuations of the accumulated
score. We relate the weighted evolution to a planted graph model, in
which a colouring is chosen first and edges are sampled between
vertices of different colours. Starting from Sly's
non-reconstruction estimates~\cite{Sly}, we derive higher moment
bounds for the conditional colour distribution at a tree root and
stability estimates that transfer these bounds to graph neighbourhoods
with potentially polynomially many edges beyond a spanning forest.
Together with estimates for the number of colourings due to Bapst,
Coja-Oghlan and Efthymiou~\cite{BCE}, these bounds control the accumulated
score and make the rejection step efficient.

\vskip 0.4cm \noindent \textbf{Use of AI.} This work has been produced with the help of ChatGPT Astra~6.0, which suggested one of the main ideas to use a helper pair for the update rule. We have also used multiple rounds of  AI interaction for designing and reviewing the proofs.   The authors  take  responsibility for the presentation and validity of the results.

\subsection{Notations}

For a graph $G=(V,E)$, we write
$\Omega_G$ for the set of proper colourings, $Z_G=|\Omega_G|$, and $\mu_G$
for the uniform distribution on $\Omega_G$, when this set is nonempty.
We write a colouring on $G$ as $\sigma\in[q]^V$, with $\sigma_v$ the colour of
$v$.
Moreover, 
for a colouring $\sigma\in[q]^V$
and a subset $B\subseteq V$,
we
write $\mu_G^{B,\sigma}$ for the uniform distribution on
$
 \{\tau\in\Omega_G:\tau_{v}=\sigma_{v}, \forall v\in V\setminus B\},
$
whenever this set is nonempty. Thus $\mu_G^{B,\sigma}$ is the conditional
distribution of $\mu_G$ with the colours outside $B$ fixed to $\sigma$.

Throughout, $d$ and $q$
are fixed constants and we consider $n\rightarrow\infty$, assuming
$n$ is sufficiently large. Unless stated otherwise, constants implicit
in asymptotic notation may depend on $d$ and $q$.
Random objects are printed in bold, and their fixed values in ordinary type:
for example, when we want to make clear the randomness of $G$, we use the more explicit notation $\rG\sim G(n,d/n)$ to denote the random graph,
and we use $G$ for a random graph that has been drawn.

\subsection{Key proof ideas}\label{sec:overview}
Here we give an overview of the algorithm and explain the motivation behind the design with the helper pair, and some key ideas behind the analysis.

\paragraph{Local repairs and the need for the helper pair.}
Choose a uniform random order $e_1,\ldots,e_m$ of the input edges and set
\begin{equation}
    \label{eq: def of G_i}
 G_0=(V,\varnothing),\qquad
 G_t=(V,\{e_1,\ldots,e_t\})\quad(1\le t\le m).
\end{equation}
Thus $G_1,\ldots,G_m$ are the successive graphs and $G_m=G$.
At $G_0$, independent uniform colours from $[q]^V$ give an exact sample.
To understand one insertion, suppose the current graph is $G_t$ and
we have a colouring $\rsigma\sim\mu_{G_t}$. We process the next edge
$e=e_{t+1}=uv$, with desired distribution
\[
 \mu_{G_{t+1}}
 =\mu_{G_t}(\,\cdot\mid\sigma_u\ne\sigma_v).
\]
Let $b_e(\sigma)$ be the indicator that $e$ is bichromatic under
$\sigma$. A natural local update leaves the colouring unchanged when
$b_e(\sigma)=1$, and otherwise recolours a neighbourhood $B$ of $u,v$ so that $e$ becomes bichromatic.
Here $B$ depends only on $G_t,e$, and we write $\mu_{G_t}^{B,\sigma}$
for the uniform distribution on proper colourings that agree with
$\sigma$ outside $B$. The repair draws from
$\mu_{G_t}^{B,\sigma}(\,\cdot\mid b_e=1)$.

Such a repair does not by itself produce a uniform colouring of
$G_{t+1}$: conditioning on $b_e=1$ also reweights the colourings outside
$B$ by the factor $\mu_{G_t}^{B,\sigma}(b_e=1)$, while the repair leaves
those colours unchanged. Assigning this factor as a score to the
update, on both branches, corrects the bias exactly. Products of these scores therefore
give an exact weight correction along the edge-insertion process.
The difficulty is to control the accumulated weight. Let
$\mathcal U(G_t)$ be the set of unused vertex pairs, let
$U=|\mathcal U(G_t)|$, and write $B_e$ for the neighbourhood used for
edge $e$. For a fixed colouring $\sigma$, the mean score of this update rule when $e$ is
uniform in $\mathcal U(G_t)$ is
\begin{equation}\label{eq:dispaver}
 \frac1U\sum_{e\in\mathcal U(G_t)}
       \mu_{G_t}^{B_e,\sigma}(b_e=1).
\end{equation}
But is this indeed a good update rule? The first edge insertion illustrates the behaviour we would like to maintain throughout the execution of the algorithm. For the local update above, the score at the start is exactly $1-1/q$,
since the two endpoint colours are independent and uniform.
We would like subsequent insertions to maintain this simple picture:
the average score should reflect the probability that the next
uniformly chosen pair is bichromatic. Consider what happens after $t$ insertions, and let $\sigma$ denote the current colouring. The corresponding probability that the next chosen pair is bichromatic is given by 
$b=b(G_t,\sigma):=U^{-1}\sum_{e\in\mathcal U(G_t)}b_e(\sigma)$.  For $\sigma$ whose colour classes have roughly equal size and $t=O(n)$,
this fraction remains close to $1-1/q$.
This suggests designing the update so that its conditional mean
score tracks this empirical fraction exactly, in order  to avoid accumulating neighbourhood-dependent errors in the
mean scores.

However, the average in \eqref{eq:dispaver}  need not equal $b(G_t,\sigma)$,
since each summand uses a different conditional distribution.
Averaging over $\sigma\sim\mu_{G_t}$ does recover the expectation of
$b(G_t,\sigma)$, but controlling the product of the scores
requires control of the mean conditioned on the current colouring.
Non-reconstruction makes individual local probabilities close
to $1-1/q$, but does not directly give a sufficiently small bound
on the difference between~\eqref{eq:dispaver} and $b(G_t,\sigma)$.
These discrepancies can accumulate over the $O(n)$ insertions.

Our update introduces an additional unused pair $e'=u'v'$, chosen
uniformly among those with $b_{e'}(\sigma)\ne b_e(\sigma)$; we call it
a \emph{helper pair}. To repair the colouring locally, we use a neighbourhood
$B=B_{G_t,L}(e\cup e')$ (see  Section~\ref{sec:block}), which depends
only on the graph and the union of the endpoints, so exchanging $e$
and $e'$ leaves $B$ unchanged. If $e$ is monochromatic, we redraw from
$\mu_{G_t}^{B,\sigma}(\,\cdot\mid b_e=1,b_{e'}=0)$, exchanging the
two statuses; otherwise we keep the colouring. The helper is not
inserted into the graph.

The update returns a score $s^{B,\sigma}_{e,e'}\in[0,1]$, computed
from the two local probabilities
$\mu_{G_t}^{B,\sigma}(b_e=1,b_{e'}=0)$ and
$\mu_{G_t}^{B,\sigma}(b_e=0,b_{e'}=1)$
(Definition~\ref{def:score}). Lemma~\ref{lem:exact} gives two
properties. First, for a fixed inserted edge $e$, if
$\rsigma\sim\mu_{G_t}$, then the expectation of the score multiplied by the indicator
that the output is $\tau$ equals $\mu_{G_t}(\tau)b_e(\tau)$.
Thus the scores again correct the bias exactly. Second, their mean
given \emph{any} fixed proper colouring $\sigma$ is exactly $b(G_t,\sigma)$
when $e$ is uniform among unused pairs. This is detailed in Lemma~\ref{lem:exact}(ii), but here is a quick way to see the cancellation:
fix $e,e'$ with $b_e(\sigma)=1$ and $b_{e'}(\sigma)=0$, and
let $B=B_{G_t,L}(e\cup e')$. Set
\[
 x=\min\left\{
 \tfrac{\mu_{G_t}^{B,\sigma}(b_e=0,b_{e'}=1)}
      {\mu_{G_t}^{B,\sigma}(b_e=1,b_{e'}=0)},1\right\}.
\]
The two scores are $s^{B,\sigma}_{e,e'}=1-(1-b)x$ and
$s^{B,\sigma}_{e',e}=bx$. The two orientations of this pair are chosen
with probabilities $1/((1-b)U^2)$ and $1/(bU^2)$, respectively.
Consequently their contributions to the mean score minus $b$ cancel:
\[
 \frac{s^{B,\sigma}_{e,e'}-b}{(1-b)U^2}
 +\frac{s^{B,\sigma}_{e',e}-b}{bU^2}
 =\frac{1-x}{U^2}-\frac{1-x}{U^2}=0.
\]
Summing over opposite-status pairs gives mean $b$ exactly. The crucial
point is that the same local probability ratio occurs in both scores,
making the cancellation possible.

 To use the mean identity, we analyse
the algorithm jointly with the random input graph, conditional on its
number of edges. When only the current prefix has been exposed, the
next edge is uniform among all unused pairs. Consequently, the product
of the normalised scores is a nonnegative supermartingale under this
joint experiment (Lemma~\ref{lem:pathlaw}). The exact weighted identity,
in contrast, holds for each fixed input graph and edge order; the
algorithm's intermediate colourings need not themselves be uniform.

\paragraph{Rejection and the loss bound.}
For convenience, we normalise each score by $\rho=\frac{n}{n-1}(1-\frac1q)$, which
satisfies $b(G_t,\sigma)\le\rho$. Let $\rS_G$ be the product of the normalised scores in one trial on
$G$. Algorithm~\ref{alg:sampler} accepts the final colouring with
probability $\min\{\rS_G,T\}/T$, for a threshold $T=n^{o(1)}$, and
otherwise starts a fresh trial. The exact identity gives
$\E[\rS_G]=\hat Z_G$, where $\hat Z_G=Z_G/(q^n\rho^{|E(G)|})$,
and weighting the final colouring by $\rS_G/\hat Z_G$ gives precisely
$\mu_G$. Thus the error introduced by the rejection step is controlled
by the loss
\[
 \delta_G=\E[\rS_G-\min\{\rS_G,T\}].
\]
Lemma~\ref{lem:randomgrapherror} shows that, with high probability over
$G$, this loss is $n^{-\omega(1)}$ and $\hat Z_G\ge1/\log n$.
These estimates give the claimed sampling error and an acceptance
probability $n^{-o(1)}$. Section~\ref{sec:mainproof} deduces the main
theorem from them and the neighbourhood-size bound in
Lemma~\ref{lem:neighbourhoods}, using standard tree dynamic programming
to implement each update.

The proof of the loss bound in Section~\ref{sec:error} separates trials
according to the fluctuations of their individual scores. More
precisely, $\mathcal V(G_t,\sigma)$ is the conditional second moment of
the next score around $\rho$, as defined in~\eqref{eq:V}.
On trials where $\mathcal V(G_t,\rsigma_t)\le\sqrt{\log n}/n$ at
every step, the mean-score identity and boundedness of the scores
control growing moments of their product, making its contribution to
the loss $n^{-\omega(1)}$. Lemma~\ref{lem:inputs} shows that trials
violating this bound contribute only $n^{-\omega(1)}$ to the expected
final weight. These bounds first average over the graph and the algorithm's
choices; Markov's inequality then gives the loss bound for typical
inputs $G$. It remains to explain the local estimates that prove
Lemma~\ref{lem:inputs}, which is the purpose of Section~\ref{sec:path}.

\paragraph{The planted model and local marginals.}
To control the variance of the score at a given step, we first need a tractable description of the graph and colouring during the execution of the algorithm. We use a \emph{planted model} $\rGhat_t(\sigma)$ where for a fixed colouring $\sigma$ with at
least $t$ bichromatic pairs, we choose a uniformly random set of $t$ such
pairs.  Lemma~\ref{lem:pathlaw} transfers estimates
from this planted model back to the algorithm, using the exact weighted identity and a supermartingale property (that comes from the expectation of the mean score that we discussed in detail earlier). We can therefore establish the needed
local estimates with $\sigma$ fixed and only the planted graph random.

The local score estimate is expressed directly through the marginal
laws $\mu_{G_t,v}^{B,\sigma}$ where $v$ ranges over the endpoints of the inserted edge and the helper pair. We perform in Section~\ref{sec:block} a joint exploration starting from these endpoints and going down to depth
$L=\lfloor\frac{11}{20}\log_{d+3}n\rfloor$. Its discovery edges form
a forest; the additional edges recorded by the exploration are called
\emph{excess edges}. We exclude their endpoints and all descendants
from $B$, and keep the colour of the vertices at depth $L$ fixed at that prescribed by $\sigma$. Each
exploration\footnote{Exploration processes starting from roots incident to excess edges are pruned and the corresponding endpoints have their colour fixed to that prescribed by $\sigma$.} stemming from a root vertex $v$ then lies in a separate component of
$G_t[B]$. Hence, for four distinct endpoints
of $e,e'$, their colours are independent under $\mu_{G_t}^{B,\sigma}$.
Writing $\unif=(1/q,\ldots,1/q)$ for the uniform vector, this gives the bichromatic-edge probability bias
\[
 \mu_{G_t}^{B,\sigma}(b_e=1)-(1-1/q)
 =-\langle\mu_{G_t,u}^{B,\sigma}-\unif,
          \mu_{G_t,v}^{B,\sigma}-\unif\rangle.
\]
Thus the deviation is quadratic in the marginal errors.
Lemma~\ref{lem:localreduction} uses this to show that
\[
 \bigl(s^{B,\sigma}_{e,e'}-s^{B,\sigma}_{e',e}\bigr)^2
 \le128q^2\sum_{v\in e\cup e'}
       \|\mu_{G_t,v}^{B,\sigma}-\unif\|_2^4.
\]
By Lemma~\ref{lem:exact}(ii), the variance of the score is an average
of these squared differences. With $O(n)$ insertions, our target is a
variance of order $1/n$, up to logarithmic factors. A convenient
target for the individual marginal errors is therefore
$n^{-1/4-\delta}$ for some $\delta>0$: this gives a score difference
of size $O(n^{-1/2-2\delta})$ and a squared difference of size
$O(n^{-1-4\delta})$. Thus it is the fourth power of the distance
from uniform that must be smaller than $1/n$.
Pairs sharing an endpoint contribute only $O(1/n)$ to this average.

\paragraph{Tree estimates and the exploration depth.}
The remaining estimates are organised into two lemmas. The
deterministic Lemma~\ref{lem:tamemarginals} bounds the actual marginals
using properties of the full discovery trees, where only colours at
depth $L$ are fixed and excess edges are ignored. These properties,
called tame marginals and bounded early excess in
Definition~\ref{def:tamemarginals}, hold for all but a $(1/n)$-fraction
of four-element root sets in a typical planted graph with failure probability that can be made $n^{-\omega(1)}$ (cf.
Lemma~\ref{lem:forestpair}).
Its proof compares the discovery forests in the planted graph with
Poisson broadcast trees, while controlling the excess edges separately.
Sections~\ref{sec:marginalproof} and~\ref{sec:forest} prove these two lemmas, respectively.

For the correlation decay estimates, we start from Sly's non-reconstruction
theorem~\cite{Sly} for Poisson trees (Lemma~\ref{lem:sly}). Near uniformity,
the expected squared distance of the root marginal from uniform decreases by a factor $\approx d/(q-1)^2$ at each level on a Poisson$(d)$ tree, and Sly derives an analogous second-moment decay on regular trees. We use the larger, convenient
rate $\zeta=D^{-19/20}$ with $D=d+3$, valid for sufficiently large $d$.
In Section~\ref{sec:tree}, we derive the higher-moment bounds of
Lemma~\ref{lem:moments}. For a Poisson tree with mean offspring at
most $D$, let $\widehat\mu$ be the marginal at the root given a
broadcast colouring at depth $L$. For every fixed even $r\ge2$,
the lemma gives
\[
 \E\|\widehat\mu-\unif\|_2^r\le C\zeta^{rL/2},
 \qquad \zeta=D^{-19/20}.
\]
The case $r=4$ gives $C\zeta^{2L}$, precisely the scale we want
smaller than $1/n$ in the squared-score bound above.
At depth $\frac12\log_D n$, this estimate
only gives $O(n^{-19/20})$, which is too weak. Our choice
$L=\lfloor\frac{11}{20}\log_D n\rfloor$ instead gives
$\zeta^{2L}=O(n^{-209/200})$, leaving a polynomial margin below
$1/n$. Even the sharper coefficient $D/(q-1)^2$ is larger than $1/D$
near $q=d/\log d$, so it also requires a depth slightly beyond
$\frac12\log_D n$ for this fourth-moment argument.

Higher moments have a separate role: they make failures of the
marginal bounds sufficiently rare. At our depth, Markov's inequality
gives
\[
 \Prb\bigl(\|\widehat\mu-\unif\|_2>n^{-0.26}\bigr)
 \le C n^{-r/800}.
\]
The fourth moment alone would give only $O(n^{-1/200})$, whereas
taking $r$ sufficiently large gives any prescribed inverse-polynomial
failure probability, as needed in Lemma~\ref{lem:forestpair} and
ultimately for the $n^{-\omega(1)}$ sampling error.

This depth creates a difficulty: an exploration reaching about
$\sqrt n$ vertices is already at the scale where collisions become
likely. Beyond it, the explored neighbourhood can contain polynomially
many excess edges. We therefore cannot discard every exploration
containing a cycle. Instead, we prune the excess endpoints and control
the effect of fixing their colours. This requires a worst-case bound
for the additional constraints, although the original boundary at
depth $L$ is sampled from the broadcast process.

The estimate in Lemma~\ref{lem:sparsecuts} gives exponential decay of
the effect of a bounded number of arbitrary compatible constraints,
provided the child marginals are sufficiently close to uniform along
the intervening tree. Here the gap between the early-cut depth
$h=\lfloor\frac9{20}\log_D n\rfloor$ and the exploration depth $L$
is useful: the fixed depth-$L$ boundary is at least
$L-h=\frac1{10}\log_D n+O(1)$ generations away from vertices at depth
at most $h$. Non-reconstruction over this remaining depth
makes the sums of child-marginal biases small, as required by the cut
estimate. Controlling just the marginals at the exploration roots
would not supply this hypothesis.

The proof of Lemma~\ref{lem:tamemarginals} first handles the many
deep cuts and verifies that the child-marginal bounds survive them,
then handles the bounded number of early cuts. With
$\lambda=D^{-19/20}$, the combined root error from the deep cuts is
at most $O(n^{1/8}\lambda^h)=O(n^{-0.3025})$, even allowing
$n^{1/8}$ excess edges. Early cuts can have a larger effect, which we need to keep track of as a sum of terms $\lambda^{\operatorname{depth}(f)}$.
Together with the tree error, these estimates give a marginal bound
$n^{-0.255}$ plus the early-cut contribution. The precise fractions
in $L$ and $h$ leave room both for the required fourth-moment decay
and for applying the cut bounds before the distant boundary is reached.

Averaging the fourth powers of these marginal bounds leads to a certain
graph statistic $\Psi(G)$, defined in~\eqref{eq:graphgeometry}, which averages
$\sum_{f\in\exc(B)}\lambda^{4\operatorname{depth}(f)}$ over all
four-element root sets. The fourth powers arise from the squared-score
bound, so $\Psi(G)$ controls the contribution of the early cuts
to the score variance.
Lemma~\ref{lem:tamemarginals} then gives
$\mathcal V(G,\sigma)\le C(1/n+\Psi(G))$ for a balanced colouring
whenever all but a $1/n$ fraction of root sets satisfy the required
properties. Lemma~\ref{lem:countlower} bounds
$\Psi(G)\le(\log n)^{1/4}/n$ on typical inputs, and the monotonicity
in Lemma~\ref{lem:psimonotonicity} gives the same bound for every
prefix. Together with the planted transfer and
Lemma~\ref{lem:forestpair}, this proves Lemma~\ref{lem:inputs} and
completes the loss estimate. The graph bounds and monotonicity of $\Psi(G)$ are
proved in Appendix~\ref{sec:countproof}; the appendix also obtains for completeness
the required lower bound on $\hat Z_G$ from the partition function
estimates of Bapst, Coja-Oghlan and Efthymiou~\cite{BCE}.

\paragraph{Running time.}
The running time is mainly determined by the size of the components in the exploration process (and hence the depth $L$) and the number of insertions. In $G(n,d/n)$, the expected number of vertices encountered in an exploration of depth $L$ from an arbitrary vertex is  $O(d^L)$, so with $L\approx \tfrac{11}{20} \log_d n$,  the volume of each 
exploration is bounded by $n^{7/12}$ vertices on typical $G$ (cf. Lemma~\ref{lem:neighbourhoods}). Tree dynamic
programming then implements each local update in
$n^{7/12}(\log n)^{O(1)}$ time. With $O(n)$ insertions per trial and at
most $2T\log n=n^{o(1)}$ trials in expectation, the expected running
time is $n^{19/12+o(1)}\le n^{8/5}$. Section~\ref{sec:mainproof} also
shows that the probability of exceeding $n^{8/5}$ is
$n^{-\Omega(\log n)}$.

We briefly sketch here how to obtain the running-time refinement of $n^{3/2+\delta}$ for any sufficiently small $\delta>0$ which was pointed out in Remark~\ref{rem:runtime}. In particular, take $\eta=\delta/3$ and get the resampling neighbourhoods by exploring at depth $L=\lfloor(\tfrac12+\eta)\log_D n\rfloor$, while setting the early cut-depth at $h=\lfloor(\tfrac12-\eta)\log_D n\rfloor$. For sufficiently large $d$, we have
$2D/(q-1)^2<D^{-1+\eta}$, so the proof of Lemma~\ref{lem:moments}
allows us to have decay controlled by $\zeta=\lambda=D^{-1+\eta}$, bounding therefore the fourth moment of the marginal bias by
$\zeta^{2L}=O(n^{-1-\eta/2})$. Also,
since $L-h=\Omega(\log n)$, Lemma~\ref{lem:moments} still gives polynomial decay for the child
marginals used in the cut bounds. We can use $n^{1/2+2\eta}$ as the bound on the size of the components in the exploration process, and $n^{3\eta}$ as the bound on the total number of excess edges.
With $\ell_0=\lfloor L/2\rfloor$, these choices give
$n^{3\eta}\lambda^{h-\ell_0}=o(1)$ and
$n^{3\eta}\lambda^h=o(n^{-1/4-\eta/16})$, so the deep cuts preserve
the child-marginal bounds and leave the required margin in the root
marginal bound. The final run-time is obtained as  before by noting that each trial has $O(n)$ insertions, and each exploration and update takes $n^{1/2+2\eta}(\log n)^{O(1)}$ time, yielding in total that the expected running time is bounded by $n^{3/2+3\eta}=n^{3/2+\delta}$; the same tail argument gives $n^{-\Omega(\log n)}$ probability of exceeding this bound.
\section{The algorithm and proof of the main theorem}\label{sec:algorithm}
In this section, we give the complete sampling algorithm and state two key 
 estimates, concerning the resampling neighbourhoods and the
random graph error. Using these, we will be able to prove Theorem~\ref{th:er} in Section~\ref{sec:mainproof}. Later Sections focus on the proofs of the two key estimates.

\subsection{The vanilla resampling step}\label{sec:insertion}
We describe a single edge insertion for a general graph $G=(V,E)$, leaving the choice of the resampling neighbourhood $B\subseteq V$ to
Section~\ref{sec:block}. Assume throughout that $|V|=n$ is sufficiently large.

Write $\mathcal{U}(G)=\binom{V}{2}\setminus E$ for the set of unused
pairs, and recall that $\Omega_G$ is the set of proper $q$-colourings of $G$, with uniform distribution
$\mu_G$ when $\Omega_G$ is nonempty. For a colouring $\sigma\in[q]^V$ and a pair $e=uv\in\binom{V}{2}$, let
\[b_e(\sigma)=\one\{\sigma_u\ne\sigma_v\}\] indicate whether $e$ is
bichromatic under $\sigma$; otherwise we call it monochromatic.  The total
number of bichromatic pairs is
\[
 N_b(\sigma)=\sum_{e\in\binom{V}{2}}b_e(\sigma)
 =\frac12\Big(n^2-\sum_{i=1}^q n_i(\sigma)^2\Big) \mbox{ where } n_i(\sigma)=|\sigma^{-1}(i)| \mbox{ is the size of colour class $i$}.
\]
When $|\mathcal{U}(G)|>0$, write
\[b(G,\sigma)=\frac{1}{|\mathcal{U}(G)|}\sum_{e\in\mathcal{U}(G)}b_e(\sigma)\] for the
fraction of unused pairs that are bichromatic. If $\sigma\in\Omega_G$,
all $|E|$ edges are bichromatic, so
\begin{equation}\label{eq:bupper}
 b(G,\sigma)=\frac{N_b(\sigma)-|E|}{\binom{n}{2}-|E|}\le\frac{N_b(\sigma)}{\binom{n}{2}}\le\rho<1,
 \qquad \rho:=\frac{n}{n-1}\Big(1-\frac1q\Big),
\end{equation}
where the inequality $N_b(\sigma)/\binom{n}{2}\le\rho$ follows from
$\sum_i n_i(\sigma)^2\ge n^2/q$.

Given $\sigma\in\Omega_G$ and a pair $e\in\mathcal{U}(G)$ to insert as a new edge into $G$,
we choose a \emph{helper pair} $e'\in \mathcal{U}(G)$ uniformly from the unused pairs
with $b_{e'}(\sigma)=1-b_e(\sigma)$. This is possible whenever
$b(G,\sigma)>0$, since~\eqref{eq:bupper} also ensures the existence of
monochromatic unused pairs. If $b(G,\sigma)=0$, we leave $\sigma$
unchanged and assign score zero.

For each choice of $e,e'$, fix a set $B=B(e\cup e')\subseteq V$
depending only on $G$ and their endpoints (but not on $\sigma$).  The resampling step keeps $\sigma$
when $e$ happens to be bichromatic under $\sigma$; otherwise it attempts to resample inside $B$
so that $e$ becomes bichromatic and $e'$ becomes monochromatic.
In both cases, we assign the following score that will be used to correct the bias of
the update.

\begin{definition}\label{def:score}
Let $G=(V,E)$ be a graph, $\sigma\in\Omega_G$ be a colouring, and  $B\subseteq V$ be a subset of vertices.  For $e,e'\in\mathcal U(G)$ with $b_e(\sigma)=1-b_{e'}(\sigma)$, we define the insertion score $s^{B,\sigma}_{e,e'}$ of the edge $e$ with helper $e'$ as follows:
\begin{itemize}
\item In the case $b_e(\sigma)=1$, set $s^{B,\sigma}_{e,e'}
 = 1-(1-b(G,\sigma))\min\left\{
  \frac{\mu_G^{B,\sigma}(b_e=0,\ b_{e'}=1)}{\mu_G^{B,\sigma}(b_e=1,\ b_{e'}=0)},\,1
  \right\}$.
\item In the case $b_e(\sigma)=0$, set $s^{B,\sigma}_{e,e'}
 =
  b(G,\sigma)\min\left\{
  \frac{\mu_G^{B,\sigma}(b_e=1,\ b_{e'}=0)}{\mu_G^{B,\sigma}(b_e=0,\ b_{e'}=1)},\,1
  \right\}$.
\end{itemize}
The insertion score is well-defined and is in $[0,1]$ ($\sigma$ certifies the denominator is non-zero in each case).
For convenience, we also set $s^{B,\sigma}_{e,e'}=0$ for $\sigma\notin\Omega_G$.
\end{definition}

We are now ready to state the main resampling step and the exact weighted identity. For a prescribed pair $e$, the algorithm will use the same update given below conditional on $\re=e$. 
\begin{lemma}\label{lem:exact}
Let $G=(V,E)$ be a $q$-colourable graph with $|V|>q$ and
$U=|\mathcal U(G)|>0$. For every two  pairs $e,e'\in \mathcal U(G)$,
fix a set $B(e\cup e')\subseteq V$.
Perform the following experiment.
\begin{enumerate}[label=\arabic*.]
\item Draw $\rsigma\sim\mu_G$ and, independently, choose $\re$ uniformly
from $\mathcal U(G)$. If $b(G,\rsigma)=0$, set
$\rs_{\re}=0$ and $\rtau=\rsigma$, and exit.
\item Conditional on $\rsigma,\re$, choose $\reprime$ uniformly from
$\{e'\in\mathcal U(G):b_{e'}(\rsigma)=1-b_{\re}(\rsigma)\}$. Let $\rB=B(\re\cup\reprime)$.

\item Let $\rs_{\re}=s^{\rB,\rsigma}_{\re,\reprime}$. If $b_{\re}(\rsigma)=0$ and $\rs_{\re}>0$, draw
$\rtau\sim
 \mu_G^{\rB,\rsigma}(\,\cdot\mid b_{\re}=1,b_{\reprime}=0)$;
otherwise set $\rtau=\rsigma$.
\end{enumerate}
Then  the following hold.
\begin{enumerate}[label=(\roman*)]
\item \label{it:sampleid} \textbf{Sample identity:}
For every $e\in\mathcal U(G)$ and $\tau\in[q]^V$, $
 \E\Big[\rs_{\re}\one_{\{\rtau=\tau\}}\mid\re=e\Big]
 =\mu_G(\tau)b_e(\tau)$.
\item \label{it:scoremean} \textbf{Score mean and variance:}
For every $\sigma\in\Omega_G$, with $b(\sigma)=b(G,\sigma)$ it holds that
\begin{equation}\label{eq:variance}
\begin{aligned}\E[\rs_{\re}\mid\rsigma=\sigma]&=b(\sigma),\quad 
 \E\big[(\rs_{\re}-b(\sigma))^2\mid\rsigma=\sigma\big]
 =\frac1{2U^2}
 \sum_{e,e'}
 \left(s^{B,\sigma}_{e,e'}-s^{B,\sigma}_{e',e}\right)^2,
 \end{aligned}
\end{equation}
where the sum ranges over $e,e'\in \mathcal{U}(G)$ with $b_e(\sigma)\neq b_{e'}(\sigma)$ and each summand uses $B=B(e\cup e')$. 
\end{enumerate}
\end{lemma}
\begin{proof} We will think of $\rsigma$ as the input colouring of the experiment and $\rtau$ as the output.

\noindent\textbf{Part (i).}
Condition on $\re=e$. The output $\rtau$ is proper for $G$, and $b_e(\rtau)=0$
can occur only on a zero-score branch. It therefore suffices to prove the claim for 
$\tau\in\Omega_G$ with $b_e(\tau)=1$. Let $b(\tau)=b(G,\tau)$ and observe that  $0<b(\tau)<1$ by~\eqref{eq:bupper}.

For colourings $\sigma$ with $b_e(\sigma)=1$, having  $\rsigma=\sigma$ gives $\rtau=\sigma$ as well, so the contribution from such $\sigma$
to $\E\Big[\rs_{\re}\one_{\{\rtau=\tau\}}\mid\re=e\Big]$ is
\begin{equation}\label{eq:sum3566}
\frac{\mu_G(\tau)}{(1-b(\tau))U}
 \sum_{e'}
 s^{B,\tau}_{e,e'},
 \end{equation}
where the sum ranges over $e'\in \mathcal{U}(G)$ with $b_{e'}(\tau)=0$ and each summand uses the shorthand $B=B(e\cup e')$. 

We next consider the contribution to $\E\Big[\rs_{\re}\one_{\{\rtau=\tau\}}\mid\re=e\Big]$
from colourings $\sigma$ such that $b_e(\sigma)=0$.
A repair that produces $\tau$ must use a helper $e'$ with $b_{e'}(\tau)=0$.
Fix one such helper, as in the sum~\eqref{eq:sum3566}, and set
$B=B(e\cup e')$ and
\[
p_{10}=\mu_G^{B,\tau}(b_e=1,b_{e'}=0),\quad p_{01}=\mu_G^{B,\tau}(b_e=0,b_{e'}=1).
\]
Here $p_{10}>0$, since $\tau$ has positive probability under $\mu_G^{B,\tau}$.
If $p_{01}=0$, no repair with this helper reaches $\tau$,
and $s^{B,\tau}_{e,e'}=1$.

Suppose $p_{01}>0$. The input colourings $\sigma\in\Omega_G$ that can be repaired to
$\tau$ with this helper are precisely those agreeing with $\tau$ outside
$B$ and satisfying $b_e(\sigma)=0$, $b_{e'}(\sigma)=1$.
For every such $\sigma$, we have $b(\sigma)>0$ and
$\mu_G^{B,\sigma}=\mu_G^{B,\tau}$, since $B$ is fixed and the exterior colours agree.
The probability of selecting $e'$ is $1/(Ub(\sigma))$, and the score is
\[
 s^{B,\sigma}_{e,e'}=b(\sigma)\min\{p_{10}/p_{01},1\}>0.
\]
Thus the conditional probability of redrawing $\tau$ is $\mu_G^{B,\tau}(\tau)/p_{10}$.
For these inputs, $\mu_G(\sigma)\mu_G^{B,\tau}(\tau)=\mu_G(\tau)\mu_G^{B,\tau}(\sigma)$,
and their total $\mu_G^{B,\tau}$-probability is $p_{01}$. Summing over precisely these
inputs gives the contribution with helper $e'$:
\[
 \begin{aligned}
 \sum_{\sigma}\mu_G(\sigma)\frac{1}{Ub(\sigma)}
       s^{B,\sigma}_{e,e'}\frac{\mu_G^{B,\tau}(\tau)}{p_{10}}&=\frac{\mu_G(\tau)}{Up_{10}}\min\{p_{10}/p_{01},1\}
       \sum_{\sigma}\mu_G^{B,\tau}(\sigma)\\
      & =\frac{\mu_G(\tau)}U\min\{p_{01}/p_{10},1\}=\frac{\mu_G(\tau)}{(1-b(\tau))U}
       \bigl(1-s^{B,\tau}_{e,e'}\bigr).
 \end{aligned}
\]
The last expression is also zero when $p_{01}=0$. Therefore, summing over
the helpers, the total contribution from inputs with $b_e(\sigma)=0$ is
\[
 \frac{\mu_G(\tau)}{(1-b(\tau))U}
 \sum_{e'\in\mathcal U(G);b_{e'}(\tau)=0}
       \bigl(1-s^{B,\tau}_{e,e'}\bigr),
\]
where each summand again uses $B=B(e\cup e')$.
Adding this to~\eqref{eq:sum3566} gives $\mu_G(\tau)$, since there
are exactly $(1-b(\tau))U$ such helpers. This proves (i).

\medskip\noindent\textbf{Part (ii).}
Fix the incoming colouring $\sigma$. If $b(\sigma)=0$, the score is zero
and the sum in \eqref{eq:variance} is empty. Otherwise fix
opposite-status pairs $e,e'$ with $b_e(\sigma)=1$, and let
$B=B(e\cup e')$. Definition~\ref{def:score} gives
\[
 \begin{aligned}
 s^{B,\sigma}_{e,e'}-b(\sigma)
 &=(1-b(\sigma))\left(s^{B,\sigma}_{e,e'}-s^{B,\sigma}_{e',e}\right),\\
 s^{B,\sigma}_{e',e}-b(\sigma)
 &=-b(\sigma)\left(s^{B,\sigma}_{e,e'}-s^{B,\sigma}_{e',e}\right).
 \end{aligned}
\]
The two orientations have probabilities $1/((1-b(\sigma))U^2)$ and $1/(b(\sigma)U^2)$.
Their centred contributions cancel, and their squared contributions
sum to $(s^{B,\sigma}_{e,e'}-s^{B,\sigma}_{e',e})^2/U^2$.
Summing over opposite-status pairs proves \eqref{eq:variance};
the factor $1/2$ counts each pair of pairs once.
\end{proof}

\subsection{The resampling neighbourhoods}\label{sec:block}
In this section, we describe the resampling neighbourhoods in one step of the algorithm. For a graph $G$ and pairs $e,e'\in\mathcal U(G)$, we define those  neighbourhoods via an exploration process having the endpoints of these pairs as roots. The only parameter of the  exploration is its depth $L>0$. For now, the choice of this parameter is not so important but we will set it to  $L=\left\lfloor\frac{11}{20}\log_D n\right\rfloor$ where $D=d+3$.

\paragraph{The exploration process $B_{G,L}(\mathcal R)$.}
We fix the graph $G=(V,E)$ and an integer depth $L>0$ of the exploration process. The input is a set $\mathcal R\subseteq V$
of at most four distinct roots. Each call starts a new breadth-first exploration. The exploration processes vertices and classifies them as discovered, processed, and undiscovered; roughly, a vertex is discovered when it is first placed in the queue for exploration, and undiscovered if it has not been placed into the queue. The vertex becomes processed after its discovery expansion in Step~2 below.

\begin{enumerate}
\item \textbf{Initialise.}
Declare all roots in $\mathcal R$ discovered and place them in the queue
in increasing vertex order. Give them depth zero and no parent;
initially the discovery forest consists of these isolated roots.

\item \textbf{Complete the discoveries.}
While the queue is nonempty and its first vertex has depth less than
$L$, remove its first vertex $v$. Query its pairs with all currently
undiscovered vertices. In increasing vertex order, declare each newly
found neighbour $w$ discovered, assign it parent $v$ and depth
$\operatorname{depth}(v)+1$, record $vw$ as a discovery edge, and append
$w$ to the queue. Declare $v$ processed and continue.
Vertices at depth $L$ are discovered but never processed. No vertex
is marked during this phase.

\item \textbf{Inspect excess edges.}
After the discovery phase, inspect every previously unqueried pair of
discovered vertices with at least one endpoint of depth less than $L$,
in lexicographic order. Each edge of $G$ found in this step is an
\emph{excess edge}; mark both of its endpoints. Inspect all these
pairs regardless of any marks already made.

\item \textbf{Retain the resampling vertices.}
Return $B_{G,L}(\mathcal R)$ consisting of the discovered vertices at
depth less than $L$ whose parent path from their root contains no marked
vertex. The path includes both the root and the vertex itself.
\end{enumerate}
The discovered vertices together with the discovery edges recorded in
Step~2 form the \emph{full discovery forest}. For every discovered vertex
$w$, let $T_w$ be its descendant tree in this forest, rooted at $w$.
In particular, the trees $T_v$ for $v\in\mathcal R$ are its components.
These trees are defined before pruning, and include vertices at depth $L$
and vertices excluded from $B$ in Step~4. We will use these full discovery
trees later to apply tree non-reconstruction estimates with the colours
at depth $L$ fixed, and then bound the effect of fixing the colours at
marked vertices.

For the exploration with output $B=B_{G,L}(\mathcal R)$, let $\exc(B)$
be the set of excess edges recorded in Step~3. For an excess edge $f=uv\in\exc(B)$, define its depth by $\operatorname{depth}(f)
 =\min\{\operatorname{depth}(u),\operatorname{depth}(v)\}$. Note that if $\operatorname{depth}(f)=\ell$, both endpoints have depth $\ell$ or $\ell+1$.

For pairs $e,e'\in\binom V2$, we are going to use this exploration with root set
$\mathcal R=e\cup e'$ to obtain $B_{G,L}(e\cup e')$.
The construction depends only on
$G,L$ and the root set, so it will be independent of the colouring. The explored neighbourhoods may contain cycles,
but the pruned neighbourhood after marking does not. We collect some relevant properties in the lemma below.
\begin{lemma}\label{lem:blockproperties}
Fix $G$, $L$ and a root set $\mathcal R$, and let $B=B_{G,L}(\mathcal R)$ be the
output of the exploration. Then,
\begin{enumerate}
\item Every $f\in\exc(B)$ connects a pair of marked vertices; neither of these belongs to $B$.
\item $G[B]$ is a forest, and every component contains exactly one root from
$\mathcal R\cap B$. Every discovered vertex is at distance at most $L$ from some root in $\mathcal R$.
\item The graph distance in $G$ between two distinct components of $G[B]$ is at least 2.
\item The set $B$ contains only vertices of discovery depth less than $L$.
\end{enumerate}
\end{lemma}
\begin{proof}
Step~3 marks both endpoints of every excess edge, and Step~4
excludes every marked vertex from $B$. This gives the first assertion.
Every discovery edge increases depth by one, so the parent path of each
discovered vertex gives its distance bound.

Every vertex of $B$ was processed during discovery, so all its
neighbours were discovered. Any edge incident to $B$ that is not a
discovery edge must therefore have been queried in Step~3, marking
its endpoint in $B$, a contradiction. Thus every edge incident to $B$
is a discovery edge. The parent path of a retained vertex is also
retained, so $G[B]$ is a forest whose components each contain their
original root and no other root. Distinct components of the induced
graph have no edge between them in $G$, proving the third assertion.
By Step~4, every vertex included in $B$ has discovery depth less than $L$.
\end{proof}

\begin{lemma}\label{lem:neighbourhoods}
\label{lem:uniformballs}
With probability $1-O(\tfrac{1}{n^2})$ over $\rG\sim G(n,d/n)$, with $L=\left\lfloor\frac{11}{20}\log_{d+3} n\right\rfloor$ 
and 
$N_{\max}=n^{7/12}$,
\begin{equation}\label{eq:uniformballs}
 \max_{v\in[n]}\bigl|\{u:\operatorname{dist}_{\rG}(u,v)\le L\}\bigr|
 \le N_{\max}/4.
\end{equation}
\end{lemma}
\begin{proof}
Let $\rVolume_{\ell}$ be the total population up to depth $\ell$ of a
Poisson$(D)$ Galton--Watson tree. The uniform exponential-moment bound
of Mossel--Sly~\cite[equation~(17)]{MS} gives constants $\kappa,C>0$
such that
\begin{equation}\label{eq:poissonvolume}
 \Prb(\rVolume_{\ell}> t D^{\ell})\le C \emm^{-\kappa t},
 \qquad \mbox{ for any integer }\ell\ge0 \mbox{ and } t>0.
\end{equation}
The same bound holds for Poisson$(\gamma)$ trees with $\gamma\le D$
by domination. Standard domination of breadth-first exploration by
a Poisson$(D)$ tree gives the bound also for the size of a radius-$\ell$
ball in $G(n,d/n)$. Since $D^L\le n^{11/20}$,
a union bound shows that~\eqref{eq:uniformballs} fails with probability
at most $C n\exp(-\kappa n^{1/30}/4)=O(n^{-2})$.
\end{proof}

\subsection{The full sampler and proof of the theorem}\label{sec:mainproof}
\label{sec:trials}
We now describe the algorithm $\mathcal{A}$, run on a fixed input graph $G$.
Let $T=n^{1/\sqrt{\log\log n}}$ and note that $T=n^{o(1)}$.
Given a graph $G$ as input, the algorithm processes the edges of $G$ one-by-one (in a random order), and uses the  update rule of Lemma~\ref{lem:exact} as the subroutine to update the current colouring. To handle exceptional cases, it will be convenient to extend the update rule of Lemma~\ref{lem:exact} to every 
colouring $\sigma\in [q]^V$: if it is not proper for the current graph $G$, return it
unchanged with score zero.

\begin{algorithm}[$\mathcal{A}$]\label{alg:sampler}
On input $G=(V,E)$ with $m$ edges, repeat independent trials until one is
accepted. In each trial do the following.
\begin{enumerate}
\item Choose a uniform order $e_1,\ldots,e_m$ of the edges of $G$.
Let $G_t=(V,\{e_1,\ldots,e_t\})$ for $0\le t\le m$.
Start at $G_0$, draw $\sigma_0\in[q]^V$ uniformly, and set the total score $S=1$.
\item \label{it:scoreupdate}  For $t=0,\ldots,m-1$, apply the update rule of Lemma~\ref{lem:exact} (extended as above) to $G=G_t$, conditioned  on $\rsigma=\sigma_t$ and
inserted edge $\re=e_{t+1}$ (using
$B=B_{G_t,L}(e_{t+1}\cup e')$ for the helper pair $e'$ chosen in the update).
Let $s_{t+1}$ denote the corresponding score.
Set $\sigma_{t+1}$ equal to the output of the update, and multiply the total score $S$ by $s_{t+1}/\rho$.
\item Accept/return $\sigma=\sigma_m$ with probability $\min\{S,T\}/T$. Otherwise begin a fresh trial.
\end{enumerate}
\end{algorithm}
A trial starts in Step~1, performs all $m$ updates, and is
\emph{successful} if it is accepted in Step~3 (note that this certainly happens if $S\geq T$).
Note that during a trial, we might get a zero score $s_{t+1}=0$; at the first such step, the newly inserted edge remains monochromatic, giving an improper colouring that remains unchanged (all subsequent scores are zero). Thus $S$ stays zero and the
trial is rejected in Step~3. At every stage, the weight is positive
iff the colouring is proper for the current prefix graph.

For the analysis of a trial on a given graph $G$, write $\rsigma_t$ for the incoming colouring at
iteration $t$ of Step~\ref{it:scoreupdate}, $\rs_{t+1}$ for the score returned for the insertion of the edge $\re_{t+1}$, and $\rsigma_m$ for the final colouring. Set
\[
 \rS_{G,0}=1,\qquad \rS_{G,t}=\prod_{t'=1}^t\frac{\rs_{t'}}{\rho},\qquad
 \rS_G=\rS_{G,m},\qquad \hat Z_G=\frac{Z_G}{q^n \rho^{|E(G)|}}.
\]
When the input graph is clear, we write $\rS_t$ for $\rS_{G,t}$; for a random input $\rG$, the final score is $\rS_{\rG}$.
Using Lemma~\ref{lem:exact}, we have that $\E[\rS_G]=\hat Z_G$ (starting from a uniform colouring of $G_0$,
iterate Lemma~\ref{lem:exact}\ref{it:sampleid}, dividing each score by $\rho$). We note here that the expectation is over all random choices of one trial of the algorithm $\mathcal{A}$ for a fixed graph $G$, i.e.,  the edge order and all random updates.

We define further the loss of one trial as
\begin{equation}\label{eq:acceptedmeasure}
 \delta_G=\E[\rS_G-\min\{\rS_G,T\}]=\hat Z_G-T\times\Prb\bigl(\text{one trial of $\mathcal{A}$ on $G$ is successful}\bigr),
\end{equation}
where the equality follows from $\E[\rS_G]=\hat Z_G$ and the acceptance probability $\min\{\rS_G,T\}/T$ in step 3 of the algorithm. Note in particular that $\delta_G\geq 0$. If $G$ is not $q$-colourable, then $\E[\rS_G]=\hat Z_G=0$, so no trial is successful and $\delta_G=0$. 

The lemma below controls this loss on a typical
random graph $\rG\sim G(n,d/n)$. The first item can be deduced fairly easily from known results in the literature (the main point there is the lower bound on $\hat Z_G$). The second item is new and is the main component that makes the analysis of the algorithm possible.

\begin{lemma}\label{lem:randomgrapherror}
Let $d$ be sufficiently large and $q\geq d/\log d$ be an integer. There is a set of graphs 
$\mathcal{Q}$ with $\Prb(\rG\in\mathcal{Q})=1-o(1)$ for
$\rG\sim G(n,d/n)$ such that:
\begin{enumerate}
    \item every $G\in\mathcal{Q}$ has
$|E(G)|\le \frac{d+1}{2}n$, $\Delta(G)\le\log n$, and $\hat Z_G\ge1/\log n$.
\item For every $G\in\mathcal{Q}$, the loss of one trial satisfies
$\delta_G\le n^{-\omega(1)}$. 
\end{enumerate}
\end{lemma}
The proof is given in Section~\ref{sec:error}.

With Lemma~\ref{lem:randomgrapherror} at hand, it is not hard to put together the proof of the main theorem.
\begin{thmer}
\statethmer
\end{thmer}
\begin{proof}
Let $\mathcal{Q}$ be the  set of graphs of Lemma~\ref{lem:randomgrapherror} and $\mathcal{Q}'$ the set of graphs of Lemma~\ref{lem:neighbourhoods}; since $\Prb\big(\rG\in\mathcal{Q}\cap \mathcal{Q}'\big)=1-o(1)$, it suffices to show the TV-distance and running time guarantees for $G\in\mathcal{Q}\cap\mathcal{Q}'$. Let $\mathcal{A}(G)$ denote the random output of Algorithm~\ref{alg:sampler}. We start by bounding the TV-distance between $\mu_G$ and the output distribution $\Prb\big(\mathcal{A}(G)\in \cdot\big)$.

Since $G\in \mathcal{Q}$, we have $\hat Z_G>0$, so $G$ is $q$-colourable. Suppose that $G$ has $m$ edges and consider an arbitrary edge order of $G$.
Starting from the uniform colouring of $G_0$,
iterate Lemma~\ref{lem:exact}\ref{it:sampleid}, dividing each score by $\rho$.
 For any $\Sigma\subseteq\Omega_G$ and $0\le t\le m$, this gives
\begin{equation}\label{eq:earlyweightedgraph}
 \E\big[\rS_t\, \one\{\rsigma_t\in\Sigma\}\big]=\hat Z_{G_t}\mu_{G_t}(\Sigma).
\end{equation}
At $t=m$, the right-hand side is independent of the edge order, so the
same identity holds after averaging it; in particular with $\rS_G=\rS_m$, we have $\E\rS_G=\hat Z_G$.
The success probability of one trial is
$p_G=\E[\min\{\rS_G,T\}]/T$, so since $\delta_G=\E[\rS_G-\min\{\rS_G,T\}]$ (by the definition in~\eqref{eq:acceptedmeasure}) we have 
\[
  p_G=(\hat Z_G-\delta_G)/T.
\]

Lemma~\ref{lem:randomgrapherror} gives
$\delta_G\le n^{-\omega(1)}=o(1/\log n)$ for every $G\in\mathcal{Q}$.
On these inputs, $\hat Z_G\ge1/\log n$, so
$p_G\ge1/(2T\log n)>0$ once $n$ is sufficiently large.
For any $\Sigma\subseteq\Omega_G$, independent repetition gives
\[
 \Prb\big(\mathcal{A}(G)\in\Sigma\big)
 =\sum_{j\geq 1}(1-p_G)^{j-1}\Prb\big(\rsigma_m\in \Sigma,\mathrm{accept}\big)=\frac{\E\big[\min\{\rS_G,T\}\one\{\rsigma_m\in\Sigma\}\big]}{\hat Z_G-\delta_G}.
\]
From~\eqref{eq:earlyweightedgraph} at $t=m$, we have $G_t=G$ and $\hat Z_{G}\mu_{G}(\Sigma)=\E\big[\rS_G\, \one\{\rsigma_m\in\Sigma\}\big]$, so we obtain
\[
 \hat Z_G\mu_G(\Sigma)
 =(\hat Z_G-\delta_G)\Prb\big(\mathcal{A}(G)\in\Sigma\big)
   +\E\big[\big(\rS_G-\min\{\rS_G,T\}\big)\one\{\rsigma_m\in\Sigma\}\big].
\]
The last term lies in $[0,\delta_G]$. We conclude that $|\mu_G(\Sigma)-\Prb\big(\mathcal{A}(G)\in\Sigma\big)|\leq 2\delta_G/\hat Z_G$, giving that the TV distance between $\mu_G$ and $\Prb\big(\mathcal{A}(G)\in \cdot\big)$ is at most
$2\delta_G/\hat Z_G\le 2n^{-\omega(1)}\log n=n^{-\omega(1)}$.  

We next bound the running-time  $\rTime_G$ on input $G$; let $\boldsymbol{A}_G$ denote the number of independent trials of Algorithm~\ref{alg:sampler} on input $G$. Recall that for $G\in \mathcal{Q}$ the acceptance probability $p_G$ of a single trial satisfies $p_G\ge\frac1{2T\log n}$. So, for any integer $k\geq 0$,
\begin{equation}\label{eq:trialtail}
 \E\big[\boldsymbol{A}_G\big]\le2T\log n,\qquad
 \Prb(\boldsymbol{A}_G>k)\le\exp\left(-\tfrac{k}{2T\log n}\right).
\end{equation}
Henceforth, we assume that uniform integer generation costs $O(\log N)$ for a uniform sample from $\{1,\ldots,N\}$. Exact implementation incurs only a polylogarithmic overhead over
polynomially many calls, except with probability
$\emm^{-\Omega((\log n)^2)}$. We absorb this overhead throughout.

Set $N_{\max}=n^{7/12}$. Since $G\in \mathcal{Q}'$ satisfies \eqref{eq:uniformballs}, for every pair $e,e'$ the  exploration $B_{G_t,L}(e\cup e')$ discovers at most $N_{\max}$ vertices since $G_t\subseteq G$, and the exploration starts from at most four roots.
Also, since $\Delta(G)\le\log n$, the whole exploration costs $N_{\max}(\log n)^{O(1)}$
time (using adjacency lists). Using the forest property of  $B_{G_t,L}(e\cup e')$ (cf. Lemma~\ref{lem:blockproperties}), we can compute the scores and the (conditional) sampling steps of Lemma~\ref{lem:exact}  by standard
tree dynamic programming in the same $N_{\max}(\log n)^{O(1)}$~time. 

To be able to choose the helper quickly,  maintain $q$ indexed colour classes of sizes $n_i=|\sigma_t^{-1}(i)|$ and a balanced search tree of edges in $G_t$. This computes the bichromatic fraction $b=b(G_t,\sigma_t)$ of unused pairs from \eqref{eq:bupper} and updates the data structure in $(\log n)^{O(1)}$ time per recoloured vertex and per edge insertion. For a monochromatic helper, we choose class $i\in [q]$
with probability proportional to  $\binom{n_i}{2}$ and a uniform pair within it; for a bichromatic helper, we choose  a pair $(i,j)$ with $i<j$ with weight
$n_i n_j$ and a uniform vertex in each class, rejecting only edges of
$G_t$. Using that $\hat n=\max_i n_i\ge n/q$, it is not hard to see that the rejection probability is at most $\frac{(n-\hat n)\Delta(G)}{\hat n(n-\hat n)}\leq q\Delta(G)/n<1/2$ for sufficiently large $n$. Thus, the number of proposals for each helper has expectation $O(1)$ and an exponential tail, uniformly over the time $t$; each proposal costs $(\log n)^{O(1)}$ time.

Each trial has $O(n)$ insertions.
Together with the random-generation overhead and
$nN_{\max}=O(n^{19/12})$, we obtain that the expected trial cost is at most
$n^{19/12}(\log n)^{O(1)}$ and  the probability of exceeding this bound is  at
most $\emm^{-\Omega((\log n)^2)}$.
Reaching a trial depends only on the preceding trials and is independent
of its cost. Thus the expected total cost is the expected cost of one
trial divided by $p_G\geq \frac{1}{2T\log n}$. Since $T=n^{o(1)}$ and $8/5>19/12$, we obtain that the running time $\rTime_G$ satisfies for all sufficiently large $n$
\[
 \E\rTime_G\le   n^{19/12}(\log n)^{O(1)}\,2T\log n\le n^{8/5}.
\]
For the tail bound, note from \eqref{eq:trialtail} that the probability we exceed $T(\log n)^3$ trials is at most $\emm^{-\Omega((\log n)^2)}$; each trial exceeds its cost bound $n^{19/12}(\log n)^{O(1)}$ with probability at most $\emm^{-\Omega((\log n)^2)}$.
Outside these events, the total cost is at most $T n^{19/12}(\log n)^{O(1)}\le n^{8/5}$. A union bound over the first $\lceil T(\log n)^3\rceil$ trials therefore gives that
$\Prb(\rTime_G>n^{8/5})$ is at most $\emm^{-\Omega((\log n)^2)}$ as well.
\end{proof}

\section{Proof of Lemma~\ref{lem:randomgrapherror}}\label{sec:error}

We collect the graph and count bounds used in the analysis;
these and the monotonicity lemma below are proved in Appendix~\ref{sec:countproof}. We first introduce a graph statistic that will be relevant for the analysis of marginal bounds later.

Let $G=(V,E)$ be a graph with $n$ vertices. Recall that $D=d+3$ and
$L=\lfloor\frac{11}{20}\log_D n\rfloor$, and set $\lambda=D^{-19/20}$.
Define the graph statistic
\begin{equation}\label{eq:graphgeometry}
 \Psi(G)=\frac1{\binom n4}
 \sum_{\mathcal R\in \binom{ V}{4}}
 \sum_{f\in\exc(B_{G,L}(\mathcal R))}\lambda^{4\operatorname{depth}(f)}.
\end{equation}

\begin{lemma}\label{lem:countlower}
Let $d$ be sufficiently large and $q\geq d/\log d$ be an integer. Then, whp over $\rG\sim G(n,d/n)$, it holds that  $|E(\rG)|\le\frac{d+1}{2}n$, $\Delta(\rG)\le\log n$, $\Psi(\rG)\le(\log n)^{1/4}/n$, and $\hat Z_{\rG}\ge\frac1{\log n}$.
\end{lemma}
For later use we also record the following monotonicity property of $\Psi$, also proved in Appendix~\ref{sec:countproof}.
\newcommand{\statelempsimonotonicity}{
Let $H\subseteq G$ be graphs on the same set of vertices.
Then $\Psi(H)\le\Psi(G)$.
}
\begin{lemma}\label{lem:psimonotonicity}
\statelempsimonotonicity
\end{lemma}

We next consider the score after one edge insertion of Algorithm~\ref{alg:sampler} on input $\rG$ in the context of  Lemma~\ref{lem:exact}. For this part of the argument, we will expose the edges of $\rG$ one-by-one as the algorithm inserts them; in particular,   each inserted edge is distributed uniformly over the remaining non-edges. At a state $(\rG_t,\rsigma_t)=(G_t,\sigma_t)$ with $\sigma_t\in[q]^V$, let $\re$  be uniform in $\mathcal{U}(G_t)$. Recall that if $\sigma_t\notin\Omega_{G_t}$ or $b(G_t,\sigma_t)=0$ we set the score $\rs_{t+1}=0$ (without choosing a helper or forest).
Otherwise, conditional on $\re$, choose $\reprime$ with  opposite
status to that of $\re$, and set
$\rB_t=B_{G_t,L}(\re\cup\reprime)$.
In this case, the score is
$\rs_{t+1}=s^{\rB_t,\sigma_t}_{\re,\reprime}$. Define
\begin{equation}\label{eq:V}
 \mathcal V(G_t,\sigma_t)=\E\bigl[(\rs_{t+1}-\rho)^2\mid \rG_t=G_t,\rsigma_t=\sigma_t\bigr].
\end{equation}

The following lemma is proved in Section~\ref{sec:path}, using
the local score and non-reconstruction estimates.

\newcommand{\stateleminputs}{
Let $d$ be sufficiently large and $q\geq d/\log d$ be an integer. Then, there is a set $\mathcal{Q}$ of graphs with
$\Prb(\rG\in\mathcal{Q})=1-o(1)$ for $\rG\sim G(n,d/n)$ such that the following holds.

Consider a trial of Algorithm~\ref{alg:sampler} on input $\rG$. Let $\mathcal H$ be the event that  $\mathcal V(\rG_t,\rsigma_t)\le\frac{1}{n}\sqrt{\log n}$ for every time $0\le t<|E(\rG)|$. Then it holds that $\E\bigl[\rS_{\rG}\,\one\{\rG\in\mathcal Q,\mathcal H^c\}\bigr]
 =n^{-\omega(1)}$,\textsuperscript{\ref{fn:omega}} where the expectation averages the graph, the edge order and all other
random choices of one trial.
}
\begin{lemma}\label{lem:inputs}
\stateleminputs
\end{lemma}

With these at hand, we can give the proof of Lemma~\ref{lem:randomgrapherror}.

\begin{proof}[Proof of Lemma~\ref{lem:randomgrapherror}]
Let $\mathcal{Q}_1$ be the set of graphs $G$ satisfying the bounds of
Lemma~\ref{lem:countlower}, and let $\mathcal{Q}_2$ be the set given by
Lemma~\ref{lem:inputs}. Set $\mathcal{Q} := \mathcal{Q}_1 \cap \mathcal{Q}_2$.
By a union bound, $\mathbb{P}(\rG \in \mathcal{Q}) = 1 - o(1)$ for
$\rG \sim G(n, d/n)$, and every $G \in \mathcal{Q}$ satisfies the desired bounds in the first item of Lemma~\ref{lem:randomgrapherror} since
$\mathcal{Q} \subseteq \mathcal{Q}_1$. 
Moreover, since $\rS_{\rG} \ge 0$ and
$\mathbf{1}\{\rG \in \mathcal{Q}\} \le \mathbf{1}\{\rG \in \mathcal{Q}_2\}$,
Lemma~\ref{lem:inputs} gives
\begin{equation}
    \label{eq: ub SG}
  \mathbb{E}\bigl[\rS_{\rG} \mathbf{1}\{\rG \in \mathcal{Q}, \mathcal{H}^c\}\bigr]
  \le \mathbb{E}\bigl[\rS_{\rG} \mathbf{1}\{\rG \in \mathcal{Q}_2, \mathcal{H}^c\}\bigr]
  = n^{-\omega(1)}.
\end{equation}

Recall from Section~\ref{sec:mainproof} that 
$\delta_{\rG}=\E\big[\rS_{\rG}-\min\{\rS_{\rG},T\}\mid \rG\big]$, so $\delta_{\rG}\geq 0$. Using the event $\mathcal H$ in Lemma~\ref{lem:inputs}, we next decompose
the expectation $\E_{\rG}\big[\delta_{\rG}\,\one\{\rG\in\mathcal{Q}\}\big]$. In fact, using the pointwise bounds $0\le \rS_{\rG}-\min\{\rS_{\rG},T\}\leq \rS_{\rG}\one\{\rS_{\rG}>T\}\le\rS_{\rG}$ and splitting according to $\mathcal H$, we obtain the upper bound
\begin{align*}
 \E_{\rG}[\delta_{\rG}\one\{\rG\in\mathcal{Q}\}]
 \le{}&\E\big[
       \rS_{\rG}\one\{\rS_{\rG}>T,\rG\in\mathcal{Q}, \mathcal{H}\}\big]
 +\E\big[\rS_{\rG}\one\{\rG\in\mathcal{Q}, \mathcal{H}^c\}\big].
\end{align*}
The right-hand expectations average the input graph and one trial.
The second term is upper bounded by $n^{-\omega(1)}$, according to \eqref{eq: ub SG}. For the first term, we replace the graph
indicator $\rG\in \mathcal{Q}$ by the upper bound $\one\{|E(\rG)|\le M\}$, where
$M=\lceil(d+1)n/2\rceil$ (this is indeed an upper bound since $\mathcal{Q}$ is also restricted to graphs satisfying the bounds in Lemma~\ref{lem:countlower}). We bound the resulting expectation \[\E\big[
       \rS_{\rG}\, \one\{\rS_{\rG}>T,\mathcal{H}, |E(\rG)|\leq M\}\big]\] by first conditioning on $|E(\rG)|=m$ for some  $m\le M$, sampling a uniform
ordered $m$-edge graph and running one trial of the algorithm; we will later average over $m$. 

Let $\mathcal F_t$ be generated by the first $t$ revealed edges and the
algorithm's other random choices up to time $t$. The remaining edges
and their order are not revealed. Conditional on this history, the next edge
is uniform among unused pairs: the updates use only the exposed graph
and fresh independent randomness. We have $\E[\rs_{t+1}\mid\mathcal F_t]\le \rho$: if $\rsigma_t\in\Omega_{\rG_t}$, this follows from Lemma~\ref{lem:exact}(ii) and~\eqref{eq:bupper}; otherwise the score is zero. 
For every real $r\ge2$ and $x\in[0,1]$, Taylor's formula for the function $f(x)=(x/\rho)^r$ around  $x_0=\rho$ gives
\[
 \left(\frac{x}{\rho}\right)^r
 \le 1+\frac r{\rho}(x-\rho)
       +\frac{r(r-1)}{2\rho^r}(x-\rho)^2.
\]
Apply this for $x=\rs_{t+1}\in [0,1]$ and condition on $\mathcal F_t$. Using  that $\E[\rs_{t+1}-\rho\mid \mathcal{F}_t]\leq 0$ and $\mathcal V(\rG_t,\rsigma_t)=\E\bigl[(\rs_{t+1}-\rho)^2\mid \mathcal{F}_t\bigr]$, we obtain
\begin{equation}\label{eq:momentt}
 \E\left[\left(\frac{\rs_{t+1}}{\rho}\right)^r\,\middle|\,\mathcal F_t\right]
 \le1+\frac{r(r-1)}{2\rho^r}\mathcal V(\rG_t,\rsigma_t).
\end{equation}

Let $\mathcal H_t$ be the event that  for every
time $t'<t$ it holds that  $\mathcal V(\rG_{t'},\rsigma_{t'})\le\frac{1}{n}\sqrt{\log n}$. Thus $\mathcal H_0$ is the
whole space, $\mathcal H_m=\mathcal H$, and $\mathcal H_{t+1}$ is
$\mathcal F_t$-measurable and contained in $\mathcal H_t$.
Multiplying~\eqref{eq:momentt} by $(\rS_{\rG,t})^r\one_{\mathcal H_{t+1}}$ gives
\[
 \E\big[(\rS_{\rG,t+1})^r\one_{\mathcal H_{t+1}}\mid\mathcal F_t\big]
 \le\left(1+\frac{r(r-1)\sqrt{\log n}}{2\rho^r n}\right)
       (\rS_{\rG,t})^r\one_{\mathcal H_t}.
\]
Iterating this inequality,
using $\rho\ge1/2$ and $m\le M=O(n)$, gives for $\rS_{\rG}=\rS_m$ that
\begin{equation}\label{eq:goodpathmoment}
\E\big[(\rS_{\rG})^r\one_{\mathcal H}\big]
 \le\exp\bigl(C r^2 2^r\sqrt{\log n}\bigr),
\end{equation}
where $C=O(1)$ is independent of $r,m,n$. Take $r=\tfrac{1}{4}\log_{2}\log n$. The exponent on the right of
\eqref{eq:goodpathmoment} is
$O((\log n)^{3/4}(\log\log n)^2)$, whereas from $T=n^{1/\sqrt{\log\log n}}$ we have 
$(r-1)\log T=\Theta(\log n\sqrt{\log\log n})$.
So, for some constant $\kappa>0$,
\begin{equation}\label{eq:weighttail}
 \E\big[\rS_{\rG}\one\{\rS_{\rG}>T,\mathcal H\}\big]
 \le \E\bigg[\frac{(\rS_{\rG})^r}{T^{r-1}}\,\one_{\mathcal H}\bigg]
 \le\emm^{-\kappa\log n\sqrt{\log\log n}}.
\end{equation}
It remains to note that the bound in \eqref {eq:weighttail} is uniform in $m\le M$. Averaging over the binomial edge
count and combining the two terms in the decomposition gives
$\E_{\rG}[\one\{\rG\in\mathcal Q\}\delta_{\rG}]
 =n^{-\omega(1)}$.
By Markov's inequality,
$ \Prb\bigl(\rG\in\mathcal{Q},\ \delta_{\rG}>n^{-\omega(1)}\bigr)=o(1)
$, so removing these graphs from $\mathcal{Q}$ yields a set of graphs $\mathcal{Q}'$ with $\Prb(\rG\in \mathcal{Q}')=1-o(1)$ which satisfies all the desired properties. This proves the lemma.
\end{proof}

\section{The planted model and the typical graph}\label{sec:path}
This section proves Lemma~\ref{lem:inputs}. We first define a
planted graph model and use its probabilities to bound expectations
involving the trial weights $\rS_t$ from Section~\ref{sec:trials}.
We then state the local score and marginal estimates and prove
the lemma. The proofs of the marginal and tameness estimates are deferred
to Section~\ref{sec:pairproof}.

\subsection{The planted graph and the transfer lemma}\label{sec:planting}
For a fixed colouring $\sigma\in[q]^V$ and an integer
$0\le t\le N_b(\sigma)$, let $\rGhat_t(\sigma)$ be the graph obtained by
choosing a uniform $t$-subset of the bichromatic pairs of $\sigma$.
We call this the \emph{planted graph} with colouring $\sigma$, and write
$\rGhat_t$ when $\sigma$ is clear. We will in fact consider  the planted model for \emph{balanced} colourings $\sigma$, i.e., $\sigma$ that satisfy
\begin{equation}\label{eq:balance}
 \max_i|n_i(\sigma)-n/q|\le\sqrt n\log n.
\end{equation}
Recall that $M=\lceil(d+1)n/2\rceil$. For sufficiently large $n$,
such a colouring has $N_b(\sigma)>M$, so its planted graph is defined
at every stage $t\le M$.

For one trial of Algorithm~\ref{alg:sampler} on input $\rG$, write $\rG_t$ for its
random prefix graph and use $\rsigma_t,\rs_t,\rS_t$ from
Section~\ref{sec:trials}.
Let $\mathcal F_t$ be generated by the first $t$ revealed edges and the
algorithm's other random choices up to time $t$. The remaining input
edges and their order are not revealed.

\begin{lemma}[Transfer from the planted model]\label{lem:pathlaw}
Let $\rG$ be a uniformly random graph on $V=[n]$ with $m$ edges, and
run one trial of Algorithm~\ref{alg:sampler} on $\rG$. The following hold.
\begin{enumerate}[label=(\roman*)]
\item For any fixed colouring $\sigma$ with $N_b(\sigma)\ge t$ and integer $t\in [0,m]$, for
any set $\mathcal E$ of graphs on $V$,
\begin{equation}\label{eq:plantedtransfer}
 \E\big[\rS_t\one\{\rG_t\in\mathcal E,\ \rsigma_t=\sigma\}\big]
 \le q^{-n}\Prb\Big(\rGhat_t(\sigma)\in\mathcal E\Big).
\end{equation}
\item For $0\leq t<m$, let  $F_t\in\mathcal F_t$ be arbitrary events. Then $\E\big[\rS_m\one\{\textstyle\bigcup_{t<m}F_t\}\big]
 \le\sum_{t<m}\E[\rS_t\one_{F_t}]$.
\end{enumerate}
\end{lemma}

\begin{proof}
Let $H$ be a $t$-edge graph on $V$ and let
$\mathcal C\subseteq[q]^V$. The weighted identity of
Lemma~\ref{lem:exact}(i), with the extended update rule, gives by
induction along an edge order of $H$ and averaging over these orders
\[
 \E\bigl[\rS_t\one\{\rsigma_t\in\mathcal C\}\mid\rG_t=H\bigr]
 =\frac{|\mathcal C\cap\Omega_H|}{q^n \rho^t}.
\]
There are $\binom{N_b(\sigma)}t$ graphs compatible with $\sigma$, each
equally likely in the planted model.
Write $(z)_t=z(z-1)\cdots(z-t+1)$, with $(z)_0=1$.
Taking $\mathcal C=\{\sigma\}$ and averaging over $H\in\mathcal E$ gives
\[
 \E\bigl[\rS_t\one\{\rG_t\in\mathcal E,\ \rsigma_t=\sigma\}\bigr]
 =q^{-n}\frac{(N_b(\sigma))_t}{\bigl(\binom{n}{2}\bigr)_t\, \rho^t}
     \Prb\bigl(\rGhat_t(\sigma)\in\mathcal E\bigr).
\]
Since $N_b(\sigma)/\binom{n}{2}\le \rho$, each factor in the ratio is at
most one, proving~\eqref{eq:plantedtransfer}.

Conditional on $\mathcal F_t$, the next edge is uniform in $\mathcal U(\rG_t)$,
since the updates use only the exposed graph and fresh randomness.
We have $\E[\rs_{t+1}\mid\mathcal F_t]\le \rho$: if $\rsigma_t\in\Omega_{\rG_t}$, this follows from Lemma~\ref{lem:exact}(ii) and~\eqref{eq:bupper}; otherwise the score is zero. Hence
\begin{equation}\label{eq:weightsupermartingale}
 \E[\rS_{t+1}\mid\mathcal F_t]
 =\frac{\rS_t}{\rho}\E[\rs_{t+1}\mid\mathcal F_t]
 \le\rS_t.
\end{equation}
Thus the weights form a nonnegative supermartingale. Iteration gives,
for every $F_t\in\mathcal F_t$,
\begin{equation}\label{eq:graphdomination}
 \E[\rS_m\one_{F_t}]
 =\E[\one_{F_t}\E(\rS_m\mid\mathcal F_t)]
 \le\E[\rS_t\one_{F_t}].
\end{equation}
A union bound proves the desired conclusion $\E\big[\rS_m\one\{\textstyle\bigcup_{t<m}F_t\}\big]
 \le\sum_{t<m}\E[\rS_t\one_{F_t}]$.
\end{proof}

\subsection{Local marginals and the proof of Lemma~\ref{lem:inputs}}
For a graph $G=(V,E)$, a proper colouring $\sigma\in\Omega_G$, a set
$B\subseteq V$, and a vertex $v\in V$, define the marginal law
\begin{equation}\label{eq:localmarginal}
 \mu_{G,v}^{B,\sigma}
 :=\bigl(\mu_G^{B,\sigma}(\tau_v=i)\bigr)_{i\in[q]},
\end{equation}
where $\mu_G^{B,\sigma}$ is the conditional distribution on $[q]^V$ where all colours on $V\backslash B$ are fixed to $\sigma$, while colours
in $B$ are free (subject to properness). We write $\mu_v$ when
$G,B,\sigma$ are clear, and let $\unif=(1/q,\ldots,1/q)$ be the uniform law.

We next state the estimates used to prove Lemma~\ref{lem:inputs}.
The score identity in Lemma~\ref{lem:exact}(ii) reduces the variance
to differences between the scores for $e,e'$ and $e',e$.
The following deterministic bound expresses those differences in terms
of the endpoint marginals defined above.

\begin{lemma}[Local score bound]\label{lem:localreduction}
Fix a graph $G$, a colouring $\sigma\in\Omega_G$, and
$e,e'\in\mathcal U(G)$ with four distinct endpoints and
$b_e(\sigma)\ne b_{e'}(\sigma)$.
Let $B=B_{G,L}(e\cup e')$ be the exploration set from  Section~\ref{sec:block}.
Then
\begin{equation}\label{eq:localreduction}
 \left(s^{B,\sigma}_{e,e'}-s^{B,\sigma}_{e',e}\right)^2
 \le128q^2\sum_{z\in e\cup e'}
       \|\mu_{G,z}^{B,\sigma}-\unif\|_2^4.
\end{equation}
\end{lemma}
\begin{proof}
Assume $e=vw$ and $e'=v'w'$; we may also assume by symmetry that $b_e(\sigma)=1$.
Abbreviate $\mu_v=\mu_{G,v}^{B,\sigma}$ and similarly for $w,v',w'$.
By Lemma~\ref{lem:blockproperties}(2), the endpoints of $e,e'$ belonging to $B$
lie in distinct components of $G[B]$, while those outside $B$ have fixed colours
(recall that an endpoint of $e,e'$ is excluded  from $B$ if it is incident to an excess edge).
Since $\mu_G^{B,\sigma}$ factors over these components, the four endpoint
colours are independent under $\mu_G^{B,\sigma}$. In particular, 
\begin{equation}
    \label{eq: bilinear form}
 \mu_G^{B,\sigma}(b_{e}=1)-(1-\tfrac{1}{q})
 =-\langle \mu_v-\unif,\mu_w-\unif\rangle,
\end{equation}
and similarly for $\mu_G^{B,\sigma}(b_{e'}=1)$. 
Let $p=\mu_G^{B,\sigma}(b_e=1)$ and
$p'=\mu_G^{B,\sigma}(b_{e'}=1)$. The equality~\eqref{eq: bilinear form} with Cauchy--Schwarz and
$2ab\le a^2+b^2$ gives
\begin{equation}\label{eq:43g5g4}
 \big(p-(1-\tfrac{1}{q})\big)^2+\big(p'-(1-\tfrac{1}{q})\big)^2
 \le\tfrac12\sum_{z\in e\cup e'}\|\mu_z-\unif\|_2^4.
\end{equation}
Independence of the colour-distributions at $v,w,v',w'$ also gives
\[
 \mu_G^{B,\sigma}(b_e=1,b_{e'}=0)=p(1-p'),\qquad
 \mu_G^{B,\sigma}(b_e=0,b_{e'}=1)=(1-p)p'.
\]
 From Definition~\ref{def:score} we have
\[
 s^{B,\sigma}_{e,e'}-s^{B,\sigma}_{e',e}
 =1-\min\bigg\{\frac{\mu_G^{B,\sigma}(b_e=0,b_{e'}=1)}{\mu_G^{B,\sigma}(b_e=1,b_{e'}=0)},1\bigg\}=\frac{\max\{p-p',0\}}{p(1-p')}.
\]
If both $|p-(1-\tfrac{1}{q})|$ and $|p'-(1-\tfrac{1}{q})|$ are at most
$1/(2q)$, then $p(1-p')\ge1/(8q)$, and hence we obtain~\eqref{eq:localreduction} from 
$(p-p')^2\le2(p-(1-\tfrac{1}{q}))^2+2(p'-(1-\tfrac{1}{q}))^2$. 
Otherwise, at least one of $|p-(1-\tfrac{1}{q})|$ or $|p'-(1-\tfrac{1}{q})|$ exceeds
$1/(2q)$, so \eqref{eq:43g5g4} gives that $\sum_{z\in e\cup e'}\|\mu_z-\unif\|_2^4>\tfrac{1}{2q^2}$,
and the bound follows from
$|s^{B,\sigma}_{e,e'}-s^{B,\sigma}_{e',e}|\le1$.
\end{proof}

Recall the exploration process $B=B_{G,L}(\mathcal{R})$. The graph $G[B]$ is a forest, whose components are trees rooted at vertices of $\mathcal R\cap B$. Note that a root  $v\in \mathcal{R}$ which is incident to an excess edge is pruned in Step 4 of the exploration, and hence is excluded from $B$ (together with all its discovered descendants).
For every discovered vertex $w$, recall its descendant tree $T_w$ from Section~\ref{sec:block}, and let
\[
 \partial T_w=\{z\in V(T_w):\operatorname{depth}(z)=L\}.
\]
For $\sigma\in\Omega_G$, define $\widehat \mu_{T_w,w}^{\sigma}$ to be the marginal distribution of the colour of $w$ under the distribution $\mu_{T_w}(\cdot \mid  \sigma_{\partial T_w})$. Thus this distribution uses only discovery edges and ignores completely the correlations induced by excess edges; it also  fixes only the colours at global depth $L$. This is to be contrasted with the marginal $\mu_{G,v}^{B,\sigma}$ at an exploration root $v\in B$ which is computed on its tree component in the pruned forest $G[B]$, with the colours outside $B$ fixed to $\sigma$; the boundary therefore includes marked endpoints of excess edges that are potentially much closer to $v$. As an extreme example, if the root $v$ is excluded from $B$, then $\mu_{G,v}^{B,\sigma}$ is the point mass at $\sigma_v$.

Recall that $D=d+3$ and $L=\lfloor\tfrac{11}{20}\log_D n\rfloor$. Set
\[
 h=\big\lfloor\tfrac9{20}\log_D n\big\rfloor \mbox{ and let } \exc_h(B)=\{f\in \exc(B)\mid \mathrm{depth}(f)\leq h\}.\]
An excess edge is early if its depth is at most $h$, and late otherwise.
Recall that the depth of an excess edge is the minimum of its endpoint depths, and that $\exc(B)$ is the set of all excess edges recorded by the exploration producing $B$. For a discovered vertex $w$ (that may not be present in $B$) we let $\exc(T_w)$ denote the excess edges $f\in \exc(B)$ that have at least one endpoint in $T_w$.

The following definition is central to the argument.
\begin{definition}\label{def:tamemarginals}
Let $G=(V,E)$ be an $n$-vertex graph and $\sigma\in\Omega_G$. For $C>0$, we say that a root-set $\mathcal{R}\in\binom{V}{4}$ has tame $\sigma$-marginals and early-excess $\leq C$ if the following hold for the exploration set $B=B_{G,L}(\mathcal{R})$ and its full discovery forest.
\begin{enumerate}[label=(\roman*)]
\item For every root $v\in\mathcal R$, $\|\widehat \mu_{T_v,v}^{\sigma}-\unif\|_2\le n^{-0.26}$.
\item For every vertex $v$ of discovery depth at most $h$,
$\sum_{w\text{ child of }v}\|\widehat \mu_{T_w,w}^{\sigma}-\unif\|_2
 \le C\frac{\log n}{n^{1/25}}$.
\item The early excess edges satisfy $|\exc_h(B)|\leq C$; similarly, for every discovered vertex $w$ with $\operatorname{depth}(w)=\lfloor L/2\rfloor$, we have $|\exc(T_w)|\leq C$. Finally, $|\exc(B)|\leq n^{1/8}$.
\end{enumerate}
A graph $G$ is $(\sigma,C)$-good if $G$ has  $\geq (1-\frac1n)\binom n4$ root sets with tame $\sigma$-marginals and early-excess $\leq C$.
\end{definition}

As noted above, for an exploration root $v$, we can obtain the marginal $\mu_{G,v}^{B,\sigma}$ from the tree measure defining $\widehat \mu_{T_v,v}^{\sigma}$ by additionally fixing the marked vertices to their colours in $\sigma$. The following lemma bounds the effect of those additional constraints. Recall that $\lambda=D^{-19/20}$ and that $\Psi(G)$ is the average over four-element root sets
$\mathcal R$ of $\sum_{f\in\exc(B_{G,L}(\mathcal{R}))}\lambda^{4\operatorname{depth}(f)}$.
\newcommand{\statelemtamemarginals}{
Let $d$ be sufficiently large and $q\ge d/\log d$ be an integer. For every fixed $C>0$, there is $C'>0$ such that the following holds for all sufficiently large $n$.

Let $G=(V,E)$ be an $n$-vertex graph with $\leq Cn$ edges  and $\sigma\in\Omega_G$ be a balanced $q$-colouring. Suppose that $\mathcal R\in\binom{V}{4}$ is a root-set with tame $\sigma$-marginals and early-excess $\leq C$, and let $B_{\mathcal R}=B_{G,L}(\mathcal R)$.
Then, simultaneously for every $v\in\mathcal R$,
\[
 \|\mu_{G,v}^{B_{\mathcal R},\sigma}-\unif\|_2\leq n^{-0.255}
       +C'\sum_{f\in\exc_h(B_{\mathcal R})}\lambda^{\operatorname{depth}(f)}.
\]
Moreover, if $G$ is $(\sigma,C)$-good, then $\mathcal V(G,\sigma)\le C'\bigl(\frac{1}{n}+\Psi(G)\bigr)$.
}
\begin{lemma}\label{lem:tamemarginals}
\statelemtamemarginals
\end{lemma}

For a fixed colouring $\sigma\in[q]^V$ and $C>0$, let $\mathcal E(\sigma,C)$ be the set of graphs $G$ on $V$ with $\sigma\in\Omega_G$ for which more than a $1/n$ fraction of the four-element root sets fail to have tame $\sigma$-marginals and early-excess $\leq C$.

\newcommand{\statelemforestpair}{
Let $d$ be sufficiently large and $q\ge d/\log d$ be an integer.
For every $K>0$, there is $C>0$ such that the
following holds for all sufficiently large $n$. 

Fix a balanced colouring $\sigma\in[q]^n$ and an integer $0\le t\le M$. Then, with probability $\geq 1-1/n^K$, the planted graph $\rGhat_t(\sigma)$ is $(\sigma,C)$-good.
}
\begin{lemma}\label{lem:forestpair}
\statelemforestpair
\end{lemma}
The proofs of Lemmas~\ref{lem:tamemarginals} and~\ref{lem:forestpair} are given in Sections~\ref{sec:marginalproof} and~\ref{sec:forest}, respectively.

We are now ready to give the proof of Lemma~\ref{lem:inputs}.
\begin{leminputs}
\stateleminputs
\end{leminputs}
\begin{proof}
Let $\mathcal Q$ be the set of graphs satisfying the bounds in Lemma~\ref{lem:countlower}, so $\Prb(\rG\in\mathcal Q)=1-o(1)$. In particular, recall that graphs in $\mathcal{Q}$ have at most $M=\lceil(d+1)n/2\rceil$ edges.

Fix arbitrary $K>0$ and condition on $|E(\rG)|=m\le M$, so $\rG$ is uniform among $m$-edge graphs. Let $C$ be the constant in Lemma~\ref{lem:forestpair} (enlarged if necessary so that $C\ge d+1$), and let $C'>0$ be the corresponding constant from Lemma~\ref{lem:tamemarginals}. For a colouring $\sigma\in [q]^n$, write $\mathcal E(\sigma)=\mathcal E(\sigma,C)$ for the set of graphs $G$ on $[n]$ with $\sigma\in\Omega_G$ which are not $(\sigma,C)$-good, see Definition~\ref{def:tamemarginals}. All expectations below are under the conditioning $|E(\rG)|=m$.

For $0\le t<m$, let $\mathcal F_t$ be generated by the first $t$ revealed edges and the algorithm's other random choices up to time $t$, and consider the events
\[
 F_t=\{\rsigma_t\text{ is unbalanced}\}\cup\{\rG_t\in\mathcal E(\rsigma_t)\}.
\]
These events belong to $\mathcal F_t$. Fix $0\le t<m$.
We bound separately the contribution from unbalanced colourings and the contribution from balanced colourings for which $\rG_t\in\mathcal E(\rsigma_t)$.

First fix any $\sigma\in[q]^n$. If $N_b(\sigma)<t$, then $\sigma$ cannot be proper for a $t$-edge graph, so $\rsigma_t=\sigma$ implies that the total score satisfies $\rS_t=0$.
Otherwise, apply Lemma~\ref{lem:pathlaw}(i) with the set $\mathcal{E}$ equal to all graphs on $[n]$ to obtain 
$\E[\rS_t\one\{\rsigma_t=\sigma\}]\le q^{-n}$.
A uniform colouring is unbalanced w.p.  $\emm^{-\Omega((\log n)^2)}$, so summing over unbalanced colourings gives
\begin{equation}\label{eq:445g4g4b}
\E[\rS_t\one\{\rsigma_t\text{ is unbalanced}\}]\le\emm^{-\Omega((\log n)^2)}.
\end{equation}
Now fix a balanced colouring $\sigma$. Its planted graph is defined for every $t\le M$, and Lemma~\ref{lem:forestpair} gives
$\Prb(\rGhat_t(\sigma)\in\mathcal E(\sigma))\le n^{-K}$.
Applying Lemma~\ref{lem:pathlaw}(i) with the graph set $\mathcal E(\sigma)$ therefore yields
\[
 \E\big[\rS_t\one\{\rsigma_t=\sigma,\ \rG_t\in\mathcal E(\sigma)\}\big]
\le q^{-n}\Prb(\rGhat_t(\sigma)\in\mathcal E(\sigma))\le q^{-n}n^{-K}.
\]
There are trivially $\leq q^n$ balanced colourings, so summing the bound over balanced $\sigma$ gives
\begin{equation}\label{eq:445g4g4}
\E[\rS_t\one\{\rsigma_t\text{ is balanced},\ \rG_t\in\mathcal E(\rsigma_t)\}]\le n^{-K}.
\end{equation}
We therefore obtain by adding \eqref{eq:445g4g4b} and \eqref{eq:445g4g4} that, uniformly over $t<m$, it holds that
\[
 \E[\rS_t\one_{F_t}]
 \le \emm^{-\Omega((\log n)^2)}+n^{-K}.
\]
On a trial with $\rG\in\mathcal Q$ and $\rS_{\rG}>0$, every incoming colouring is proper. Whenever $F_t$ does not hold, Lemma~\ref{lem:tamemarginals} applies, since $|E(\rG_t)|\le M\le Cn$ and Lemma~\ref{lem:psimonotonicity} gives $\Psi(\rG_t)\le\Psi(\rG)$. Hence
\[
 \mathcal V(\rG_t,\rsigma_t)\le C'\bigl(1/n+\Psi(\rG)\bigr)
 \le C'\frac{1+(\log n)^{1/4}}n
 \le\frac{\sqrt{\log n}}n
\]
for sufficiently large $n$.
Thus failure of $\mathcal H$ on such a trial requires some $F_t$, and Lemma~\ref{lem:pathlaw}(ii) gives
\[
 \E\bigl[\rS_{\rG}\one\{\rG\in\mathcal Q,\mathcal H^c\}\bigr]
 \le\sum_{t<m}\E[\rS_t\one_{F_t}]
 \le M\bigl(\emm^{-\Omega((\log n)^2)}+n^{-K}\bigr).
\]
The bound is uniform in $m\le M$, and every graph in $\mathcal Q$ has at most $M$ edges. Averaging over the binomial edge count gives the same bound for $G(n,d/n)$. Since $K>0$ is arbitrary and both $\mathcal Q$ and $\mathcal H$ are independent of $K$, the expectation $\E\bigl[\rS_{\rG}\one\{\rG\in\mathcal Q,\mathcal H^c\}\bigr]$ is $n^{-\omega(1)}$, proving the lemma.
\end{proof}

\section{Proofs of Lemmas~\ref{lem:tamemarginals} and~\ref{lem:forestpair}}\label{sec:pairproof}
We prove Lemma~\ref{lem:tamemarginals} and Lemma~\ref{lem:forestpair}.
Section~\ref{sec:sensitivity} controls the effect of cuts and proves
Lemma~\ref{lem:tamemarginals}. Section~\ref{sec:tree} collects the tree
non-reconstruction estimates; Section~\ref{sec:forest} transfers these
estimates to the planted graph to prove Lemma~\ref{lem:forestpair}.

\subsection{The colouring recursion and cut bounds}\label{sec:sensitivity}
Let $T$ be a finite tree rooted at $o$, and write $\partial T$ for
its leaf set (including leaves at depths smaller than the height of $T$).
For a family $\alpha=\{\alpha_x\}_{x\in\partial T}$ of probability
vectors on $[q]$, consider the measure $\mu_T^\alpha(\cdot)$ over colourings: for any $\tau \in [q]^{V(T)}$, 
\[
 \mu_T^{\alpha}(\tau)
 \propto\one\{\tau\in \Omega_T\}
          \prod_{x\in\partial T}\alpha_x(\tau_x).
\]
We say that $\alpha$ is \emph{compatible with $T$} if there is a
proper colouring of $T$ with positive weight, or equivalently,
if $\mu_T^\alpha$ is well-defined.
We write $
 \mu_{T,o}^{\alpha}(i)=\mu_T^{\alpha}(\tau_o=i)$, for every $ i\in[q]$. 
 As special cases, taking $\alpha_x$ to be a point mass fixes the colour of $x$ and taking $\alpha_x=\unif$ leaves it free.
We also write $\mu_{T,o}^{\xi}$ when $\alpha_x$ is the point mass at $\xi_x$ on the specified boundary leaves and is uniform at the remaining leaves.

For each vertex $v$ of $T$, let $T_v$ be its descendant subtree.
We let $\mu_{T_v,v}^{\alpha}$ denote the marginal distribution
of the colour of $v$ under $\mu_{T_v}^{\alpha}$, where $\alpha$
is restricted to the leaves of $T_v$.
This distribution includes the constraints below $v$, but not
its edge to its parent, and $\mu_{T_x,x}^{\alpha}=\alpha_x$ at a leaf $x$.
If $v$ is not a leaf
and has children $w_1,\ldots,w_k$, then fixing $\tau_v=i$ leaves a fraction
$1-\mu_{T_{w_j},w_j}^{\alpha}(i)$ of the total colouring weight in each $T_{w_j}$.
These weights factorise over the disjoint child subtrees. Consequently
$\mu_{T_v,v}^{\alpha}=F(\mu_{T_{w_1},w_1}^{\alpha},\ldots,\mu_{T_{w_k},w_k}^{\alpha})$ computes the parent's colour distribution from those of its
child subtrees, where for probability vectors $z_1,\ldots,z_k$ on $[q]$ we define
\begin{equation}\label{eq:colouringrecursion}
 Z(z_1,\ldots,z_k)=\sum_{i=1}^q\prod_{j=1}^k(1-z_{j,i}),\qquad
 F_i(z_1,\ldots,z_k)
 =\frac{\prod_{j=1}^k(1-z_{j,i})}{Z(z_1,\ldots,z_k)}.
\end{equation}
For concreteness, in the case where $Z(z_1,\ldots,z_k)=0$,
we set $F(z_1,\ldots,z_k)=\unif$.

We use the following two properties of $F$. The first controls changes
of a child marginal, and the second will give the higher-moment bounds
in Section~\ref{sec:tree}. The proof is given in Appendix~\ref{sec:recursionproof}.
\newcommand{\statelemrecursion}{
Let $q\ge2$ be an integer. Then the following hold.
\begin{enumerate}[label=(\roman*)]
\item For every $\lambda>1/(q-1)$, there is $\eta>0$ such that, for
 every integer $k\ge1$ and probability vectors $z_2,\ldots,z_k$ with
 $\sum_{v=2}^k\|z_v-\unif\|_2\le2\eta$, every two probability vectors
 $z_1,z'_1$ satisfy
 \[
  \|F(z'_1,z_2,\ldots,z_k)-F(z_1,z_2,\ldots,z_k)\|_2
  \le\lambda\|z'_1-z_1\|_2.
 \]
\item For every even integer $r\ge4$, there is a constant $C=C(q,r)>0$ such
 that, for every integer $k\ge0$ and probability vectors
 $z_1,\ldots,z_k$,
 \[
 \begin{aligned}
  \frac{Z(z_1,\ldots,z_k)}{Z(\unif,\ldots,\unif)}
       \big\|F(z_1,\ldots,z_k)-\unif\big\|_2^r
  &\le\tfrac{2}{(q-1)^r}\sum_{v\in[k]}\|z_v-\unif\|_2^r+C^{k+1}\sum_{\substack{v,w\in[k]\\v\ne w}}
       \|z_v-\unif\|_2^{r-2}\|z_w-\unif\|_2^2.
 \end{aligned}
 \]
\end{enumerate}
}
\begin{lemma}\label{lem:recursion}
\statelemrecursion
\end{lemma}

We call a set $\Upsilon\subseteq V(T)$ an \emph{antichain} of $T$ if no
vertex of $\Upsilon$ is an ancestor of another vertex of $\Upsilon$.
For an antichain $\Upsilon$ of $T$, let $T_{\Upsilon}$ be the tree
obtained from $T$ by deleting the strict descendants of every
$v\in\Upsilon$; its leaf set is
$\partial T_{\Upsilon}=\Upsilon\cup(\partial T\cap V(T_{\Upsilon}))$.
For a family $\beta=(\beta_v)_{v\in\Upsilon}$ of probability vectors on
$[q]$, let $\mu_{T_{\Upsilon}}^{\alpha,\beta}$ be the probability
measure on $[q]^{V(T_{\Upsilon})}$ given by
\[
 \mu_{T_{\Upsilon}}^{\alpha,\beta}(\tau)
 \propto\one\{\tau\in\Omega_{T_{\Upsilon}}\}
 \prod_{x\in\partial T_{\Upsilon}\setminus\Upsilon}\alpha_x(\tau_x)
 \prod_{v\in\Upsilon}\beta_v(\tau_v),
\]
whenever the right-hand side is not identically zero, and write
$\mu_{T_{\Upsilon},o}^{\alpha,\beta}(i)=\mu_{T_{\Upsilon}}^{\alpha,\beta}(\tau_o=i)$
for $i\in[q]$.

For a vertex $w$ of $T_{\Upsilon}$, the descendant subtree of $w$ in
$T_{\Upsilon}$ is $(T_w)_{\Upsilon\cap V(T_w)}$, which we abbreviate
by $(T_{\Upsilon})_w$. Applying the definition above to the tree $T_w$
with root $w$, the antichain $\Upsilon\cap V(T_w)$, and the families
$(\alpha_x)_{x\in\partial T_w}$ and $(\beta_v)_{v\in\Upsilon\cap V(T_w)}$
gives the root distribution $\mu^{\alpha,\beta}_{(T_{\Upsilon})_w,w}$.

\begin{lemma}\label{lem:sparsecuts}
Let $q\ge2$ and $\lambda>1/(q-1)$. There is $\eta=\eta(q,\lambda)>0$ and,
for every integer $K\ge0$, a constant $C=C(q,\lambda,K)>0$ such that the
following holds for every tree $T$ with root $o$ and a compatible boundary condition
$\alpha=\{\alpha_x\}_{x\in\partial T}$.

Suppose that every nonleaf vertex $v$ of $T$ satisfies
\[\sum_{w\, \mbox{{\scriptsize  child of }}v}\|\mu_{T_w,w}^{\alpha}-\unif\|_2\le\eta.\] 
Then every antichain $\Upsilon$ of $T$ with $|\Upsilon|\le K$ and every
$\beta=\{\beta_v\}_{v\in \Upsilon}$ for which $\mu_{T_{\Upsilon}}^{\alpha,\beta}$ is defined satisfy
\[
 \|\mu_{T_{\Upsilon},o}^{\alpha,\beta}-\mu_{T,o}^{\alpha}\|_2
 \le C\sum_{v\in\Upsilon}\lambda^{\operatorname{dist}(o,v)}.
\]
\end{lemma}
\begin{proof}[Proof of Lemma~\ref{lem:sparsecuts}]
Choose $\eta$ as in Lemma~\ref{lem:recursion}(i).
Let $z_1,\ldots,z_k,z'_1,\ldots,z'_k$ be probability vectors on $[q]$
such that $Z(z'_1,\ldots,z'_k)>0$, and let $\lambda^*=\max\{\lambda,\sqrt2/\eta\}$. If $\sum_j\|z_j-\unif\|_2\le\eta$, then
\begin{equation}\label{eq:severalchildren}
 \|F(z'_1,\ldots,z'_k)-F(z_1,\ldots,z_k)\|_2\le\lambda^*\sum_j\|z'_j-z_j\|_2.
\end{equation}
Indeed, if $\sum_j\|z'_j-z_j\|_2\ge\eta$, then the left side is at most $\sqrt2\le(\sqrt2/\eta)\sum_j\|z'_j-z_j\|_2$. If $\sum_j\|z'_j-z_j\|_2<\eta$, replacing $z_j$ by $z'_j$ one at a time keeps the other coordinates within total distance $\eta+\sum_j\|z'_j-z_j\|_2<2\eta$ of $\unif$, so each step costs at most $\lambda\|z'_j-z_j\|_2$ by Lemma~\ref{lem:recursion}(i), and summing over $j$ gives \eqref{eq:severalchildren}.

For $v\in V(T_{\Upsilon})$ let $\mu_v=\mu_{T_v,v}^{\alpha}$, $\widetilde\mu_v=\mu_{(T_{\Upsilon})_v,v}^{\alpha,\beta}$, $k(v)=|\Upsilon\cap V(T_v)|$ and $S_v=\sum_{x\in\Upsilon\cap V(T_v)}\lambda^{\operatorname{dist}(v,x)}$. We will show by induction from the leaves that \[\|\widetilde\mu_v-\mu_v\|_2\le\sqrt2(\lambda^*/\lambda)^{k(v)}S_v.\]
 Note that $v=o$ gives the lemma with $C=\sqrt2(\lambda^*/\lambda)^K$. 
 
 If $k(v)=0$, then $\widetilde\mu_v=\mu_v$; if $v\in\Upsilon$, then $k(v)=S_v=1$. Otherwise both $\mu_v$ and $\widetilde\mu_v$ are given by $F$ applied to the children of $v$. Let $J$ be the set of children $w$ with $k(w)\ge1$; then $S_v=\lambda\sum_{w\in J}S_w$. If $J=\{w\}$, the other children are unchanged and satisfy the hypothesis, so Lemma~\ref{lem:recursion}(i) gives $\|\widetilde\mu_v-\mu_v\|_2\le\lambda\|\widetilde\mu_w-\mu_w\|_2$. If $|J|\ge2$, then $k(w)<k(v)$ for $w\in J$, and~\eqref{eq:severalchildren} gives
\[
 \|\widetilde\mu_v-\mu_v\|_2\le\lambda^*\sum_{w\in J}\sqrt2(\lambda^*/\lambda)^{k(v)-1}S_w=\sqrt2(\lambda^*/\lambda)^{k(v)}S_v.
\]
This completes the induction and proves the lemma.
\end{proof}

\subsubsection{Proof of Lemma~\ref{lem:tamemarginals}}\label{sec:marginalproof}
We use the full discovery trees from Section~\ref{sec:block}, with their depth-$L$ boundaries. When counting excess endpoints, count each vertex once for each incident excess edge; an incidence is early or late according to its edge. 
\begin{lemtamemarginals}
\statelemtamemarginals
\end{lemtamemarginals}
\begin{proof}
Fix $G,\sigma$. We first prove the marginal bound for a root set $\mathcal R$ with tame $\sigma$-marginals and early-excess $\leq C$, and abbreviate the exploration set as $B=B_{\mathcal R}$. For $v\in \mathcal{R}$, recall that $T_v$ is the discovery tree corresponding to $v$ (with $B$ obtained from $\cup_{v} T_v$ after pruning endpoints of excess edges and their descendants, and removing the depth-$L$ vertices). 

For every discovered vertex $v$, write
$\widehat \mu_v=\widehat \mu_{T_v,v}^{\sigma}$ for the discovery-tree marginal at $v$.
Let $\ell_0=\lfloor L/2\rfloor$ and
$K_0=2\lceil C\rceil$. By the early-excess assumption,  the number of early excess endpoint incidences and the number of excess endpoint incidences in each depth-$\ell_0$ subtree are bounded by $K_0$, cf.  Definition~\ref{def:tamemarginals}(iii).
For sufficiently large $d$, we have $\lambda=D^{-19/20}>1/(q-1)$, as required by Lemma~\ref{lem:sparsecuts}, and recall that $h=\lfloor \tfrac9{20}\log_D n\rfloor$.

For $v\in \mathcal{R}$ and its discovery-tree $T_v$, we will consider a suitable boundary condition on $T_v$ (as in Section~\ref{sec:sensitivity}) to account for the presence of excess edges; to capture the pruning step that leads to the exploration set $B$ it will be useful to think of excess edges as cuts. A cut at an endpoint $x$ of an excess edge $f$ fixes $x$ to
$\sigma_x$ and deletes its strict descendants.
The cut is \emph{deep} if $\operatorname{depth}(f)>h$, and \emph{early}
otherwise. The classification is by the edge's depth so one of its early endpoints
can have depth $h+1$; also, a vertex can have many cuts incident to it (but they have no further effect on the boundary condition). 

For a given collection of cuts, keep only endpoints with no other
cut endpoint as a strict ancestor; these form an antichain. Let $T$ be
the resulting tree, and let $\alpha=\{\alpha_x\}_{x\in\partial T}$ be
its boundary condition, where $\partial T$ is its full leaf set and, for $i\in [q]$,
\[
 \alpha_x(i)=\begin{cases}
 \one\{i=\sigma_x\},&x\text{ is a cut endpoint or }\operatorname{depth}(x)=L,\\
 1/q,&\text{otherwise}.
 \end{cases}
\]

For each discovered vertex $v$, let $T_v'$ be the tree obtained from $T_v$ by
applying only deep cuts, and let  $\mu'_v=\mu_{T_v',v}^{\alpha}$ be its marginal in $T_v'$
where $\alpha$ is applied to  the leaf set $\partial T_v'$. 
Note that for a root vertex $v$, applying all cuts instead gives the marginal $\mu_{G,v}^{B,\sigma}$, since every edge incident to $B$ is a discovery
edge by the proof of Lemma~\ref{lem:blockproperties}.
We will prove the following bounds for a constant $K$ depending only on $C,d,q$.

\begin{enumerate}
\item \label{eq:lateerror} Deep cut bounds: For every root $v\in\mathcal R$, $\|\mu'_v-\widehat \mu_v\|_2
       \le Kn^{1/8}\lambda^h=O(n^{-.3025})$.
\item \label{eq:lateerrorb} Child biases after deep cuts: For every vertex $v$ of depth less than $\ell_0$,
 $\sum_{w\, \mbox{{\scriptsize  child of }}v}\|\mu'_w-\unif\|_2
 =o(1)$.

\item \label{eq:lateerrorc} Early cut bounds: For every root $v\in\mathcal R$,
$ \|\mu_{G,v}^{B,\sigma}-\mu'_v\|_2
 \le K\sum_{f\in\exc_h(B)}
       \lambda^{\min\{\operatorname{depth}(f),\ell_0\}}$.
\end{enumerate}
From these, the desired marginal estimate follows by the triangle inequality. Indeed, using $|\exc_h(B)|\le C$ and the tameness bound on $\|\widehat \mu_v-\unif\|_2$ from
Definition~\ref{def:tamemarginals}(i), Items~\ref{eq:lateerror} and~\ref{eq:lateerrorc} give
\[
\begin{aligned}
 \|\mu_{G,v}^{B,\sigma}-\unif\|_2
 &\le\|\mu_{G,v}^{B,\sigma}-\mu'_v\|_2
      +\|\mu'_v-\widehat \mu_v\|_2+\|\widehat \mu_v-\unif\|_2\\
 &\le n^{-.26}+Kn^{1/8}\lambda^h+CK\lambda^{\ell_0}
      +K\sum_{f\in\exc_h(B)}\lambda^{\operatorname{depth}(f)}.
\end{aligned}
\]
Recall that $L=\lfloor\tfrac{11}{20}\log_D n\rfloor$,
$\ell_0=\lfloor L/2\rfloor$, $h=\lfloor\tfrac9{20}\log_D n\rfloor$,
and $\lambda=D^{-19/20}$, where $D=d+3$, so 
\[
\begin{aligned}
 n^{1/8}\lambda^h
 =O(n^{1/8-(19/20)(9/20)})=O(n^{-.3025}),\qquad 
 \lambda^{\ell_0}
 =O(n^{-(19/20)(11/40)})=O(n^{-.26125}).
\end{aligned}
\]
This proves the desired marginal
bound with $C'=K$.

It remains to prove Items~\ref{eq:lateerror} to~\ref{eq:lateerrorc}. For the rest of the proof, write $T_v^{\le\ell}$ for the subtree of $T_v$ induced by vertices of
global depth at most $\ell$; note in particular that $T_v=T_v^{\leq L}$. When $T=T_v^{\le\ell}$, $\partial T$
denotes its full leaf set, including any leaves of depth less than $\ell$.
\paragraph{Proof of Items~\ref{eq:lateerror} and~\ref{eq:lateerrorb}: deep cuts.}

Let $C_0$ be the constant in Lemma~\ref{lem:sparsecuts} for $K=K_0$.
For a vertex $v$ at  $\operatorname{depth}(v)\le h$, define
\[
 N_v=\sum_{f\in\exc(B)\setminus\exc_h(B)}|f\cap V(T_v)|.
\]
Thus $N_v$ counts late endpoint incidences below $v$. By
Definition~\ref{def:tamemarginals}(iii),
\[
 N_v\le K_0\quad\mbox{if }\ell_0\le\operatorname{depth}(v)\le h,
 \qquad \sum_{v\in\mathcal R}N_v\le2n^{1/8}.
\]
Also, $N_v=\sum_wN_w$ over the children $w$ of $v$ when
$\operatorname{depth}(v)<h$, since every late endpoint has depth greater than $h$.

Fix $v$ of depth $\ell\in[\ell_0,h]$ and apply Lemma~\ref{lem:sparsecuts}
to $T=T_v^{\le h}$ with
\[
 \alpha_x=\begin{cases}
 \widehat \mu_x,&\operatorname{depth}(x)=h,\\
 \unif,&\operatorname{depth}(x)<h,
 \end{cases}\quad(x\in\partial T),
 \qquad
 \Upsilon=\{x\in V(T):\operatorname{depth}(x)=h,\ N_x>0\},
\]
and $\beta_x=\mu'_x$ for $x\in\Upsilon$.
Summing over the descendants of each depth-$h$ vertex gives
\[
 \mu_{T,v}^{\alpha}=\widehat \mu_v,
 \qquad \mu_{T_{\Upsilon},v}^{\alpha,\beta}=\mu'_v.
\]
Indeed, the number of extensions with $\tau_x=i$ is proportional to
$\widehat \mu_x(i)$ before imposing the late constraints, and to
$\mu'_x(i)$ afterwards. Shallower leaves are unconstrained, giving
$\alpha_x=\unif$. All weights are positive at the corresponding colour
in $\sigma$, so the boundary conditions are compatible with $T$.
The set $\Upsilon$ is an antichain of size at most $N_v\le K_0$,
and the original child-bias sums are at most $C\log n/n^{1/25}=o(1)$
by Definition~\ref{def:tamemarginals}(ii). Lemma~\ref{lem:sparsecuts} therefore gives
\begin{equation}\label{eq:latepropagation}
 \|\mu'_v-\widehat \mu_v\|_2\le C_0N_v\lambda^{h-\ell}
       \quad \mbox{ for all vertices $v$ at }\operatorname{depth}(v)=\ell\in [\ell_0,h].
\end{equation}
We extend~\eqref{eq:latepropagation} to depths $\ell<\ell_0$ by induction.
Suppose it holds for the children of $v$; in the next two displays,
$w$ ranges over those children. Their total change satisfies
\[
 \sum_{w\, \mbox{{\scriptsize  child of }}v}\|\mu'_w-\widehat \mu_w\|_2
 \le C_0\sum_{w\, \mbox{{\scriptsize  child of }}v} N_w\lambda^{h-\ell-1}
 =C_0N_v\lambda^{h-\ell-1}
 \le 2C_0n^{1/8}\lambda^{h-\ell_0}=O(n^{-.04125}).
\]
Their original bias sum is at most $C\log n/n^{1/25}$, so every
intermediate family obtained by replacing the child distributions one
at a time has total bias $o(1)$. Applying Lemma~\ref{lem:recursion}(i)
at each replacement gives
\[
 \|\mu'_v-\widehat \mu_v\|_2
 \le\lambda\sum_{w\, \mbox{{\scriptsize  child of }}v}\|\mu'_w-\widehat \mu_w\|_2
 \le C_0N_v\lambda^{h-\ell},
\]
with the same constant $C_0$. At a root, $N_v\le2n^{1/8}$ proves
Item~\ref{eq:lateerror} for $K\ge2C_0$.
Adding the original child biases to their total change gives
\[
 \sum_{w\, \mbox{{\scriptsize  child of }}v}\|\mu'_w-\unif\|_2
 \le C\frac{\log n}{n^{1/25}}+2C_0n^{1/8}\lambda^{h-\ell_0}=o(1)
 \qquad(\operatorname{depth}(v)<\ell_0),
\]
which is Item~\ref{eq:lateerrorb}.

\paragraph{Proof of item~\ref{eq:lateerrorc}: early cuts.}
Fix a root $v$ and let $T=T_v^{\le\ell_0}$, with
\[
 \alpha_x=\begin{cases}
 \mu'_x,&\operatorname{depth}(x)=\ell_0,\\
 \unif,&\operatorname{depth}(x)<\ell_0,
 \end{cases}\qquad(x\in\partial T).
\]
Then $\mu_{T,v}^{\alpha}=\mu'_v$, and similarly the descendant marginal
at each $w\in V(T)$ is $\mu'_w$. Item~\ref{eq:lateerrorb} supplies
the child-bias hypothesis of Lemma~\ref{lem:sparsecuts}.

For each early endpoint $x\in V(T_v)$, let $\bar x$ be its ancestor
(possibly $x$ itself) at depth $\min\{\operatorname{depth}(x),\ell_0\}$.
Let $\bar X=\{\bar x: x\in V(T_v)\text{ is an early endpoint}\}$, and let
$\Upsilon$ be the set of $z\in\bar X$ that have no strict ancestor in $\bar X$.
 Then $\Upsilon$
is an antichain with $|\Upsilon|\le K_0$. For $z\in\Upsilon$, let
\[
 P_z=\{x\in V(T_z):x\text{ is marked or }\operatorname{depth}(x)=L\},
\]
and define the replacement boundary condition by
\[
 \beta_z(i)=\mu_{T_z}(\tau_z=i\mid\tau_{P_z}=\sigma_{P_z}),
 \qquad z\in\Upsilon,\ i\in[q].
\]
A deleted vertex lies below an early marked vertex whose colour is
already fixed. The vector $\beta_z$ incorporates all constraints inside
$T_z$, with the edge to its parent omitted. If $z$ is marked, in
particular whenever $\operatorname{depth}(z)<\ell_0$, then
$\beta_z(i)=\one\{i=\sigma_z\}$.
Summing over the removed descendants therefore gives
\[
 \mu_{T_{\Upsilon},v}^{\alpha,\beta}=\mu_{G,v}^{B,\sigma}.
\]
The boundary conditions are compatible, since every weight is positive
at its colour in $\sigma$. Lemma~\ref{lem:sparsecuts} now yields
\[
 \|\mu_{G,v}^{B,\sigma}-\mu'_v\|_2
 \le C_0\sum_{z\in\Upsilon}\lambda^{\operatorname{depth}(z)}
 \le 2C_0\sum_{f\in\exc_h(B)}
       \lambda^{\min\{\operatorname{depth}(f),\ell_0\}}.
\]
For the last inequality, assign each $z\in\Upsilon$ to an early endpoint
incidence $x\in f$ with $\bar x=z$. Then
$\operatorname{depth}(z)=\min\{\operatorname{depth}(x),\ell_0\}
\ge\min\{\operatorname{depth}(f),\ell_0\}$, so $\lambda<1$ bounds its
contribution by $\lambda^{\min\{\operatorname{depth}(f),\ell_0\}}$.
Distinct vertices $z$ use distinct incidences, and each edge has only
two endpoints, giving the factor $2$. Since $K\ge2C_0$, this proves
Item~\ref{eq:lateerrorc}, completing the marginal estimate.

It remains to prove the second part of the lemma, the score bound. By hypothesis, $|E(G)|\le Cn$; assume that at least $(1-\frac1n)\binom n4$ root sets have tame $\sigma$-marginals and early-excess $\leq C$.
Set $t=|E(G)|$ and $b=b(G,\sigma)$. Using the definition of $\mathcal V(G,\sigma)$ from \eqref{eq:V} and applying Lemma~\ref{lem:exact}(ii) gives
\begin{align}
\mathcal V(G,\sigma)&=\E\big[(\rs_{t+1}-\rho)^2\mid \rG_t=G,\rsigma_t=\sigma\big]\notag\\
&=(\rho-b)^2+\E[(\rs_{t+1}-b)^2\mid \rG_t=G,\rsigma_t=\sigma]\notag\\
&=(\rho-b)^2+\frac1{2|\mathcal U(G)|^2}
 \sum_{e,e'}
 \left(s^{B,\sigma}_{e,e'}-s^{B,\sigma}_{e',e}\right)^2,\label{eq:varianceb}
\end{align}
where the last sum ranges over ordered pairs
$e,e'\in\mathcal U(G)$ with $b_e(\sigma)\neq b_{e'}(\sigma)$, and each summand uses
$B=B_{G,L}(e\cup e')$. By~\eqref{eq:bupper} and the balancedness of $\sigma$, we have
\[
 \rho-b=\frac{\frac12\sum_{i=1}^q(n_i(\sigma)-n/q)^2+(1-\rho)t}{\binom n2-t}
 =O\left(\frac{(\log n)^2}{n}\right),
\]
and hence $(\rho-b)^2=O((\log n)^4/n^2)=o(1/n)$.
Pairs sharing an endpoint contribute $O(1/n)$ to~\eqref{eq:varianceb}: there are $O(n^3)$ of them, each squared score difference is at most one, and $|\mathcal U(G)|=\Theta(n^2)$.
For pairs with four distinct endpoints, apply Lemma~\ref{lem:localreduction} and enlarge the resulting nonnegative sum to all ordered disjoint pairs. This costs only a bounded factor, since $(\binom n2/|\mathcal U(G)|)^2=1+O(1/n)$.
Each four-element root set gives exactly six ordered disjoint pairs, and the exploration and local marginals depend only on their union. With $B_{\mathcal R}=B_{G,L}(\mathcal R)$, we obtain
\[\mathcal V(G,\sigma)\leq C'\bigg(\frac1n+\frac{1}{\binom{n}{4}}\sum_{\mathcal{R}\in \binom{V}{4}}\sum_{v\in \mathcal{R}} \Big\|\mu_{G,v}^{B_{\mathcal R},\sigma}-\unif\Big\|_2^4\bigg). \]

For a root set $\mathcal R$ with tame $\sigma$-marginals and early-excess $\leq C$, let $B=B_{\mathcal R}$.
Definition~\ref{def:tamemarginals}(iii) gives $|\exc_h(B)|\le C$, so the marginal bound of Lemma~\ref{lem:tamemarginals} and $(\sum_{j=1}^a z_j)^4\le a^3\sum_{j=1}^a z_j^4$ for nonnegative $z_j$ give, for every $v\in\mathcal R$,
\[
 \|\mu_{G,v}^{B,\sigma}-\unif\|_2^4
 \le C'\bigg(n^{-1.02}
       +\sum_{f\in\exc_h(B)}\lambda^{4\operatorname{depth}(f)}\bigg).
\]
Enlarge the sum to all $f\in\exc(B)$. Summing over the four roots and averaging over the tame root sets bounds their contribution by $C'[n^{-1.02}+\Psi(G)]$.
By hypothesis, the remaining root sets form at most a $1/n$ fraction. Each fourth power is at most one, so they contribute at most $4/n$. This proves the score bound and completes the proof.
\end{proof}

\subsection{Revisiting tree non-reconstruction}\label{sec:tree}

Let $T$ be a rooted tree with root $o$. The $q$-colouring broadcasting
process on $T$ chooses the colour of the root uniformly from $[q]$ and
chooses each child's colour independently and uniformly from the colours
different from its parent's colour.
Write $T_{\ell}$ for the tree truncated at depth $\ell$ and 
$\partial_{\ell} T$ for the vertices at distance $\ell$ from $o$.
Given a boundary colouring $\xi\in[q]^{\partial_{\ell} T}$ that could extend to
a proper colouring of $T_{\ell}$, define the conditional distribution
of the root by
\[
 \mu_{T_{\ell},o}^{\xi}(i)=\mu_{T_{\ell}}(\tau_o=i\mid\tau_{\partial_{\ell} T}=\xi),
 \qquad i\in[q].
\]
The following lemma is the Poisson Galton--Watson analogue of the
non-reconstruction part of \cite[Theorem~1]{Sly}; the extension from
regular trees to Poisson offspring is sketched in the remarks following
\cite[Section~2.3]{Sly}.

\begin{lemma}[{\cite{Sly}}]\label{lem:sly}
For every $\eta>0$, there is $q_0(\eta)$ such that the following
holds for every integer $q\ge q_0(\eta)$ and every $\gamma>0$ with $\gamma\le q\bigl(\log q+\log\log q+1-\log2-\eta\bigr)$.

Let $\rTree$ be a Poisson$(\gamma)$ Galton--Watson tree with root $o$ and consider the
$q$-colouring broadcasting process on~$\rTree$. For an integer $\ell\geq 0$, let $\rTree_{\ell}$
be its $\ell$-depth truncation and $\rv{\xi}_{\ell}$ be the broadcast colouring
on $\partial_{\ell}\rTree_{\ell}$. 
Then
\[
 \lim_{\ell\to\infty}\E\big\|\mu_{\rTree_{\ell},o}^{\rv{\xi}_{\ell}}-\unif\big\|_2^2=0,
 \qquad \unif=(1/q,\ldots,1/q),
\]
where the expectation is over the tree and the broadcast colouring.
\end{lemma}
We use only the qualitative convergence in
Lemma~\ref{lem:sly} and strengthen it to quantitative bounds.
Similar quantitative bounds have been studied on the regular tree in~\cite{MSW04,MSW07, SZ}.
In particular, Sly and Zhang~\cite{SZ} show that, under non-reconstruction and
strictly below the Kesten--Stigum bound, deviations of the conditional
root distribution from uniform larger than $\emm^{-\xi\ell}$ have probability at most
$\exp(-\emm^{\xi\ell})$ for some $\xi>0$ and all sufficiently large depths $\ell$.
Here we require explicit moment decay for Poisson offspring, which we
derive below. Sly's proof~\cite{Sly} already contains a quantitative
second-moment contraction; the argument below also controls every fixed
even moment, giving arbitrarily small inverse-polynomial failure
probabilities at logarithmic depths, as required in the sampling analysis to control the TV error.

For the quantitative estimates, set
\[
 \theta=-\frac1{q-1},\qquad \zeta=D^{-19/20}.
\]
For $q \ge d/\log d$ we have $D/q \le \log q + \log\log q + o(1)$ as
$d \to \infty$. Since $1 - \log 2 > 0$, the Poisson mean $\gamma = D$
therefore lies strictly inside the range of Lemma~\ref{lem:sly} for
sufficiently large $d$. Moreover, $2D\theta^2/\zeta = O((\log d)^2/d^{1/20})
= o(1)$, so $2D\theta^2 < \zeta$ for sufficiently large $d$.

We now present moment bounds for the conditional colour distribution of
the root. 
Let $0\le\gamma\le D$, let $\rTree$ be a Poisson$(\gamma)$ Galton--Watson
tree with root $o$, and run the $q$-colouring broadcasting process
on~$\rTree$. For an integer $\ell\ge0$, let $\rv{\xi}_{\ell}$ be the
broadcast colouring on $\partial_{\ell}\rTree$, and for an even integer
$r\ge2$ set
\[
 \mathcal{M}_r(\ell)=\E\big\|\mu_{\rTree_{\ell},o}^{\rv{\xi}_{\ell}}-\unif\big\|_2^r,
\]
where the expectation averages the tree and its boundary colouring. For
$\gamma=D$, Lemma~\ref{lem:sly} states that $\mathcal{M}_2(\ell)\to0$ as
$\ell\to\infty$. Every probability vector $\mu$ on $[q]$ satisfies
$\|\mu-\unif\|_2^2=\sum_i\mu(i)^2-1/q\le1$, so $\mathcal{M}_r(\ell)\le1$
for all $r$ and $\ell$.

\begin{lemma}\label{lem:moments}
Let $d$ be sufficiently large and $q\ge d/\log d$.
For every even integer $r\ge2$, there is a constant $C$  such that $\mathcal{M}_r(\ell)\le C\zeta^{r\ell/2}$
for every integer $\ell\ge0$ and every $0\le\gamma\le D$.
\end{lemma}
\begin{proof}[Proof of Lemma~\ref{lem:moments}]
Fix $0\le\gamma\le D$. We first obtain a recursion relating
$\mathcal{M}_2(\ell)$ to $\mathcal{M}_2(\ell-1)$ and use
Lemma~\ref{lem:sly} to deduce exponential decay in $\ell$. We then bound
each higher even moment in terms of the preceding ones. All constants
below depend only on $d$ and $q$ and, in the higher-moment argument, on
the fixed moment order $r$; none depends on $\ell$ or $\gamma$.

Fix $\ell\ge1$. To describe the recursion, consider the following
auxiliary experiment. Let $\rv{k}\sim\operatorname{Poisson}(\gamma)$
be the number of children of the root. Independently of $\rv{k}$,
generate independent trees $\rTree_v$, indexed by $v=1,2,\ldots$,
each a Poisson$(\gamma)$ Galton--Watson tree with root $v$ and
truncated at depth $\ell-1$. On each tree, run an independent
broadcast colouring, starting with a uniform colour at $v$, and let
$\rv{\xi}_v$ be the resulting colouring of
$\partial_{\ell-1}\rTree_v$. We use the first $\rv{k}$ trees as the
child subtrees. For each $v$, define the vector
\[
 \rz_v=\bigl(\mu_{\rTree_v,v}^{\rv{\xi}_v}(i)-1/q\bigr)_{i\in[q]}.
\]
Thus $\rz_v$ is the deviation from uniform of the conditional
distribution of the colour of $v$, given the tree $\rTree_v$ and the
colouring $\rv{\xi}_v$ of $\partial_{\ell-1}\rTree_v$. When $\ell=1$, we
have $\partial_0\rTree_v=\{v\}$, so $\rv{\xi}_v$ is the colour of $v$ and
this conditional distribution is a point mass.

Write $\widehat{\E}$ and $\widehat{\Prb}$ for expectation and
probability in this auxiliary experiment; they average over $\rv{k}$, the independent child trees,
and their independent broadcast colourings. In particular, the
vectors $\rz_v$ are independent of one another and of $\rv{k}$, and
\[
 \widehat{\E}\rz_v=0,\qquad
 \widehat{\E}\|\rz_v\|_2^r=\mathcal{M}_r(\ell-1).
\]
The first equality follows by averaging the conditional colour
distribution of $v$, whose unconditional colour is uniform. The
second is precisely the definition of $\mathcal{M}_r(\ell-1)$.

In the broadcast on $\rTree$, the colours of the children of $o$ depend
on the colour of $o$, whereas in the auxiliary experiment they are
independent and uniform. Let
\[
 \rv{O}=\bigl(\rv{k},(\rTree_v,\rv{\xi}_v)_{v\le\rv{k}}\bigr)
\]
consist of the number of children, the first $\rv{k}$ trees $\rTree_v$,
and the colourings $\rv{\xi}_v$ of $\partial_{\ell-1}\rTree_v$. Joining
$o$ to the roots of $\rTree_1,\ldots,\rTree_{\rv{k}}$ gives a tree
through depth $\ell$, and the colourings $\rv{\xi}_1,\ldots,\rv{\xi}_{\rv{k}}$
together form a colouring of its vertices at depth $\ell$. Applied to the
children of $o$ in $\rTree$, their descendant trees through depth $\ell$
of $\rTree$, and the broadcast colours on $\partial_\ell\rTree$, the same
construction gives the pair $(\rTree_\ell,\rv{\xi}_\ell)$. We compare the
distribution of $(\rTree_\ell,\rv{\xi}_\ell)$ with the distribution of $\rv{O}$ under
$\widehat{\Prb}$.

Write $\mu_v=\mu_{\rTree_v,v}^{\rv{\xi}_v}$ and $\rv{\tau}_v$ for the
colour of $v$ in the auxiliary distribution. Since
$\widehat{\Prb}(\rv{\tau}_v=c)=1/q$ and
$\widehat{\Prb}(\rv{\tau}_v=c\mid\rTree_v,\rv{\xi}_v)=\mu_v(c)$, Bayes'
rule shows that, conditional on $\rv{\tau}_v=c$, the law of
$(\rTree_v,\rv{\xi}_v)$ has density $q\mu_v(c)$ with respect to its law
under $\widehat{\Prb}$. In the broadcast on $\rTree$, if $o$ has colour
$i$, then the colour of a child $v$ of $o$ is uniform on
$[q]\setminus\{i\}$. Given that colour, the descendant tree of $v$
through depth $\ell$ of $\rTree$, together with the broadcast colours on
its vertices at depth $\ell$ of $\rTree$, has the law of
$(\rTree_v,\rv{\xi}_v)$ conditional on $\rv{\tau}_v$ being that colour.
Its density with respect to the law of $(\rTree_v,\rv{\xi}_v)$ under
$\widehat{\Prb}$ is therefore
\[
 \frac1{q-1}\sum_{c\ne i}q\mu_v(c)
 =\frac{q}{q-1}\bigl(1-\mu_v(i)\bigr)
 =1+q\theta\rz_{v,i}.
\]
Conditional on the colour of $o$, the  descendant trees and their
broadcast colours are independent over the children of $o$; in both
settings the number of children is Poisson$(\gamma)$ and independent of
everything else. Averaging over the uniform colour of $o$, we obtain with $\rDensity=\frac1q\sum_{i=1}^q \prod_{v=1}^{\rv{k}}(1+q\theta\rz_{v,i})$ that
\begin{equation}\label{eq:changeofmeasure}
 \E f(\rTree_\ell,\rv{\xi}_\ell)=\widehat{\E}\bigl[\rDensity f(\rv{O})\bigr]
\end{equation}
for every nonnegative function $f$ of a tree through depth $\ell$ and a
colouring of its vertices at depth $\ell$, where $f(\rv{O})$ is evaluated at the joined tree and its boundary colouring. For convenience, let
\[
 \rLike_i=\prod_{v=1}^{\rv{k}}(1+q\theta\rz_{v,i}) \quad  (i\in [q]), \mbox{ so that }
 \rDensity=\frac1q\sum_{i=1}^q\rLike_i.
\]
Write $\rLike=(\rLike_i)_{i\in[q]}$.
Since $\widehat{\E}(1+q\theta\rz_{v,i})=1$ and the factors in $\rLike_i$
are independent of one another and of $\rv{k}$, we have
$\widehat{\E}\rDensity=1$.

Let $\rMarginal=\rLike/(q\rDensity)$ when $\rDensity>0$ and
$\rMarginal=\unif$ otherwise. By~\eqref{eq:changeofmeasure} with
$f=\one\{\rDensity=0\}$, we have $\rDensity>0$ at
$(\rTree_\ell,\rv{\xi}_\ell)$ almost surely. There the tree recursion
gives $\mu_{\rTree_\ell,o}^{\rv{\xi}_\ell}(i)\propto\prod_v(1-\mu_v(i))$,
the product being over the children $v$ of $o$ (with $\mu_v$ computed
from the descendant tree of $v$ and the colours $\rv{\xi}_\ell$ on it).
Since $1+q\theta\rz_{v,i}=\frac{q}{q-1}(1-\mu_v(i))$, this is
$\rMarginal(i)$. Applying
\eqref{eq:changeofmeasure} to $f=\|\rMarginal-\unif\|_2^r$ therefore
gives
\begin{equation}\label{eq:weightedmoment}
 \mathcal{M}_r(\ell)
 =\E\big\|\mu_{\rTree_\ell,o}^{\rv{\xi}_\ell}-\unif\big\|_2^r
 =\widehat{\E}\bigl[\rDensity\|\rMarginal-\unif\|_2^r\bigr].
\end{equation}
The vectors $\rz_v$ are independent under $\widehat{\Prb}$, and the
factor $\rDensity$ accounts for their dependence through the colour of
$o$.

We now derive a recursion for the second moment. Colour symmetry and
$\sum_i\rz_{v,i}=0$ give
\[
 \widehat{\E}\rz_{v,i}^2=\frac{\mathcal{M}_2(\ell-1)}q,\qquad
 \widehat{\E}\rz_{v,i}\rz_{v,j}=-\frac{\mathcal{M}_2(\ell-1)}{q(q-1)}\quad(i\ne j).
\]
Set $\chi=q\theta^2\mathcal{M}_2(\ell-1)$. Conditional on $\rv{k}=k$, the factors in $\rLike_i$ are independent and
identically distributed, so using  $\widehat{\E}\rz_{v,i}=0$ we obtain for $j\neq i$
\[
 \widehat{\E}[\rLike_i^2\mid\rv{k}=k]
 =\prod_{v=1}^k\widehat{\E}(1+q\theta\rz_{v,i})^2
 =(1+\chi)^k, \quad \widehat{\E}[\rLike_i\rLike_j\mid\rv{k}=k]
=\bigl(1+q^2\theta^2\widehat{\E}\rz_{v,i}\rz_{v,j}\bigr)^k
=\bigl(1-\tfrac{\chi}{q-1}\bigr)^k.
\]
Since $\rv{k}\sim\operatorname{Poisson}(\gamma)$, averaging over $\rv{k}$ gives
\begin{equation*}
 \widehat{\E}\rLike_i^2=\widehat{\E}(1+\chi)^{\rv{k}}=\emm^{\gamma\chi},
 \qquad
 \widehat{\E}\rLike_i\rLike_j=\widehat{\E}\bigl(1-\tfrac{\chi}{q-1}\bigr)^{\rv{k}}
 =\emm^{-\gamma\chi/(q-1)}\quad(i\ne j).
\end{equation*}
From these it follows that
\[
 \widehat{\E}\rDensity^2
 =\tfrac1{q^2}\sum_{i,j\in[q]}\widehat{\E}\rLike_i\rLike_j
 =\tfrac1{q}\Bigl(\emm^{\gamma\chi}+(q-1)\emm^{-\gamma\chi/(q-1)}\Bigr).
\]
Since $\sum_i(\rLike_i-\rDensity)^2=\sum_i\rLike_i^2-q\rDensity^2$, it follows that
\begin{align*}
 \widehat{\E}\sum_i(\rLike_i-\rDensity)^2
   &=q\emm^{\gamma\chi}-q\widehat{\E}\rDensity^2
    =(q-1)\bigl(\emm^{\gamma\chi}-\emm^{-\gamma\chi/(q-1)}\bigr).
\end{align*}
Since $\gamma\le D$ and $\mathcal{M}_2(\ell-1)\le1$, we have
$0\le\gamma\chi\le qD\theta^2$. Using  
$\widehat{\E}\rDensity=1$ and expanding the exponentials to second order with $\gamma\chi=q\gamma\theta^2\mathcal{M}_2(\ell-1)$, we obtain
\begin{equation}
    \begin{aligned}
 \widehat{\E}\sum_i(\rLike_i-\rDensity)^2
   &\le q^2\gamma\theta^2\mathcal{M}_2(\ell-1)+C\mathcal{M}_2(\ell-1)^2,
     \label{eq:numeratormoment}\\
 \widehat{\E}(\rDensity-1)^2
   &=\widehat{\E}\rDensity^2-1\le C\mathcal{M}_2(\ell-1)^2, 
\end{aligned}
\end{equation}
where $C$ depends only on $q$ and $D$; note that in the second expression the
first-order terms cancel.

We split~\eqref{eq:weightedmoment} with $r=2$ according to whether
$\rDensity\ge1/2$. On $\{\rDensity\ge1/2\}$, since
$\rMarginal(i)-1/q=(\rLike_i-\rDensity)/(q\rDensity)$ for every $i\in[q]$, we have
\[\rDensity\|\rMarginal-\unif\|_2^2=\sum_i(\rLike_i-\rDensity)^2/(q^2\rDensity)
\le2q^{-2}\sum_i(\rLike_i-\rDensity)^2.\] 
On
$\{\rDensity<1/2\}$, we instead have $\rDensity\|\rMarginal-\unif\|_2^2\le1$.
Chebyshev's inequality then gives
\[
 \mathcal{M}_2(\ell)
 \le2q^{-2}\widehat{\E}\sum_i(\rLike_i-\rDensity)^2
    +\widehat{\Prb}(\rDensity<1/2)
 \le2q^{-2}\widehat{\E}\sum_i(\rLike_i-\rDensity)^2
    +4\widehat{\E}(\rDensity-1)^2.
\]
The bounds in~\eqref{eq:numeratormoment} 
show that, for every $\ell\ge1$,
\begin{equation}\label{eq:secondrecursion} \mathcal{M}_2(\ell)\le2\gamma\theta^2\mathcal{M}_2(\ell-1)+C \mathcal{M}_2(\ell-1)^2.
\end{equation}

We next show that $\mathcal{M}_2(\ell)$ decays exponentially in $\ell$.
We will write $\mathcal{M}_2^{(\gamma)}(\ell)$ for
$\mathcal{M}_2(\ell)$ defined with the Poisson$(\gamma)$ Galton--Watson tree,
to make the dependence on $\gamma$ explicit. Lemma~\ref{lem:sly} implies
\begin{equation}\label{eq:slyD}
 \lim_{\ell\to\infty}\mathcal{M}_2^{(D)}(\ell)=0.
\end{equation}
We claim that, for every $\ell\ge0$ and $0\le\gamma\le D$,
\begin{equation}\label{eq:gammacomparison}
 \mathcal{M}_2^{(\gamma)}(\ell)\le\mathcal{M}_2^{(D)}(\ell).
\end{equation}
Indeed, let $\rTree'$ be a Poisson$(D)$ Galton--Watson tree with
broadcast colouring $\rv{\sigma}$, and delete each of its edges
independently with probability $1-\gamma/D$. The component $\rTree$ of
$o$ is a Poisson$(\gamma)$ Galton--Watson tree, $\rv{\sigma}$ restricts to
a broadcast colouring of $\rTree$, and
$\partial_\ell\rTree\subseteq\partial_\ell\rTree'$. Hence
$\mu_{\rTree_\ell,o}^{\rv{\sigma}_{\partial_\ell\rTree}}$ is the
conditional expectation of
$\mu_{\rTree'_\ell,o}^{\rv{\sigma}_{\partial_\ell\rTree'}}$ given
$\rTree_\ell$ and $\rv{\sigma}_{\partial_\ell\rTree}$, and
\eqref{eq:gammacomparison} follows from conditional Jensen's inequality. Choose $y_0>0$ so that $2D\theta^2+Cy_0<\zeta$, with $C$ as in
\eqref{eq:secondrecursion}. By~\eqref{eq:slyD}, there is $\ell_*$,
depending only on $d$ and $q$, such that
$\mathcal{M}_2^{(D)}(\ell)\le y_0$ whenever $\ell\ge\ell_*$.
By~\eqref{eq:gammacomparison}, $\mathcal{M}_2^{(\gamma)}(\ell)\le y_0$
for every $\ell\ge\ell_*$ and $0\le\gamma\le D$. For $\ell>\ell_*$,
\eqref{eq:secondrecursion} and $\gamma\le D$ then give
\[
 \mathcal{M}_2^{(\gamma)}(\ell)
 \le\bigl(2D\theta^2+C\mathcal{M}_2^{(\gamma)}(\ell-1)\bigr)\mathcal{M}_2^{(\gamma)}(\ell-1)
 \le(2D\theta^2+Cy_0)\mathcal{M}_2^{(\gamma)}(\ell-1)
 \le\zeta\mathcal{M}_2^{(\gamma)}(\ell-1).
\]
Iterating down to $\ell_*$ and using
$\mathcal{M}_2^{(\gamma)}(\ell)\le1\le\zeta^{\ell-\ell_*}$ for $\ell\le\ell_*$
gives $\mathcal{M}_2^{(\gamma)}(\ell)\le\zeta^{-\ell_*}\zeta^{\ell}$ for
every $\ell\ge0$ and $0\le\gamma\le D$. This completes the proof for the case $r=2$ of the lemma.

We now turn to the higher moments. Fix an even $r\ge4$. We first show
that, for every $\ell\ge1$,
\begin{equation}\label{eq:higherrecursion}
 \mathcal{M}_r(\ell)\le2\gamma|\theta|^r \mathcal{M}_r(\ell-1)
       +C \mathcal{M}_{r-2}(\ell-1)\mathcal{M}_2(\ell-1).
\end{equation}
By~\eqref{eq:weightedmoment},
$\mathcal{M}_r(\ell)=\widehat{\E}[\rDensity\|\rMarginal-\unif\|_2^r]$.
Condition on $\rv{k}=k$ and recall that the child probability vectors are
$\mu_v=\unif+\rz_v$. By~\eqref{eq:colouringrecursion}, we have 
$Z(\unif,\ldots,\unif)=q(1-1/q)^k$ and $
 \rDensity=\frac{Z(\mu_1,\ldots,\mu_k)}{Z(\unif,\ldots,\unif)}$.
When $\rDensity>0$, we also have
$\rMarginal=F(\mu_1,\ldots,\mu_k)$. Thus Lemma~\ref{lem:recursion}(ii)
gives
\[
 \rDensity\|\rMarginal-\unif\|_2^r
 \le 2|\theta|^r\sum_{v\in[k]}\|\rz_v\|_2^r
    +C^{k+1}\sum_{v,w\in[k];\, v\ne w}
          \|\rz_v\|_2^{r-2}\|\rz_w\|_2^2.
\]
The same inequality holds trivially when $\rDensity=0$.
Conditional on $\rv{k}=k$, the vectors $\rz_v$ are independent under
$\widehat{\Prb}$, so the two sums on the right have expectations
$k\mathcal{M}_r(\ell-1)$ and
$k(k-1)\mathcal{M}_{r-2}(\ell-1)\mathcal{M}_2(\ell-1)$, respectively.
Since $\widehat{\E}\rv{k}=\gamma$ and
$\widehat{\E}[C^{\rv{k}+1}\rv{k}(\rv{k}-1)]$ is bounded uniformly for
$\gamma\le D$, averaging over $\rv{k}$ proves~\eqref{eq:higherrecursion}.

We finish by induction on the even moment order $r$. The case $r=2$
was proved above. For an even $r\ge4$, assume that the lemma holds for
the order $r-2$. Together with the case $r=2$, this gives, for every
$\ell\ge1$ and $0\le\gamma\le D$,
$\mathcal{M}_{r-2}(\ell-1)\le C\zeta^{(r-2)(\ell-1)/2}$ and
$\mathcal{M}_2(\ell-1)\le C\zeta^{\ell-1}$, so the last term
of~\eqref{eq:higherrecursion} is at most $C\zeta^{r(\ell-1)/2}$. Also, $\gamma\le D$ and
$2D|\theta|^r\le(2D\theta^2)^{r/2}<\zeta^{r/2}$, where we used $2D\ge1$
and $2D\theta^2<\zeta$.
We also have $\mathcal{M}_r(0)\le1$.
Iterating~\eqref{eq:higherrecursion} down to $\ell=0$ gives, for every
$\ell\ge0$,
\[
 \mathcal{M}_r(\ell)
 \le (2D|\theta|^r)^{\ell}
    +C\sum_{j=0}^{\ell-1}(2D|\theta|^r)^j\zeta^{r(\ell-1-j)/2}
 \le\zeta^{r\ell/2}\biggl(1+\frac{C\zeta^{-r/2}}
       {1-2D|\theta|^r\zeta^{-r/2}}\biggr).
\]
The last inequality uses $(2D|\theta|^r)^\ell\le\zeta^{r\ell/2}$ and sums
the geometric series with ratio $2D|\theta|^r\zeta^{-r/2}<1$; for $\ell=0$
the sum is empty. Although the constants may depend on $r$,
the condition on $d,q$ is the same for every fixed moment order.
\end{proof}

\subsection{Tameness of planted neighbourhoods}\label{sec:forest}\label{sec:pair-conclusion}

\begin{lemforestpair}
\statelemforestpair
\end{lemforestpair}
\begin{proof}
Let $\ell_0=\lfloor L/2\rfloor$.
Fix $K>0$ and fix a balanced colouring $\sigma$. Independently of $\rGhat_t(\sigma)$, choose $\rRoots$ uniformly from $\binom V4$ and let $\rB=B_{\rGhat_t,L}(\rRoots)$. We first establish tameness of $\rRoots$ with  probability $1-O(n^{-K-2})$, then apply Markov's inequality to show $\rGhat_t(\sigma)$ satisfies tameness.

It is useful to compare three random models, with laws
$\Prb_t$, $\Prb_p$ and $\Prb_\gamma$, respectively.
To focus on the randomness over these random models, we condition on their input $\rRoots=\mathcal R$.
Let $N=N_b(\sigma)$, $p=t/N$ and $\gamma=2t/n$.
If $t=0$, the roots are isolated and there are no excess
edges, so all the assertions are immediate. Henceforth assume $t>0$.
Since $t\le M$, we have $\gamma\le d+1+2/n<D-1$ for sufficiently
large $n$. 

\begin{enumerate}[label=\arabic*.]
\item \textbf{Planted graph $\Prb_t$.}
Sample $\rGhat_t(\sigma)$ and run the exploration of
Section~\ref{sec:block} from $\mathcal R$, recording the colour
$\sigma_v$ of every discovered vertex $v$.

\item \textbf{Binomial planted graph $\Prb_p$.}
Let $\rGhat_p(\sigma)$ include each bichromatic pair of $\sigma$
independently with probability $p$, and run the same exploration.
Conditioning $\rGhat_p(\sigma)$ to have exactly $t$ edges gives
$\rGhat_t(\sigma)$.

\item \textbf{Poisson broadcast forest $\Prb_\gamma$.}
Let $\rv{\mathcal T}_\gamma(\sigma)$ consist of four independent
Poisson$(\gamma)$ Galton--Watson trees through depth $L$, one for each
$v\in\mathcal R$, each carrying the broadcast process of
Section~\ref{sec:tree} started from the root colour $\sigma_v$.
Explore $\rv{\mathcal T}_\gamma(\sigma)$ breadth-first with a single
queue, as in Steps~1--2 of Section~\ref{sec:block}. As the forest is
explored, we label each tree vertex by a distinct vertex of $V$: the
roots by the corresponding vertices of $\mathcal R$, and each new child
of colour $i$ by a uniformly random vertex of $\sigma^{-1}(i)$ not yet
used as a label.
\end{enumerate}
Set $N_{\max}=n^{7/12}$. We say the discovery phase, i.e.\ Steps~1--2
of the exploration  of
Section~\ref{sec:block}, \emph{overflows} if it discovers more than $N_{\max}$
vertices. In each model we stop the discovery phase as soon as it
overflows, after finishing the current parent's expansion, and then skip
Step~3. If a Poisson child count exceeds the number of available labels
of its colour, we instead declare overflow and stop immediately,
recording the count and an overflow flag in the transcript.
Such transcripts have probability zero under $\Prb_t$ and $\Prb_p$.
The \emph{discovery transcript} is the resulting labelled coloured
forest (for $\rGhat_t(\sigma)$ and $\rGhat_p(\sigma)$ each vertex is its
own label), and for these two graphs the \emph{full transcript} also
records the outcome of the excess queries in Step~3.
Without overflow, the discovery transcript contains the full discovery
forest through depth $L$. For a vertex $v$ of this forest, let $T_v$ be
its descendant tree and let $\xi_v$ be the colours on $\partial T_v$, the
vertices of $T_v$ at depth $L$. For $\rGhat_t(\sigma)$ and
$\rGhat_p(\sigma)$ we have $\xi_v=\sigma_{\partial T_v}$, so
$\mu_{T_v,v}^{\xi_v}=\widehat\mu_{T_v,v}^{\sigma}$; for
$\rv{\mathcal T}_\gamma(\sigma)$ the colours $\xi_v$ come from the broadcast.

Let $\mathcal G_{\rm disc}$ be the event that the discovery phase does not
overflow and the full discovery forest satisfies all the following bounds (D1)--(D6).
\begin{center}
\begin{tabular}{lll}
 & Quantity & Upper bound\\[3pt]
(D1) & Total population through depth $L$ & $n^{1/200}D^L$\\[3pt]
(D2) & Total population through depth $h+2$ & $n^{1/200}D^{h+2}$\\[3pt]
(D3) & Population of $T_v$ for $\operatorname{depth}(v)=\ell_0$
 & $n^{1/200}D^{L-\ell_0}$\\[3pt]
(D4) & Number of children at any vertex through depth $h+1$ & $C_1\log n$\\[3pt]
(D5) & Bias $\|\mu_{T_v,v}^{\xi_v}-\unif\|_2$ at every root $v\in\mathcal R$ & $n^{-0.26}$\\[3pt]
(D6) & Bias $\|\mu_{T_v,v}^{\xi_v}-\unif\|_2$ through depth $h+1$ & $n^{-0.04}$
\end{tabular}
\end{center}
Here $C_1$ depends only on $K$, $d$ and $q$. 
This event $\mathcal G_{\rm disc}$ is determined by the discovery transcript. 
Let $\mathcal G_{\rm exc}$ be the event that Step~3 of the exploration is
run and finds at most $n^{1/8}$ excess edges in total, at most $C_2$ early
excess edges, and, for every vertex $w$ at depth $\ell_0$, at most $C_2$
excess edges with an endpoint in $T_w$, where $C_2$ also depends only on $K$,
$d$ and $q$. This event is determined by the full transcript.
The proof rests on the following
four properties.
\begin{enumerate}[label=(\roman*)]
\item\label{it:forest-fixedcomparison}
For every event $A$ determined by the full transcript,
$\Prb_t(A)=O\big(\Prb_p(A)\big)$.

\item\label{it:forest-treecomparison}
For every event $A$ determined by the discovery transcript,
$\Prb_p(A)=O\big(\Prb_\gamma(A)^{1/2}\big)$.

\item\label{it:forest-discoverybounds}
$\Prb_\gamma(\mathcal G_{\rm disc}^c)=O(n^{-2K-4})$.

\item\label{it:forest-excess}
$\Prb_p(\mathcal G_{\rm disc}\setminus\mathcal G_{\rm exc})=O(n^{-K-2})$.
\end{enumerate}
The implicit constants in~\ref{it:forest-fixedcomparison}
and~\ref{it:forest-treecomparison} depend only on $d$ and $q$.
Combining the four properties, we obtain
\[
 \Prb_t\big((\mathcal G_{\rm disc}\cap\mathcal G_{\rm exc})^c\big)
 \le\Prb_t(\mathcal G_{\rm disc}^c)
   +\Prb_t(\mathcal G_{\rm disc}\setminus\mathcal G_{\rm exc})
 =O\big(\Prb_\gamma(\mathcal G_{\rm disc}^c)^{1/2}\big)
   +O\big(\Prb_p(\mathcal G_{\rm disc}\setminus\mathcal G_{\rm exc})\big)
 =O(n^{-K-2}),
\]
where the first term uses~\ref{it:forest-fixedcomparison},
\ref{it:forest-treecomparison} and~\ref{it:forest-discoverybounds}, and
the second uses~\ref{it:forest-fixedcomparison}
and~\ref{it:forest-excess}.

On $\mathcal G_{\rm disc}\cap\mathcal G_{\rm exc}$, the root set
$\mathcal R$ has tame $\sigma$-marginals and early-excess at most
$C=\max\{C_1,C_2\}$: (D5) gives Definition~\ref{def:tamemarginals}(i),
(D4) and (D6) give Definition~\ref{def:tamemarginals}(ii) 
and $\mathcal G_{\rm exc}$ gives Definition~\ref{def:tamemarginals}(iii). Since all bounds are uniform in
$\mathcal R$, the uniformly random root set $\rRoots$, chosen independently
of $\rGhat_t$, fails to be tame with probability $O(n^{-K-2})$. This probability is the expected fraction of root sets
that fail to be tame in $\rGhat_t$, so by Markov's inequality the
probability that this fraction exceeds $1/n$, that is, that $\rGhat_t$
is not $(\sigma,C)$-good, is at most $n\cdot O(n^{-K-2})\le n^{-K}$ for
sufficiently large $n$.  We next prove
\ref{it:forest-fixedcomparison}--\ref{it:forest-excess}.

\paragraph{Proof of~\ref{it:forest-fixedcomparison}.}
Step~2 of the exploration expands at most $N_{\max}$ parents, since every
expansion starts with at most $N_{\max}$ discovered vertices, and each
expansion queries at most $n$ pairs. Step~3 of the exploration is run only
without overflow, and then queries at most $N_{\max}^2$ pairs among the
discovered vertices. Hence a full transcript $\tau$ queries at most
$N_{\max}n+N_{\max}^2=o(n^2)$ pairs, and leaves a set of
$N_\tau\ge N/2$ bichromatic pairs unqueried. Let $s$ be the number of
queried pairs that $\tau$ records as edges. Since $\Prb_t$ is $\Prb_p$
conditioned on having $t$ edges, and under $\Prb_p$ the unqueried pairs
are independent of $\tau$, we have
\[
 \frac{\Prb_t(\tau)}{\Prb_p(\tau)}
 =\frac{\Prb\big(\operatorname{Bin}(N_\tau,p)=t-s\big)}
       {\Prb\big(\operatorname{Bin}(N,p)=t\big)}
 \le\frac{\max_k\Prb\big(\operatorname{Bin}(N_\tau,p)=k\big)}
         {\Prb\big(\operatorname{Bin}(N,p)=t\big)}
 =O\left(\frac{(1+N_\tau p)^{-1/2}}{(1+t)^{-1/2}}\right)
 =O(1)
\]
by standard estimates for binomial point probabilities, using $t=Np$ and
$N_\tau p\ge t/2$. Summing over $\tau\in A$
proves~\ref{it:forest-fixedcomparison}.

\paragraph{Proof of~\ref{it:forest-treecomparison}.}
Let $\rLikelihood$ be the density of $\Prb_p$ with respect to $\Prb_\gamma$
on discovery transcripts. By Cauchy--Schwarz,
$\Prb_p(A)=\E_\gamma[\rLikelihood\one_A]\le(\E_\gamma\rLikelihood^2)^{1/2}\Prb_\gamma(A)^{1/2}$,
so it suffices to show $\E_\gamma\rLikelihood^2=O(1)$. Both models build
the discovery forest one expansion at a time, and they differ only in the
number of children of each colour, which is binomial in $\rGhat_p(\sigma)$
and Poisson in $\rv{\mathcal T}_\gamma(\sigma)$. Since $p\le M/N=O(1/n)$,
and before overflow only few vertices are used, these laws have nearly
equal means and are close. Each step then multiplies
$\E_\gamma\rLikelihood^2$ by $1+o(1/N_{\max})$, and over the at most
$qN_{\max}$ steps this gives $O(1)$. We now make this precise.

Consider an expansion of a parent $v$ in Step~2 of the exploration, and
for $i\ne\sigma_v$ let $N_i$ be the number of undiscovered vertices of
colour $i$ at its start. Under $\Prb_p$, conditional on the discovery
transcript so far, the numbers of new children of the colours
$i\ne\sigma_v$ are independent $\operatorname{Bin}(N_i,p)$, and given
these numbers the children of colour $i$ form a uniform subset of the
$N_i$ undiscovered vertices of colour $i$. Under $\Prb_\gamma$ the
numbers are independent Poisson$(\nu)$ with $\nu=\gamma/(q-1)$, and the
labels are chosen in the same way. Let $\rv Z_1,\rv Z_2,\ldots$ be the
revealed counts in the order they are revealed, and let $\mathcal D_j$ be
generated by $\rv Z_1,\ldots,\rv Z_j$ and the labels chosen with them; in
particular, $\mathcal D_{j-1}$ determines the parent, the colour $i$ and
the number $N_i$ belonging to $\rv Z_j$. Set
$\rv R_j=\Prb(\operatorname{Bin}(N_i,p)=\rv Z_j)/\Prb(\operatorname{Poisson}(\nu)=\rv Z_j)$,
and $\rv R_j=1$ for all $j$ beyond the last revealed count. Since the two
models differ only in the law of these counts,
$\rLikelihood=\prod_{j\ge1}\rv R_j$.

Every expansion starts with at most $N_{\max}$ discovered vertices, so
$n_i(\sigma)-N_{\max}\le N_i\le n_i(\sigma)$. Since $\sigma$ is balanced,
$n_i(\sigma)=n/q+O(\sqrt n\log n)$ and
$2N=(1-1/q)n^2(1+O(\log^2n/n))$. As $p/\nu=(q-1)n/(2N)$, it follows
that $N_ip/\nu=(N_iq/n)(1+O(\log^2n/n))$, and hence, uniformly in $t>0$,
$|N_ip/\nu-1|\le\delta:=O(N_{\max}/n+\log n/\sqrt n)$.

It follows that each $\rv R_j$ has conditional second moment close to
one. Indeed, given $\mathcal D_{j-1}$, under $\Prb_\gamma$ the count
$\rv Z_j$ is Poisson$(\nu)$, so using
$\Prb(\operatorname{Bin}(N_i,p)=z)\le \emm^{pz}\,\Prb(\operatorname{Poisson}(N_ip)=z)$,
\[
\begin{aligned}
 \E_\gamma\big[\rv R_j^2\mid\mathcal D_{j-1}\big]
 &=\sum_{z\ge0}
   \frac{\Prb\big(\operatorname{Bin}(N_i,p)=z\big)^2}
        {\Prb\big(\operatorname{Poisson}(\nu)=z\big)}
 \le\sum_{z\ge0}\emm^{2pz}\,
   \frac{\Prb\big(\operatorname{Poisson}(N_ip)=z\big)^2}
        {\Prb\big(\operatorname{Poisson}(\nu)=z\big)}\\
 &=\exp\Big(\nu\Big[\Big(\frac{N_ip}{\nu}\Big)^2(\emm^{2p}-1)
      +\Big(\frac{N_ip}{\nu}-1\Big)^2\Big]\Big)
 \le \emm^{C(\delta^2+p)},
\end{aligned}
\]
where the last step uses $\nu\le D$ and $|N_ip/\nu-1|\le\delta$.

Finally, at most $N_{\max}$ parents are expanded, each revealing $q-1$
counts, so $\rv R_j=1$ for $j>qN_{\max}$. Since
$\rv R_1,\ldots,\rv R_{j-1}$ are $\mathcal D_{j-1}$-measurable, iterating
the bound above gives
\[
 \E_\gamma\rLikelihood^2
 =\E_\gamma\prod_{j\le qN_{\max}}\rv R_j^2
 \le \emm^{CqN_{\max}(\delta^2+p)}=O(1),
\]
since $N_{\max}\delta^2=O(N_{\max}^3/n^2+N_{\max}\log^2n/n)=o(1)$ and
$N_{\max}p=O(N_{\max}/n)=o(1)$. This proves~\ref{it:forest-treecomparison}.

\paragraph{Proof of~\ref{it:forest-discoverybounds}.}
We show that the whole forest $\rv{\mathcal T}_\gamma(\sigma)$ through
depth $L$ satisfies (D1)--(D6) with probability $1-O(n^{-2K-4})$. By (D1)
its population is then at most $n^{1/200}D^L\le n^{0.555}<N_{\max}$, so the
discovery phase does not overflow and its transcript is the whole forest;
hence $\Prb_\gamma(\mathcal G_{\rm disc}^c)=O(n^{-2K-4})$.

We first prove (D1)--(D4). Since $\gamma<D$, the exponential
tail~\eqref{eq:poissonvolume} bounds the population through depth $\ell$
by $n^{1/200}D^\ell$, with failure probability $n^{-\omega(1)}$. Applying~\eqref{eq:poissonvolume}
with $t=n^{1/200}/4$ to each of the four trees, through depths $L$ and
$h+2$, gives (D1) and (D2). For (D3), apply~\eqref{eq:poissonvolume},
conditional on presence of $w$, to the descendant tree of each vertex $w$ at depth
$\ell_0$, through depth $L-\ell_0$, with a union bound over the expected
$O(D^{\ell_0})$ such vertices. Each of these failure probabilities is
$O(D^L\emm^{-\kappa n^{1/200}/4})$. For (D4),
$\Prb(\operatorname{Poisson}(\gamma)>C_1\log n)=O(n^{-2K-5})$ for $C_1$
large enough depending on $K$, and a union bound over the expected
$O(n^{9/20})$ vertices through depth $h+1$ gives failure probability
$O(n^{-2K-4})$.

It remains to prove (D5) and (D6). Every vertex through depth $h+1$ is at
distance at least $\tfrac1{10}\log_Dn-2$ from the depth-$L$ boundary, so
Lemma~\ref{lem:moments} makes all moments of its bias polynomially small,
and Markov's inequality with a large moment order controls all such
vertices at once. Precisely, let $v$ be a vertex of
$\rv{\mathcal T}_\gamma(\sigma)$ at depth $\ell<L$. Conditional on the
presence of $v$ and on its colour, its descendant tree $T_v$ through
depth $L$ is a Poisson$(\gamma)$ Galton--Watson tree through depth
$L-\ell$, carrying the broadcast process started from the colour of $v$.
By colour symmetry and Lemma~\ref{lem:moments}, the conditional $r$th
moment of $\|\mu_{T_v,v}^{\xi_v}-\unif\|_2$ is
$\mathcal M_r(L-\ell)\le C\zeta^{r(L-\ell)/2}=C(D^{L-\ell})^{-19r/40}$.
At a root, $\ell=0$ and $D^L\ge n^{11/20}/D$, so this moment is
$O(n^{-209r/800})$; through depth $h+1$, $D^{L-\ell}\ge n^{1/10}/D^2$, so
it is $O(n^{-19r/400})$. Since $n^{-0.26}=n^{-208/800}$,
$n^{-0.04}=n^{-16/400}$, and the expected number of vertices at depth
$\ell$ is $4\gamma^\ell$, Markov's inequality and a union bound over the
four roots and over all vertices through depth $h+1$ give
\begin{align*}
 \Prb_\gamma(\text{(D5) fails})
 &\le4\cdot O\big(n^{-209r/800}\big)\,n^{208r/800}
 =O\big(n^{-r/800}\big),\\
 \Prb_\gamma(\text{(D6) fails})
 &\le\sum_{\ell\le h+1}4\gamma^\ell\cdot
   O\big(n^{-19r/400}\big)\,n^{16r/400}
 =O\big(n^{9/20}\big)\,O\big(n^{-3r/400}\big)
 =O\big(n^{9/20-3r/400}\big).
\end{align*}
For a fixed even $r\ge800(2K+4)$ we have $r/800\ge2K+4$ and
$3r/400-9/20\ge6(2K+4)-9/20\ge2K+4$, so both are $O(n^{-2K-4})$.
This proves~\ref{it:forest-discoverybounds}.

\paragraph{Proof of~\ref{it:forest-excess}.}
Condition on a discovery transcript $\tau$ in $\mathcal G_{\rm disc}$. Then
Step~3 of the exploration is run, and the pairs it tests were not
queried in Step~2, so under $\Prb_p$ they are edges independently, each
with probability at most $p=O(1/n)$. Since the discovered forest is small,
the expected numbers of excess edges are small, and Chernoff bounds
control them.

We bound the expected number of excess edges of each kind by the number
of pairs that could produce them, times $p$. By (D1), the discovered
forest has at most $n^{1/200}D^L=O(n^{111/200})$ vertices, so Step~3
tests $O(n^{111/100})$ pairs. An early excess edge has both endpoints at
depth at most $h+1$, so by (D2) at most
$O(n^{1/100}D^{2h+4})=O(n^{91/100})$ tested pairs can be early excess
edges. For $w$ at depth $\ell_0$, every excess edge with an endpoint in
$T_w$ is a tested pair with one endpoint in $T_w$; since
$D^{L-\ell_0}=O(n^{11/40})$, (D3) and (D1) bound their number by
$O(n^{1/200+11/40+111/200})=O(n^{167/200})$. Hence, given $\tau$, the expected values of
$|\exc(B)|$, $|\exc_h(B)|$ and $|\exc(T_w)|$ are $O(n^{11/100})$,
$O(n^{-9/100})$ and $O(n^{-33/200})$, respectively.

Each count is a sum of independent indicators, so if its mean is at most
$\mu$, it is at least $u\ge\mu$ with probability at most $(\emm\mu/u)^u$.
There are at most $N_{\max}=n^{7/12}$ vertices at depth $\ell_0$, so a
union bound gives
\begin{align*}
 \Prb_p\big(\mathcal G_{\rm exc}^c\mid\tau \big)
 &\le\big(O(n^{-3/200})\big)^{n^{1/8}}
   +O\big(n^{-9C_2/100}\big)
   +n^{7/12}\,O\big(n^{-33C_2/200}\big),
\end{align*}
using $u=n^{1/8}=n^{25/200}$ for the total count and $u=C_2$ for the
other two. For $C_2\ge12(K+2)$ every term is $O(n^{-K-2})$. Averaging
over $\tau$ in $\mathcal G_{\rm disc}$
proves~\ref{it:forest-excess}.
\end{proof}

\section{Extension to other models}
\label{sec:extension-other-models}

Our approach suggests extensions to other sparse models that exhibit symmetry (uniform
single-site marginals). Examples of interest here are NAE-SAT and the antiferromagnetic Ising model. The main ingredients involved in obtaining a sampling algorithm are the exact weighted local
insertion rule, cancellation of first-order boundary effects around the
uniform marginals, and quantitative non-reconstruction estimates that establish strong decay to control the accumulated score fluctuations over linearly
many insertions. The first two ingredients remain true for both models; it is the last ingredient that requires more attention. In particular, the marginal bias must be $n^{-1/4-\Omega(1)}$ for exploration depths $\approx\frac{1}{2}\log_D n$ where $D$ is the branching factor of the relevant tree. At these depths, early excess edges are unlikely, while the effect of deeper excess edges can be controlled by stability estimates.

As for colourings, the decay is strong enough in the case of NAE-SAT, and a suitable adaptation of our approach gives a sampling algorithm up to non-reconstruction. For the antiferromagnetic Ising model, the decay becomes too slow near the reconstruction threshold for the present analysis, and further ideas appear to be needed to resolve the full non-reconstruction range. We discuss the similarities and differences with colourings in more detail below.

\paragraph{NAE-SAT.}
Consider a random signed NAE-$k$-SAT formula $\Phi$ with
$m=\lfloor\alpha n\rfloor$ independent uniform clauses, each involving $k$
distinct variables, and let $\mu_\Phi$ be uniform on its satisfying
assignments. As noted in Remark~\ref{rem:nae}, for sufficiently large fixed $k$, our approach can be adapted
to nearly uniform sampling from $\mu_\Phi$ at every fixed density
$0<\alpha<\alpha_{\mathrm{rec}}(k)$, where
$\alpha_{\mathrm{rec}}(k)=(1+o_k(1))2^{k-1}\log k/k$ is the reconstruction
threshold~\cite{MRT}. The argument uses clause insertions and helper clauses
in place of edges, together with analogous moment, cut and solution-count
estimates; we omit the full adaptation from this version and focus on
the decay estimate.

On the factor tree with Poisson$(k\alpha)$ child clauses per variable
and $k-1$ child variables per clause, once the marginals have small bias from uniform, the decay towards uniform is governed by the linearised second-moment
coefficient $O(k\log k/2^k)$ which is bounded away from 1. The analogue of
Lemma~\ref{lem:moments} holds with
$\hat\zeta=\hat D^{-19/20}$, where
$\hat D=k^2 2^k$ bounds the mean number $k\alpha(k-1)$ of child
variables. Taking
$L=\lfloor\frac{11}{20}\log_{\hat D}n\rfloor$ gives $\hat\zeta^{L}\leq n^{-26/50}$.

The probability that the inserted clause is satisfied deviates from
$1-2^{1-k}$ (the value at uniform marginals) quadratically in the biases of its variables, by the same
argument as for~\eqref{eq: bilinear form} in the proof of
Lemma~\ref{lem:localreduction}. Its expected squared deviation is therefore
controlled by fourth moments, and $\hat \zeta^{2L}$ in particular. The latter is smaller than $1/n^{1+\Omega(1)}$, allowing therefore for a linear number of clause insertions.

\paragraph{Antiferromagnetic Ising.}
For the zero-field antiferromagnetic Ising model on $G(n,d/n)$, fix $d>1$
and write the edge interaction as $1-\vartheta\sigma_u\sigma_v$ with
$\sigma_u,\sigma_v\in\{-1,+1\}$ and $0<\vartheta<1$. On the Poisson$(d)$ tree, the strict
non-reconstruction region is $\tilde\zeta=d\vartheta^2<1$~\cite{EKPS}.
For fixed $d,\vartheta$, the expected squared distance of the
root marginal from uniform, given a random boundary at depth $L$
under the free Ising measure, is $\Theta(\tilde\zeta^L)$.
As for NAE-SAT, the conditional expected edge interaction differs
from $1$ (its value at uniform marginals) quadratically in the
endpoint biases, so in the corresponding calculation on two
independent trees its variance is $\Theta(\tilde\zeta^{2L})$.
The key point here is that $\tilde\zeta$ is nearly $1$ close to the reconstruction threshold, therefore giving variance at $L=\frac{1}{2}\log_d n$ that is too large for the present analysis to accommodate $\Theta(n)$ insertions. Hence, covering the full Ising non-reconstruction region appears to require
a different way to control the accumulated weights.

\bibliographystyle{alphaurl}
\bibliography{references}
\appendix

\section{Proofs of Lemmas~\ref{lem:countlower} and~\ref{lem:psimonotonicity}}\label{sec:countproof}
For integers $0\le m\le\binom n2$, let $\cG(n,m)$ denote the uniform
distribution on simple graphs on $[n]$ with exactly $m$ edges.
We use the following part of the colour-count concentration theorem
of Bapst, Coja-Oghlan and Efthymiou.

\begin{theorem}[{\cite[Theorem~1.1]{BCE}}]\label{th:countconcentration}
Fix an integer $q\ge3$ and a real number
$0<d'\le2(q-1)\log(q-1)$. Let $m=\lceil d'n/2\rceil$ and
$\rG\sim\cG(n,m)$. For every function $\omega(n)\to\infty$,
\[
 \Prb\left(\emm^{-\omega(n)}\E Z_{\rG}\le Z_{\rG}
       \le \emm^{\omega(n)}\E Z_{\rG}\right)=1-o(1)
 \qquad\text{as }n\to\infty.
\]
\end{theorem}

\begin{proof}[Proof of Lemma~\ref{lem:countlower}]
The edge-count and maximum-degree bounds follow from standard
binomial tail estimates. We hence focus on bounding $\Psi(\rG)$ and $\hat Z_{\rG}$.

We start with the statistic $\Psi(\rG)$, defined in~\eqref{eq:graphgeometry}.
Recall that $D=d+3$, $L=\lfloor\frac{11}{20}\log_D n\rfloor$
and $\lambda=D^{-19/20}$. Let $\rRoots$ be a uniform set of four distinct roots in $\rG\sim G(n,d/n)$, chosen independently, and consider the exploration $B_{\rG,L}(\rRoots)$.
For $0\le\ell<L$, the number of vertices discovered through depth $\ell+1$ has second moment $O(D^{2\ell})$, by domination by four Poisson$(D)$ branching processes. Both endpoints of every excess edge of depth $\ell$ lie among these vertices.
Conditional on the discovery phase, each unqueried pair is an edge with probability $d/n$. Hence the expected number of excess edges of depth $\ell$ is $O(D^{2\ell}/n)$. Consequently,
\begin{equation}\label{eq:geometrymean}
 \E\Psi(\rG)\le\frac{C}{n}
       \sum_{\ell=0}^{L-1}(D^2\lambda^4)^{\ell}=O(1/n).
\end{equation}
Markov's inequality gives $\Psi(\rG)\le(\log n)^{1/4}/n$ with failure probability $O((\log n)^{-1/4})$.

Recall that $\hat Z_G=Z_G/(q^n\rho^{|E(G)|})$,
where $\rho=\frac{n}{n-1}(1-\frac1q)$.
To prove the bound on the normalised partition function $\hat Z_{\rG}$, we will use the exact-edge model $\cG(n,\boldsymbol{m})$ for $\boldsymbol{m}\sim\operatorname{Bin}(\binom n2,d/n)$ and average appropriately over the relevant values of $\boldsymbol{m}=m$.  

In particular, consider the random graph process $\rG_t\sim\cG(n,t)$ for $0\leq t\leq\binom n2$, where $\rG_{t+1}$ is obtained from $\rG_t$ by adding a uniformly random edge from $\binom{V}{2}\backslash E(\rG_t)$. With $M=\lceil(d+1)n/2\rceil$, we will apply Theorem~\ref{th:countconcentration}
 to $\rG_M$ with $d'=d+1$ (note that its conditions are satisfied for $q\geq d/\log d$ and $d$ large), so we just need to estimate $\E Z_{\rG_{M}}$. This is standard, but we give the details for completeness.  For a uniform random colouring $\rsigma$ and integer $0\le m\le M$, we have 
\begin{equation}\label{eq:expect13}
 \E_{\rG_m}\hat Z_{\rG_m}
=\E_{\rsigma}\bigg[\frac1{\rho^m}\prod_{t=0}^{m-1}
       \frac{N_b(\rsigma)-t}{\binom n2-t}\bigg]\le1.
\end{equation}
For the lower bound, consider the contribution to \eqref{eq:expect13} by colourings $\rsigma=\sigma$ that satisfy $\max_i|n_i(\sigma)-\frac{n}{q}|\le\sqrt{2n}$. Chebyshev's inequality and a union bound give $\geq q^n/2$ such colourings $\sigma$; for each of them,  we have
$N_b(\sigma)=\rho\binom n2-\frac12\sum_i(n_i(\sigma)-\frac{n}{q})^2
 =\rho\binom n2-O(n)$. Hence, for $0\le t<m\le M$, we can bound
$\frac{N_b(\sigma)-t}{\rho(\binom n2-t)}
 \ge1-O(1/n)$, and the normalised product in \eqref{eq:expect13} is bounded below by a positive constant, so 
$\E_{\rG_m}\hat Z_{\rG_m}=\Theta(1)$ uniformly over $m\le M$.
Taking $\omega(n)=\tfrac14\log\log n$ in
Theorem~\ref{th:countconcentration} therefore gives
\[
 \Prb\left(\hat Z_{\rG_M}<1/{\sqrt{\log n}}\right)=o(1).
\]

We next obtain a lower bound on $\hat Z_{\rG_m}$ uniformly over $m\le M$.
The process $\{\hat Z_{\rG_t}\}_{t=0}^M$ is a nonnegative supermartingale satisfying $\E[\hat Z_{\rG_{t+1}}\mid \rG_t]\leq \hat Z_{\rG_t}$. Indeed, when $\rG_t=G_t$ for some $q$-colourable $G_t$, using that for any proper colouring $\sigma\in\Omega_{G_t}$ the fraction of bichromatic unused pairs satisfies $b(G_t,\sigma)\leq\rho$ by~\eqref{eq:bupper}, we have 
\[
 \E[\hat Z_{\rG_{t+1}}\mid \rG_t=G_t]
 =\hat Z_{G_t}\frac{\E_{\rsigma\sim\mu_{G_t}}[b(G_t,\rsigma)]}{\rho}\le \hat Z_{G_t}.
\]
When $G_t$ is uncolourable, then instead $\hat Z_{G_t}=\hat Z_{\rG_{t+1}}=0$. 

For any $z\in(0,1)$ and integer $0\le m\le M$, we therefore obtain by iterating that $\E[\hat Z_{\rG_M}\mid\rG_m]\le\hat Z_{\rG_m}$. It follows that $\E\big[\hat Z_{\rG_M}\one\{\hat Z_{\rG_m}<z\}\mid \rG_m\big]\leq z$, so $\E\big[\hat Z_{\rG_M}\one\{\hat Z_{\rG_m}<z\}\big]\leq z$ as well.  Markov's inequality then gives
\[
 \begin{aligned}
 \Prb\big(\hat Z_{\rG_m}<z\big)
 &\le\Prb\big(\hat Z_{\rG_M}<\sqrt z\big)
   +\Prb\big(\hat Z_{\rG_M}\ge\sqrt z,\,\hat Z_{\rG_m}<z\big)\\
   &\leq \Prb\big(\hat Z_{\rG_M}<\sqrt z\big)+\frac1{\sqrt z}\E\big[\hat Z_{\rG_M}\one\{\hat Z_{\rG_m}<z\}\big]\\
 &\le\Prb\big(\hat Z_{\rG_M}<\sqrt z\big)+\sqrt z.
 \end{aligned}
\]
Taking $z=1/\log n$ gives $\Prb(\hat Z_{\rG_m}<1/\log n)=o(1)$ uniformly over $m\le M$.
Finally, let $\rv{m}\sim\operatorname{Bin}(\binom n2,d/n)$ be independent
of this graph process. Then $\rG_{\rv{m}}\sim G(n,d/n)$ and
$\Prb(\rv{m}>M)=o(1)$. Averaging the uniform bound over $\rv{m}$ proves the lemma.
\end{proof}

\begin{lempsimonotonicity}
\statelempsimonotonicity
\end{lempsimonotonicity}
\begin{proof}
Fix a root set $\mathcal R\in\binom V4$. For $0\le\ell<L$, let
\[
 N_G(\ell)=\big|\{f\in\exc(B_{G,L}(\mathcal R)):
                 \operatorname{depth}(f)\le\ell\}\big|.
\]
Define similarly $N_H(\ell)$. 
We first show that $N_H(\ell)\le N_G(\ell)$ for every $\ell$.

Consider first $N_G(\ell)$. The discovery phase of $B_{G,L}(\mathcal{R})$ is breadth-first, so discovery depth equals distance
from $\mathcal R$. Consider the graph  on vertices at $G$-distance at most
$\ell+1$ from $\mathcal R$, keeping only edges with an endpoint at distance
at most $\ell$. Then, let $\hat G$ be the multigraph obtained by merging the four roots into one vertex, keeping all loops
and parallel edges. The discovery edges through depth $\ell+1$ form a
spanning tree of the resulting multigraph. The remaining edges are exactly
the excess edges of depth at most $\ell$, so $N_G(\ell)$ equals the cycle
rank of $\hat G$, namely its number of edges minus its number of vertices plus one.

We define similarly $\hat H$, so that $N_H(\ell)$ equals the cycle
rank of $\hat H$. Adding edges cannot increase distances from $\mathcal R$, so $\hat H$ is a subgraph of  $\hat G$. Cycle rank cannot decrease under this inclusion: both
multigraphs are connected, so each additional vertex requires at least
one additional edge. This proves $N_H(\ell)\le N_G(\ell)$. Summation by parts then gives
\[
 \sum_{f\in\exc(B_{G,L}(\mathcal R))}\lambda^{4\operatorname{depth}(f)}
 =\lambda^{4(L-1)}N_G(L-1)
  +\sum_{\ell=0}^{L-2}
      (\lambda^{4\ell}-\lambda^{4(\ell+1)})N_G(\ell),
\]
and the same identity holds with $H$ in place of $G$.
All coefficients on the right are nonnegative, since $\lambda\in (0,1)$.
Hence the weighted sum for $H$ is at most the one for $G$.
Averaging over $\mathcal R$ proves $\Psi(H)\le\Psi(G)$.
\end{proof}

\section{Proof of Lemma~\ref{lem:recursion}}\label{sec:recursionproof}
Recall the functions $F=(F_1,\hdots, F_q)$ and $Z$ from Section~\ref{sec:sensitivity}, where  for probability vectors $z_1,\ldots,z_k$ on $[q]$ we have
\begin{equation*}
Z(z_1,\ldots,z_k)=\sum_{i=1}^q\prod_{j=1}^k(1-z_{j,i}),\qquad
 F_i(z_1,\ldots,z_k)
 =\frac{\prod_{j=1}^k(1-z_{j,i})}{Z(z_1,\ldots,z_k)}.\tag{\ref{eq:colouringrecursion}}
\end{equation*}
Recall also that in the case where $Z(z_1,\ldots,z_k)=0$,
we set $F(z_1,\ldots,z_k)=\unif$.
We now give the deferred proof of Lemma~\ref{lem:recursion}, which we restate here for convenience.
\begin{lemrecursion}
\statelemrecursion
\end{lemrecursion}
\begin{proof}
Throughout this proof, $\|\cdot\|$ denotes the Euclidean norm.

\noindent\textbf{Part (i).}
Let $z_2,\ldots,z_k$ be probability vectors with $\sum_{j\ge2}\|z_j-\unif\|\le2\eta$, let $a_i=\prod_{j\ge2}\frac{1-z_{j,i}}{1-1/q}$, and, for a probability vector $y$ on $[q]$,
use~\eqref{eq:colouringrecursion} to write
\[F^a(y)=F(y,z_2,\ldots,z_k),\mbox{ so that $F^a_i(y)=a_i(1-y_i)/Z^a(y)$ with $Z^a(y)=\sum_ia_i(1-y_i)$.}\]
For each $j\ge2$, the factor $\frac{1-z_{j,i}}{1-1/q}=1-\frac{q}{q-1}(z_{j,i}-\frac1q)$ differs from $1$ by at most $2\|z_j-\unif\|\le1/4$ when $\eta\le1/16$. For each factor apply the inequality $\emm^{-2|t|}\le1-t\le \emm^{|t|}$ (valid for $|t|\le1/2$), and multiply over $j\ge2$ to obtain
\[\emm^{-4\sum_{j\ge2}\|z_j-\unif\|}\le a_i\le \emm^{2\sum_{j\ge2}\|z_j-\unif\|}.\]
Thus $\emm^{-8\eta}\le a_i\le\emm^{4\eta}$ and hence $\varepsilon=\max_i|a_i-1|\le8\eta$. Since $\sum_i(1-y_i)=q-1$ for every probability vector $y$, it follows that $Z^a(y)\ge(\min_ia_i)(q-1)\ge(1-\varepsilon)(q-1)>0$. In particular, $F^a(y)$ is well-defined for every such $y$.
Write $a=(a_i)_{i\in[q]}$ and $\mathbf1=(1,\ldots,1)\in\mathbb R^q$.
For probability vectors $z,z'$ we have $Z^a(z)-Z^a(z')=\langle a,z'-z\rangle=\langle a-\mathbf1,z'-z\rangle$, as $\sum_i(z'_i-z_i)=0$. Hence, for every $i\in[q]$,
\begin{align*}
 F^a_i(z')-F^a_i(z)
 =\frac{a_i(1-z'_i)Z^a(z)-a_i(1-z_i)Z^a(z')}{Z^a(z)Z^a(z')}
 =\frac{F^a_i(z)\langle a-\mathbf1,z'-z\rangle-a_i(z'_i-z_i)}{Z^a(z')}.
\end{align*}
Using $\|F^a(z)\|\le1$, $\|a-\mathbf1\|\le\sqrt q\,\varepsilon$
and $\|(a_i(z'_i-z_i))_{i\in[q]}\|\le(1+\varepsilon)\|z'-z\|$,
we obtain
\[
 \|F^a(z')-F^a(z)\|\le\frac{1+(1+\sqrt q)\varepsilon}{(1-\varepsilon)(q-1)}\|z'-z\|\le\lambda\|z'-z\|,
\]
where the last inequality holds for all $\eta$ sufficiently small, depending only on $q$ and $\lambda$, since $\lambda>1/(q-1)$. This proves part (i).

\medskip\noindent\textbf{Part (ii).}
Fix an even $r\ge4$ and probability vectors $z_1,\ldots,z_k$. If $Z(z_1,\ldots,z_k)=0$, the left-hand side of the desired inequality is zero, so suppose that $Z(z_1,\ldots,z_k)>0$. Set $\theta=-1/(q-1)$, let $x_j=z_j-\unif$, and define
\begin{equation}\label{eq:Qwt466}
 A_i=\prod_{j=1}^k\frac{1-z_{j,i}}{1-1/q}
     =\prod_{j=1}^k(1+q\theta x_{j,i}),\qquad
 W=\frac1q\sum_{i=1}^qA_i
   =\frac{Z(z_1,\ldots,z_k)}{Z(\unif,\ldots,\unif)}.
\end{equation}
Write $A=(A_i)_{i\in[q]}$ and abbreviate $F=F(z_1,\ldots,z_k)$, so~\eqref{eq:colouringrecursion} gives $F=A/(qW)$. For $k=0$, the products are empty and $F=\unif$, so the inequality holds. For $k=1$, we have $W=1$ and $F-\unif=\theta x_1$, which also gives the inequality. Suppose henceforth that $k\ge2$, and relabel the vectors so that $\|x_1\|\ge \|x_2\|\ge\cdots\ge \|x_k\|$.

Let $A^{(1)}$ and $W^{(1)}$ be obtained from $A$ and $W$ by replacing $z_2,\ldots,z_k$ with $\unif$. Then
\begin{equation}\label{eq:recursion-onechild}
 A_i^{(1)}=1+q\theta x_{1,i},\qquad
 W^{(1)}=1,\qquad
 A^{(1)}/q-\unif=\theta x_1.
\end{equation}
Fix $i\in[q]$ and write $a_j=1+q\theta x_{j,i}$, so that $A_i=a_1\cdots a_k$ and $A_i^{(1)}=a_1$. Since $a_j=\frac{q}{q-1}(1-z_{j,i})$, we have $0\le a_j\le q/(q-1)$, and $|a_j-1|\le q|\theta|\|x_j\|\le q|\theta|\|x_2\|$ for $j\ge2$. Hence, writing the difference as a telescoping sum and using the triangle inequality,
\[
 |A_i-A_i^{(1)}|
 =\Bigl|\sum_{m=2}^k\bigl(a_1\cdots a_m-a_1\cdots a_{m-1}\bigr)\Bigr|
 \le\sum_{m=2}^k a_1\cdots a_{m-1}|a_m-1|
 \le(k-1)\Bigl(\frac{q}{q-1}\Bigr)^{k-1}q|\theta|\|x_2\|.
\]
We also have $\|A-A^{(1)}\|\le\sqrt q\max_i|A_i-A_i^{(1)}|$ and $|W-1|=|W-W^{(1)}|\le\max_i|A_i-A_i^{(1)}|$. Since  $q|\theta|=q/(q-1)$, it follows that with $K=2(\sqrt q+1)q/(q-1)$, it holds that
\begin{equation}\label{eq:recursion-productchange}
 \|A-A^{(1)}\|+|W-1|\le K^{k+1}\|x_2\|.
\end{equation}
Since $WF=A/q$ and $A^{(1)}/q-\unif=\theta x_1$ by~\eqref{eq:recursion-onechild}, we have
\[
 W^{1/r}(F-\unif)-\theta x_1
 =\frac{A-A^{(1)}}q+(W^{1/r}-W)F+(1-W^{1/r})\unif.
\]
Since $W^{1/r}$ lies between $W$ and $1$, we have
$|W^{1/r}-W|+|1-W^{1/r}|=|W-1|$.
Both $F$ and $\unif$ are probability vectors, so their Euclidean norms are at most $1$.
The triangle inequality and~\eqref{eq:recursion-productchange} therefore give
\begin{equation}\label{eq:recursion-singlechild}
 \begin{aligned}
 \|W^{1/r}(F-\unif)-\theta x_1\|
 &\le\tfrac1q\big\|A-A^{(1)}\big\|+|W-1|\le K^{k+1}\|x_2\|.
 \end{aligned}
\end{equation}
By convexity, $(a+b)^r\le2a^r+(2r)^r b^r$ for all $a,b\ge0$.
Thus the triangle inequality and~\eqref{eq:recursion-singlechild}  give
\begin{align*}
 W\|F-\unif\|^r
 &\le2|\theta|^r\|x_1\|^r+(2rK)^{r(k+1)}\|x_2\|^r\\
 &\le2|\theta|^r\sum_{v\in[k]}\|x_v\|^r
   +(2rK)^{r(k+1)}\sum_{v,w\in[k];\,v\ne w}
          \|x_v\|^{r-2}\|x_{w}\|^2.
\end{align*}
Here the second inequality uses
$\|x_2\|^r\le\|x_1\|^{r-2}\|x_2\|^2$.
Substituting $x_v=z_v-\unif$, $|\theta|=1/(q-1)$ and the
expression for $W$ from \eqref{eq:Qwt466} proves part~(ii) with $C=(2rK)^r$.
\end{proof}

\end{document}